\documentclass[a4paper,fleqn,usenatbib]{mnras}

\usepackage{color,soul}
\usepackage{deluxetable}
\usepackage{amsmath,amssymb}
\usepackage{graphicx}
\usepackage{url}
\usepackage[T1]{fontenc}
\usepackage{ae,aecompl}

\newcommand{\ie}{{\it i.e.}}
\newcommand{\eg}{{\it e.g.}}
\newcommand{\be}{\begin{equation}}
\newcommand{\ee}{\end{equation}}

\newcommand{\ferengi}{\textsc{ferengi}}
\newcommand{\hst}{\textit{HST}}
\newcommand{\hubble}{\textit{Hubble Space Telescope}}
\newcommand{\subaru}{\textit{Subaru}}
\newcommand{\sextractor}{\textsc{SExtractor}}
\newcommand{\galapagos}{\textsc{Galapagos}}
\newcommand{\galfit}{\textsc{GALFIT}}

\newcommand{\sersic}{S\'{e}rsic}

\definecolor{titlecol3}{rgb}{1,0,0}
\newcommand{\ffeatures}{$f_{\rm features}$}
\newcommand{\ffeaturesz}{$f_{\mathrm{features,}z}$}
\newcommand{\ffeaturesrest}{$f_{\mathrm{features,}z=0.3}$}
\newcommand{\ffeaturesdebiased}{$f_{\mathrm{features,debiased}}$}
\newcommand{\fbest}{$f_{\mathrm{features,best}}$}
\newcommand{\fodd}{$f_{\mathrm{odd}}$}
\newcommand{\fbar}{$f_{\mathrm{bar}}$}

\newcommand{\fsmooth}{$f_{\rm smooth}$}
\newcommand{\fartifact}{$f_{\rm artifact}$}

\newcommand{\zsim}{$z_{\mathrm{sim}}$}

\newcommand{\Bband}{$B_{435W}$}
\newcommand{\Vband}{$V_{606W}$}
\newcommand{\iband}{$i_{775W}$}
\newcommand{\Iband}{$I_{814W}$}
\newcommand{\zband}{$z_{850LP}$}

\newcommand{\main}{\texttt{main}}
\newcommand{\faded}{\texttt{faded}}
\newcommand{\recolored}{\texttt{recoloured}}
\newcommand{\goods}{\texttt{goods-shallow}}
\newcommand{\stripe}{\texttt{stripe-82-single}}
\newcommand{\coadd}{\texttt{stripe-82-coadd}}
\newcommand{\redshifted}{\texttt{redshifted}}
\newcommand{\simagn}{\texttt{simulated-agn}}

\title[Galaxy Zoo: Hubble]{Galaxy Zoo: Morphological Classifications for 120,000 Galaxies in HST Legacy Imaging}

\author[Willett et al.]{
Kyle W. Willett$^{1,2}$,
Melanie A. Galloway$^{1}$,
Steven P. Bamford$^{3}$,
Chris J. Lintott$^{4}$,
\newauthor
Karen L. Masters$^{5,6}$,
Claudia Scarlata$^{1}$,
B.D. Simmons$^{4,7}$\footnote{Einstein Fellow},
Melanie Beck$^{1}$,
\newauthor
Carolin N. Cardamone$^{8}$,
Edmond Cheung$^{9}$,
Edward M. Edmondson$^{5}$,
Lucy F. Fortson$^{1}$,
\newauthor
Roger L. Griffith$^{10,11}$,
Boris H\"au{\ss}ler$^{4,12,13}$,
Anna Han$^{14}$,
Ross Hart$^{3}$,
\newauthor
Thomas Melvin$^{5}$,
Michael Parrish$^{15}$,
Kevin Schawinski$^{16}$,
R.J. Smethurst$^{4}$,
\newauthor
Arfon M. Smith$^{4,15,17}$
\\
$^{1}$School of Physics and Astronomy, University of Minnesota, 116 Church St. SE, Minneapolis, MN 55455, USA \\
$^{2}$Department of Physics and Astronomy, University of Kentucky, 505 Rose St., Lexington, KY 40506, USA \\
$^{3}$School of Physics and Astronomy, The University of Nottingham, University Park, Nottingham, NG7 2RD, UK \\
$^{4}$Oxford Astrophysics, University of Oxford, Denys Wilkinson Building, Keble Road, Oxford, OX1 3RH, UK \\
$^{5}$Institute for Cosmology and Gravitation, University of Portsmouth, Dennis Sciama Building, Burnaby Road, Portsmouth, PO1 3FX, UK \\
$^{6}$SEPnet, South East Physics Network, UK \\
$^{7}$Center for Astrophysics and Space Sciences, Department of Physics, University of California, San Diego, CA 92093, USA \\
$^{8}$Department of Mathematics and Science, Wheelock College, Boston, MA 02215, USA \\
$^{9}$Kavli Institute for the Physics and Mathematics of the Universe (WPI), \\ Todai Institutes for Advanced Study, The University of Tokyo, Kashiwa 277-8583, Japan \\
$^{10}$Infrared Processing and Analysis Center, California Institute of Technology, Pasadena, CA 91125, USA \\
$^{11}$Department of Astronomy \& Astrophysics, 525 Davey Lab, The Pennsylvania State University, University Park, PA 16802, USA \\
$^{12}$Centre for Astrophysics, Science \& Technology Research Institute, University of Hertfordshire, Hatfield, AL10 9AB, UK \\
$^{13}$European Southern Observatory, Alonso de C\'ordova 3107, Vitacura, Casilla 19001, Santiago, Chile \\
$^{14}$Department of Astronomy, Yale University, New Haven, CT 06520, USA \\
$^{15}$Adler Planetarium, 1300 S. Lake Shore Drive, Chicago, IL 60605, USA \\
$^{16}$Institute for Astronomy, Department of Physics, ETH Z\"urich, Wolfgang-Pauli-Strasse 27, CH-8093 Z\"urich, Switzerland \\
$^{17}$GitHub
}

\date{Accepted XXX. Received YYY; in original form ZZZ}
\pubyear{2016}

\begin{document}
\label{firstpage}
\pagerange{\pageref{firstpage}--\pageref{lastpage}}
\maketitle

\begin{abstract}
We present the data release paper for the Galaxy Zoo: Hubble (GZH)
project.\footnote{This publication has been made possible by the participation
of more than 200,000~volunteers in the Galaxy Zoo project. Their contributions
are individually acknowledged at
\url{http://authors.galaxyzoo.org/authors.html}.} This is the third phase in a
large effort to measure reliable, detailed morphologies of galaxies by using
crowdsourced visual classifications of colour composite images. Images in GZH
were selected from various publicly-released \textit{Hubble Space Telescope}
Legacy programs conducted with the Advanced Camera for Surveys, with filters that probe
the rest-frame optical emission from galaxies out to $z\sim1$.  The bulk of the
sample is selected to have $m_{I814W}<23.5$, but goes as faint as
$m_{I814W}<26.8$ for deep images combined over 5~epochs. The median redshift of
the combined samples is $\langle z\rangle=0.9\pm0.6$, with a tail extending out
to $z\simeq4$. The GZH morphological data include measurements of both bulge-
and disk-dominated galaxies, details on spiral disk structure that relate to
the Hubble type, bar identification, and numerous measurements of clump
identification and geometry.  This paper also describes a new method for
calibrating morphologies for galaxies of different luminosities and at
different redshifts by using artificially-redshifted galaxy images as a
baseline. The GZH catalogue contains both raw and calibrated morphological vote
fractions for $119,849$~galaxies, providing the largest dataset to date
suitable for large-scale studies of galaxy evolution out to $z\sim1$. 
% Number = hubble + goods_shallow + stripe82_coadd
\end{abstract}

\begin{keywords}
galaxies:structure --- galaxies:high redshift --- galaxies:evolution
--- methods:data analysis --- catalogues
\end{keywords}

\section{Introduction}\label{sec:intro}

The morphology of galaxies encodes information on the orbital parameters and
assembly history of their contents, including gas, dust, stars, and the central
black hole. The morphology is also closely related to the local environment of
the galaxy, as mutual interactions such as tides, shocks in cluster
environments, and direct mergers can all change the shape of the galaxy's
gravitational potential. For $M^\star$~galaxies in the local Universe, this
interplay between the physical development of a galaxy and its external
appearance typically manifests at the most basic level as the difference
between bulge-dominated systems with no/little spiral structure (early-types)
and disk-dominated, rotationally-supported galaxies (late-types) frequently
exhibiting spiral arms.  This dichotomy has been used to explore much of the
astrophysics governing galaxy formation and evolution, and has been shown to be
closely linked with other galactic properties such as stellar mass, halo mass,
bolometric luminosity, black hole activity, effective radius, and the relative
ages of the stellar populations.

The advent of larger telescopes sensitive to a full range of observing wavelengths
has revealed that the distribution and properties of galaxy morphology have
strongly evolved over the lifetime of the Universe. At redshift $z\simeq1$
(roughly 6~Gyr after the Big Bang), many galaxies are still in the process of
assembling the baryonic mass required to reproduce the mature, coherent
structures seen in the present day. This growth occurs in a variety of
ways, including accretion of baryons from large-scale galactic filaments
onto halos via streaming, mergers of individual dark matter halos along with
their baryons, conversion of gas into stars through gravitational collapse and
star formation, etc. The process can also be slowed or even reversed via
feedback from stellar winds, supernovae, and active black holes.  Each of these
processes affects the galaxy morphology in different ways, and so an accurate
measurement of the demographics as a function of redshift provides an extremely
powerful observational constraint on the physics involved \citep[for recent
reviews see][]{but13,con14}. 

Theoretical predictions for the morphology of galaxies as a function of
redshift are primarily computed within the $\Lambda$CDM cosmological framework.
Full treatments model gravitational interactions between baryons and dark
matter, hydrodynamics of the gas, and baryonic physics related to star
formation and evolution. The most advanced simulations now span volumes up to
$\sim100$~Mpc$^3$ while simultaneously resolving the smaller ($<1$ kpc) scales
necessary to reproduce the influence of baryonic physics \citep{vog14a,sch15}.
Such simulations predict clustering of galaxies on large scales in a
hierarchical assembly model \citep{sil12}. The structure of individual galaxies
is affected by their merger history \citep{too72,ste02,hop10,kav14a,kav14}, local
environment \citep[\eg, the morphology-density relation;][]{dre80}, initial
dark halo mass, secular evolution rate, and many other factors. Morphologies of
individual simulated high-mass galaxies at $z\sim2-3$ commonly show kpc-scale
``clumpy'' structures, with few galaxies that are either smooth or well-ordered
spirals; asymmetric galaxies with strong density contrasts dominate simulated
populations in the early Universe until at least $z\sim1$ \citep{bel12,gen14}. 

Observational studies of galaxies at high-redshift also display a wide range of
morphological types, many of which are rare or absent at $z\sim0$.  These
include spheroids and disks (akin to the ellipticals and spirals seen in the
local Universe), but also a significant population of massive, more irregular
galaxies, including mergers, tadpoles, chains, double-clumps, and
clump-clusters \citep{elm05,elm07,cam11a,for11a,kar15}.  In contrast, while
grand-design spirals have been observed as far back as $z=2.18$ \citep{law12},
their spatial density suggests that they are exceedingly rare at these high
redshifts, with a very low overall disk fraction \citep{mor13}. Current
observational data thus strongly suggests that the classical Hubble
sequence/tuning fork \citep{hub36} is not a suitable framework for
characterising high-redshift morphology. 

Space-based observatories, particularly the \hubble{} (\hst), have been
responsible for the bulk of imaging studies of high-redshift galaxies.
Observations of fields with very deep imaging
\citep[eg,][]{wil96,gia04,bec06,dav07,sco07,gro11,koe11} give the photometric
sensitivity necessary to detect $L^\star$ galaxies at $z>1$, while also
providing the angular resolution to distinguish internal structure and
characterise the morphology. While these measurements are helped by the fact
that the angular diameter distance is relatively flat beyond $z>1$ in a flat
$\Lambda$CDM cosmology, the relevant angular scales are only of the order
$\sim5-10~\mathrm{kpc}/^{\prime\prime}$ \citep{wri06}. \hst{} can thus resolve
much of the structure for a Milky Way-sized galaxy out to moderately high
redshifts (at least distinguishing a disk from a bulge), but is limited for
more compact structures. Since the size of galaxies evolves as roughly
$r\propto(1+z)^{-1}$ \citep{mao98,law12a}, the compact sizes of high-redshift
galaxies make detailed morphologies a challenge even for \hst{} \citep{che12}.
However, the public availability of more than $10^5$~galaxies in archival
imaging has generated a sample with the potential
for statistically-robust studies of galaxy demographics and evolution. 

One of the major difficulties in studying the morphologies of galaxies lies in
the techniques used for measurement. Visual classification by experts has been
used for many decades \citep[eg,][]{hub26,dev59,san61,van76,nai10,bai11,kar15}.
These methods have the advantage of using the significant processing power of
the human brain to identify patterns, but suffer from issues such as lack of
scaling to large surveys and potential issues with replicability and
calibration (e.g., see \citealt{lah95} for a discussion on the extent to which
eight expert classifiers agree with each other). Automated measurements, both
parametric \citep{pen02a,sim11,lac12} and non-parametric
\citep{abr03,con03,lot04,sca07,bam08,fre13}, scale well to very large sample
sizes, but do not always fully capture the relevant features, especially for
asymmetric galaxies that become increasingly common at high redshifts. The
Galaxy Zoo project \citep{lin08,for12} utilises crowdsourced visual classifications
to measure galaxies in colour-composite images. The efforts of more than
200,000~classifiers allow for multiple independent classifications of
each image which are combined and calibrated to give a distribution of vote
fractions proportional to the probability of a feature being visible. While
crowdsourced data require extensive calibration \citep{bam09,wil13}, they have
a proven reliability and have been used for a wide variety of scientific
studies \citep[eg,][]{lan08,bam09,dar10,mas11c,ski12,sim13,sch14,wil15,sme16}. 

This paper presents the classifications collected from the Galaxy Zoo Hubble
(GZH) project.\footnote{\url{http://zoo3.galaxyzoo.org/}} GZH was the third
phase of Galaxy Zoo, following its initial results classifying
$\sim900,000$ Sloan Digital Sky Survey (SDSS)~images into primarily early/late~types \citep{lin11} and
Galaxy~Zoo~2, which covered a subset of $\sim250,000$~images using a more detailed
classification scheme that included bars, spiral arms, and galactic bulges
\citep{wil13}. GZH used a similarly detailed classification scheme, but focused
for the first time on images of high-redshift galaxies taken with \hst. The
Galaxy Zoo: CANDELS project has also classified morphologies of galaxies at
high redshift, but using ACS and WFC3 rest-frame infrared imaging (Simmons
et~al., submitted).

The sample selection and creation of the images used for GZH is described in
Section~\ref{sec:data}. Section~\ref{sec:interface} describes the GZH interface
and the collection of classifications. Section~\ref{sec:debiasing} outlines the
process used to calibrate and correct the crowdsourced vote fractions for
redshift-dependent bias. Section~\ref{sec:results} gives the main catalogue of
results, with several examples of how the data may be queried in
Section~\ref{sec:cookbook}. Section~\ref{sec:analysis} gives a short overview
of the observed morphological demographics and compares them to several other
catalogues, with a summary in Section~\ref{sec:summary}.

This paper assumes the WMAP9 cosmological parameters of
$(\Omega_m,\Omega_\Lambda,h)=(0.286,0.714,0.693)$ \citep{hin13}.

\section{Sample and Data}\label{sec:data}

The GZH project contains images drawn from a number of different dedicated
surveys and sample selection criteria. The majority of the data (as implied by
the project name) were sourced directly from \hst{} Legacy Surveys, all of which
primarily used imaging from the Advanced Camera for Surveys (ACS). In addition
to \hst{} images, the project uses images from SDSS Stripe~82, as well as
simulated \hst{} images from multiple sources. Below we provide details on each
source of imaging and describe the creation process for images shown to classifiers. 
We also detail the sources of metadata for galaxies in the sample, such as 
photometry and redshifts.

\subsection{Hubble Legacy Surveys}\label{ssec:legacy_surveys}

Image information from
multiple \hst{} surveys were combined into a single photometric and
morphological database, the Advanced Camera for Surveys General Catalog (ACS-GC) by 
\citet{gri12}. A summary of the key parameters of the ACS-GC is given in
Table~\ref{tbl:gzh_numbers}.

\begin{figure*}
\includegraphics[width=160mm]{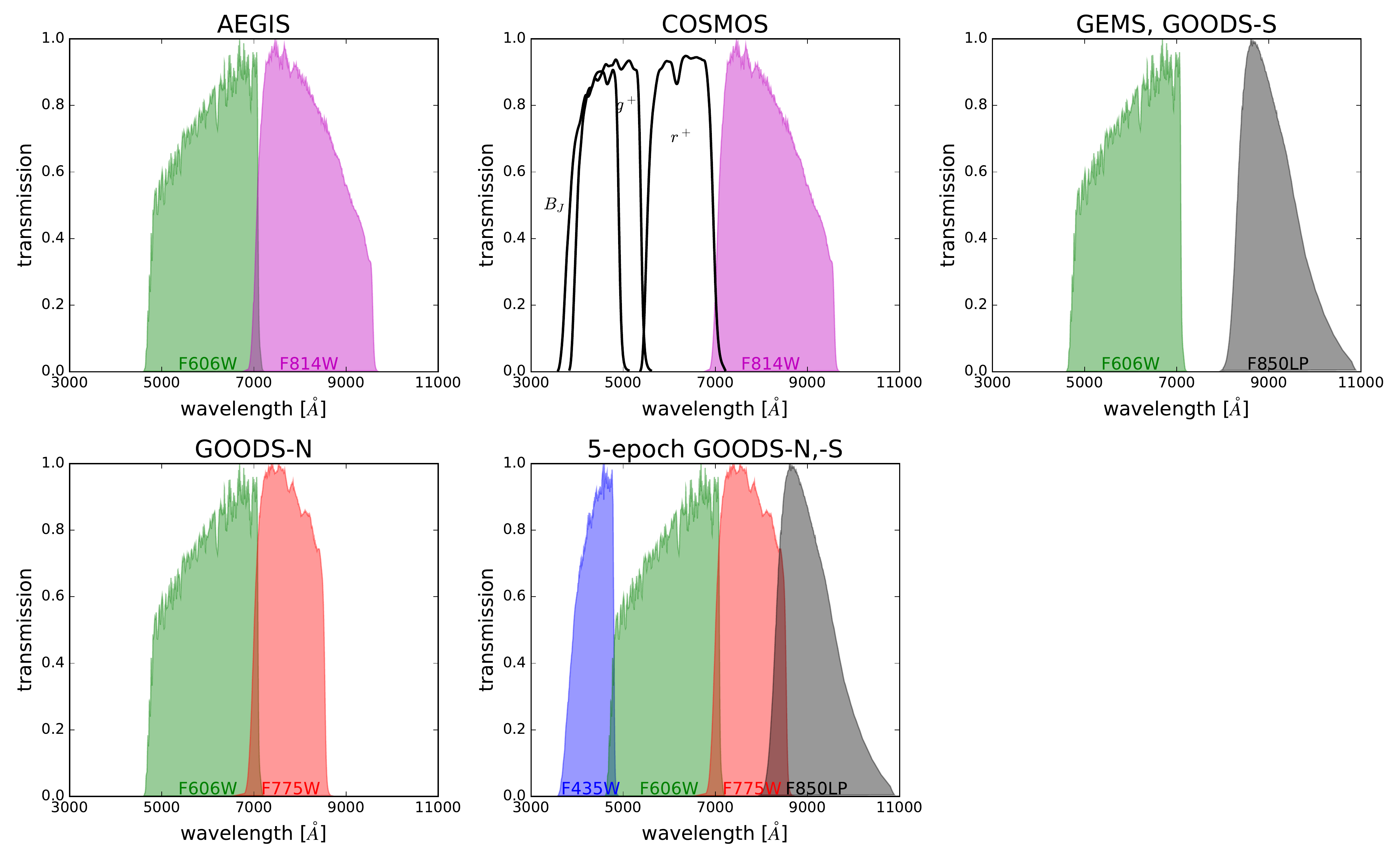}
\caption{Transmission curves of the filters used by \hst{} Advanced Camera for
Surveys (ACS) in wide-field channel mode for the various surveys in GZH. The
unfilled black curves show the filters for the Suprime Camera on \subaru, which
was used to create colour gradients in the GZH COSMOS
images.}
\label{fig:filtercurves}
\end{figure*}

The properties of the individual surveys are as follows:

\begin{itemize}

\item The All-Wavelength Extended Groth Strip International Survey
    \citep[AEGIS;][]{dav07} covers a strip centred at
    $\alpha=14^\textrm{h}17^\textrm{m}, \delta=+52^\circ30^\prime$. This area
    of the sky was selected for a deep survey due to a combination of low
    extinction and low Galactic/zodiacal emission. The ACS images covered 63
    separate tiles over a total area of $\sim710$~arcmin$^2$. The two ACS bands
    for AEGIS had exposure times of 2300~seconds in F606W (\Vband) and
    2100~seconds in F814W (\Iband). The final mosaic images were dithered
    to a resolution of 0.03~\arcsec/pixel. For extended objects, the
    limiting magnitude of sources was 26.23~(AB) in \Vband{} and 25.61 (AB) in
    \Iband. 

\item The Great Observatories Origins Deep Survey \citep[GOODS;][]{gia04}
    covered two separate fields in the Northern and Southern Hemispheres: the Hubble
    Deep Field-North ($\alpha=12^\textrm{h}36^\textrm{m},
    \delta=+62^\circ14^\prime$) and the Chandra Deep Field-South
    ($\alpha=03^\textrm{h}32^\textrm{m}, \delta=-27^\circ48^\prime$). The \hst{}
    ACS imaging data from the two fields are referred to as GOODS-N and GOODS-S,
    respectively. ACS imaging in GOODS fields used 4~filters -- F435W (\Bband),
    \Vband, F775W (\iband), and F850LP (\zband). The mean exposure times for each
    epoch varied by band, from 1050--2100~seconds. The \Bband{} images were completed
    in a single epoch at the beginning of the survey, but the \Vband, \iband, and
    \zband{} images were taken in five separate epochs separated by 40--50~days
    each. GZH includes co-added images from GOODS at both 2-epoch and 5-epoch
    depths. Images were dithered to a pixel scale of 0.03~\arcsec/pixel and covered a
    total area of $\sim320$~arcmin$^2$ (160~arcmin$^2$ each for the north and south fieldss). The
    $5\sigma$ limiting magnitude for extended sources was 25.7 for \Vband{} and
    25.0 for \iband. 

\item The Galaxy Evolution from Morphologies and SEDs
    \citep[GEMS;][]{rix04,cal08} survey was centred on the Chandra Deep
    Field-South. The GEMS data covered $\sim800$~arcmin$^2$, completely surrounding
    the area covered by GOODS-S. Images from ACS in GEMS had 1~orbit per pointing
    for a total of 63~pointings. The exposure times were 2160 and 2286~seconds in
    \Vband{} and \zband{}, respectively. The image resolution had a pixel scale
    of 0.03~\arcsec/pixel. The $5\sigma$ limiting magnitude for source
    detection was 25.7~AB in \Vband{} and 24.2~AB in \zband. 

\item The Cosmic Evolution Survey \citep[COSMOS;][]{sco07,koe07} covered an area of
    $\sim1.8$~deg$^2$ centred at $\alpha=10^\textrm{h}00^\textrm{m},
    \delta=+02^\circ12^\prime$. Its location near the celestial equator was
    designed to enable coverage by ground-based telescopes in both the Northern and
    Southern Hemispheres, in addition to space-based observatories. The ACS data
    for COSMOS consisted of 1~orbit per pointing with an exposure time of
    2028~seconds in \Iband; 590~total pointings were used to cover the entire
    field. The image resolution was dithered to 0.05~\arcsec/pixel. The 50\%
    completeness magnitude for a galaxy with a half-light radius of
    $0.50^{\prime\prime}$ in \Iband{} was 24.7~mag. 

\end{itemize}

%Selection of galaxies from these samples is described in Section \ref{ssec:galselect} below. 
In the ACS-GC, individual galaxies were identified using a
combination of \sextractor{} \citep{ber96} and the galaxy-profile fitting
framework \galapagos{} \citep{bar12}. GZH included all galaxies with $m<23.5$,
where $m$ is in the \Iband, \zband, or \iband{} for the AEGIS \& COSMOS, GEMS \&
GOODS-S (2-epoch), and GOODS-N (2-epoch) surveys, respectively. The full-depth
GOODS images from both fields included galaxies with $m<26.8$. This yielded a
total of $119,849$~images (Table~\ref{tbl:gzh_numbers}).

Images from \hst{} legacy surveys were used to create multiple different colour
and greyscale images using different depths and filter combinations; these are
described further in Section \ref{ssec:images}. Additionally, the use of a
small set of \hst{} images to create simulated images of active galactic
nucleus (AGN) host galaxies is described in Section \ref{ssec:sim_agn}.

% Note: galaxy counts remove the duplicate observations from gz_hst_table

\begin{table*}
\center
\caption{Summary of GZH \hubble{} imaging \label{tbl:gzh_numbers}}
\begin{tabular}{lllcrr}
\hline\hline
Survey &  Total $t_{\rm exp}$ & Filters & Resolution & Area & $N_{\rm galaxies}$ \\
 & [sec] & & [\arcsec/pix] & [arcmin$^2$] & \\
\hline
AEGIS                                   & 2100$-$2300  & \Vband, \Iband{}                 & 0.03 & 710                & 8507    \\
COSMOS                                  & 2028         & \Iband{}                         & 0.05 & 6480               & 84954   \\
GEMS                                    & 2160$-$2286  & \Vband, \zband{}                 & 0.03 & 800                & 9087    \\
GOODS                                   & $-$          & $-$                              & $-$  & $-$                & $-$     \\
\hspace{10pt} \emph{GOODS-N~2~epoch}    & 2100$-$4200  & \Vband, \iband                   & 0.03 & 320                & 2551    \\
\hspace{10pt} \emph{GOODS-S~2~epoch}    & 2100$-$4200  & \Vband, \zband                   & 0.03 & 320                & 3593    \\
\hspace{10pt} \emph{GOODS-N~5~epoch}    & 5100$-$10500 & \Bband, \Vband, \iband, \zband{} & 0.03 & $^{\prime\prime}$  & 6015    \\
\hspace{10pt} \emph{GOODS-S~5~epoch}    & 5100$-$10500 & \Bband, \Vband, \iband, \zband{} & 0.03 & $^{\prime\prime}$  & 5142    \\
\hline
all \hst{} surveys                      & $-$          & $-$                              & $-$  & 8630               & 119849  \\
\hline\hline
\end{tabular}
\end{table*}

\subsection{The Sloan Digital Sky Survey}\label{ssec:sdss}

The GZH project also includes images from the SDSS \citep{yor00,str02}, in particular
those from Data~Release~7 \citep{aba09}. A small number of images were used in
the creation of simulated \hst{} images, described in Sections
\ref{ssec:ferengi} and \ref{sec:debiasing} below. The majority of SDSS images
used in GZH were from Stripe~82, including both
single-epoch and co-added images. The single-epoch images provide a local
sample for comparison to higher-redshift galaxies (including the new
measurements of clumpy structure), while the co-added images allow for analysis
of morphological properties as a function of image depth. 

%The creation of 51,861 images from these sources is described in Section
%\ref{ssec:galselect} below.
Single-epoch images from SDSS Stripe~82 were selected using the criteria
from \citet{wil13}, which required limits of \texttt{petroR90\_r}$ >
3$\arcsec~(where \texttt{petroR90\_r} is the radius containing 90\% of the
$r$-band Petrosian flux) and a magnitude brighter than $m_r < 17.77$.
21,522~galaxies in SDSS met these criteria.
Co-added images from Stripe~82 were selected from the union of galaxies with
co-added magnitudes brighter than $17.77$~mag, and the galaxies detected in the
\stripe{} images and matched to a co-add source. This resulted in a total
set of 30,339~images. Of the images in the co-added sample, 5144 (17~percent)
were dimmer than the initial cut of 17.77~mag.

\subsection{Simulated \hst{} images}\label{ssec:simulatedimages}

To facilitate correction of classifications in the presence of known redshift
bias (Section~\ref{sec:debiasing}), GZH includes two different samples of
simulated \hst{} images: real \hst{} galaxy images (Section
\ref{ssec:legacy_surveys}) with nuclear emission added to simulate AGN host
galaxies, and lower-redshift galaxy images from SDSS (Section \ref{ssec:sdss}),
artificially redshifted from $0.3 \leq z \leq 1$.  Each is described separately
below. 

\subsubsection{Images with simulated nuclear point sources}\label{ssec:sim_agn}

As active galactic nuclei often have bright, unresolved optical emission, AGN
have the potential to mimic or distort the identification of a bulge component.
GZH thus includes a set of images designed to measure the effect of AGN on
morphological classifications.  The presence of an AGN was simulated by
modelling the point spread function (PSF) of the telescope and then inserting a
bright source near the centre of a real galaxy. For each image, the simulated
AGN was assigned one of three colours -- either blue, red, or flat (white) as
seen in the colour images -- and a range of brightnesses such that
$L_\mathrm{ratio} \equiv L_\mathrm{galaxy}/L_\mathrm{AGN}$ is in
$(0.2,1.0,5.0,10.0,50.0)$. Combining these parameters generated 15~images
with different simulated AGN for each host, in addition to the original galaxy
image. 

Two sets of simulated AGN were generated in GZH. The first set (version~1) was
assembled from 95~galaxies from GOODS-S imaging and empirical PSFs made by
combining stars in the GOODS fields using the PSF creation tools in
\texttt{daophot} \citep{ste87}.  The second set (version~2) was assembled from 96~galaxies in
GOODS-S; this version used simulated PSFs from \texttt{TinyTim} \citep{kri93},
drizzled using the same procedures as those used in the reduction of the GOODS-S
images \citep{koe02,koe03,gia04}. The use of these two versions facilitates
comparisons between these different PSF creation methods, which are widely used
in AGN host galaxy morphology studies
\citep[e.g.,][]{san04,sim08,pie10a,simm11}.  Each PSF creation method has
advantages and disadvantages: the empirical PSFs better represent the nuances
of the PSF in the specific data being used and look more realistic at lower
luminosities, but the extended features of the noiseless \texttt{TinyTim} PSFs
are visually more realistic at higher luminosities.
% okay so I'm a cheeky self-citing biyotch - BDS

Images with simulated AGN were classified in the interface in an identical
manner and were evenly distributed with unaltered images of the galaxies.
Volunteers were not explicitly told that the images had been altered during
classification, as the goal was to measure the effect on normal classifications
using the same technique as closely as possible. Following classification, a
classifier could view a page with additional details about each galaxy; where
applicable, these pages contained further information regarding image
modifications. \citet{sim13} used the simulated AGN host galaxy images for
more reliable identification of bulgeless galaxies hosting AGN in SDSS data.

\subsubsection{Generating images of artificially-redshifted galaxies}\label{ssec:ferengi}

The dimming and resolution effects of redshift can significantly affect galaxy
classifications derived from any method. To facilitate corrections due to
redshift, we include a sample of 288~galaxies with SDSS imaging that can be
transformed to simulate \hst{} imaging out to $z = 1$. 
The SDSS images were redshifted and processed to mimic simulated \hst{} imaging 
parameters using the \ferengi{} code \citep{bar08a}.

%The selection criteria for the different morphological categories is
%summarised in Table \ref{tbl:morphologies}. 
 
%\input{morphologies.tex}

\begin{figure*}
\center
\includegraphics[width=160mm]{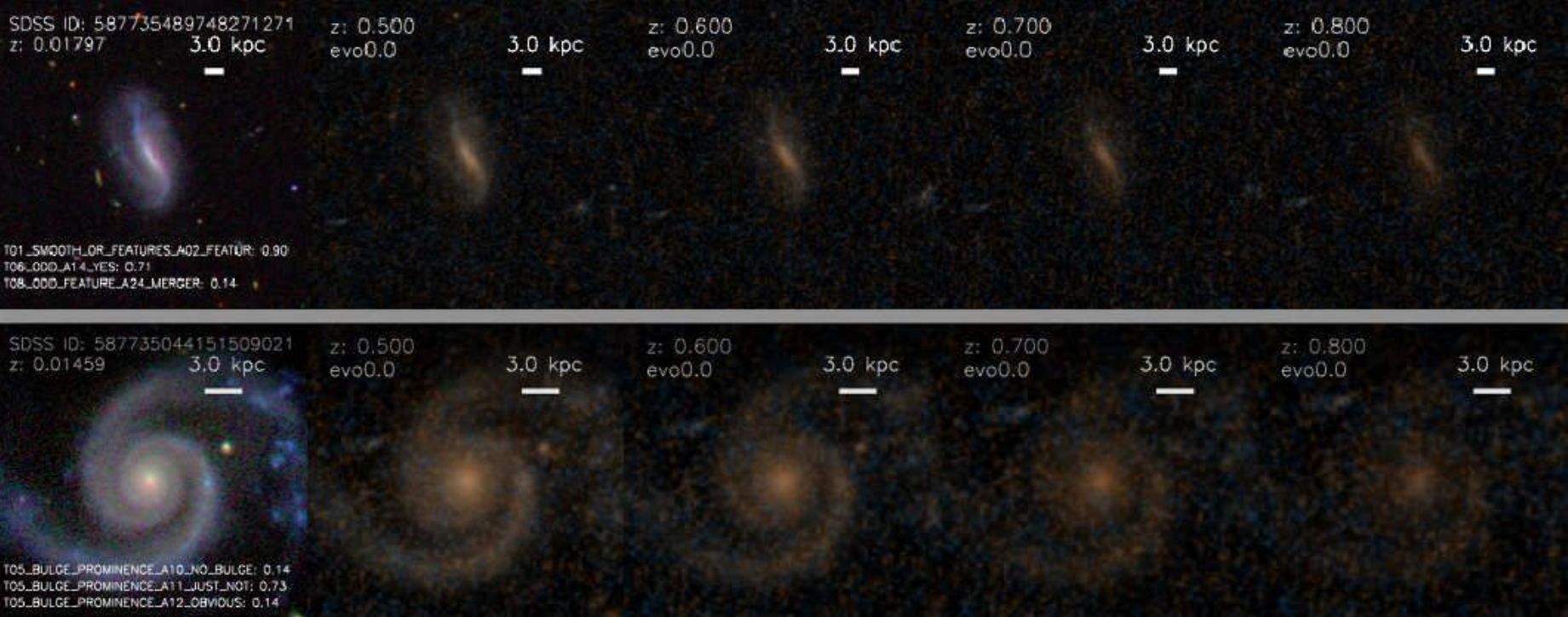}
\caption{Examples of two galaxies which have been run through the \ferengi{}
code to produce simulated \hst{} images. The measured value of \ffeatures{}
from GZH for the images in each panel are (1) Top row: \ffeatures~= (0.900,
0.625, 0.350, 0.350, 0.225) and (2) Bottom row: \ffeatures~= (1.000, 0.875,
0.875, 0.625, 0.375). \label{fig:exampleFERENGI}}
\end{figure*}

In addition to the physical parameters of the input images, the \ferengi{}
output depends on assumptions of the global galaxy evolution model.
This evolution is parameterised by a crude model that mimics the brightness increase of galaxies
with increasing redshift \citep[\eg,][]{lil98,lov12}. The effect on the
redshifted images is simply an empirical addition to the magnitude of a galaxy
of the form $M' = e\times z + M$, where $M'$ is the corrected magnitude, and
$e$ is the evolutionary correction in magnitudes ($e=-1$ essentially
brightens the entire galaxy by 1~magnitude by $z=1$). \ferengi{} was run on the images
for values of $e$ starting from $e=0$ and decreasing to $e=-3.5$ in increments
of $\Delta e = 0.5$. Figure~\ref{fig:exampleFERENGI} shows several examples
of the effects of ``losing'' spiral/disk features with increasing redshift
for two galaxies with no evolution corrections ($e=0$); as the signal-to-noise ratio in the images
decreases and the galaxies become fainter, the contrast between features
like spiral arms goes down and \ffeatures{} drops to the point where both
galaxies would have been classified as likely ellipticals. 

The final number of \ferengi{} images produced for each galaxy is ultimately a
function of the galaxy's redshift (since the new images cannot be resampled at
better angular resolution than the original SDSS data), as well as the number
of $e$ values selected. The use of these images to correct for redshift-dependent
bias is described in Section \ref{sec:debiasing}.

\subsection{Creating colour images}\label{ssec:images}

The images used for classification in GZH were colour-composite JPEGs made from
multi-band data. These were created following the method of \citet{lup04}, which
preserves colour information irrespective of intensity.  An asinh intensity
mapping was applied to enhance the appearance of faint features while avoiding
saturating galaxy centres. The relative scalings of the filter bands were
chosen to reproduce the colour appearance of the SDSS images in
previous iterations of Galaxy Zoo.

Many of the Legacy surveys described in Section~\ref{ssec:legacy_surveys}
provided \hst{} images in only two filters. For these, the shorter-wavelength band was
mapped to the blue channel, the longer-wavelength band to the red channel, and
the green channel created by taking the arithmetic mean of the red and blue. The bands
used in each of the surveys are listed in Table~\ref{tbl:gzh_numbers}.
Although four bands were available for the GOODS survey, only two bands were
used to create the original 2-epoch images, for consistency with AEGIS and
GEMS. The 2-epoch GOODS-N and GOODS-S images were created using differing
filters --- this was a deliberate choice made so that the GEMS images could be
directly compared with the overlapping coverage of GOODS-S
(Figure~\ref{fig:filtercurves}).

Only 2-epoch GOODS images were included at the launch of GZH.  Deeper, 5-epoch 
GOODS images were added into GZH in March 2015. The deeper images made
use of the full four-band data by using the arithmetic mean of \Bband{} and
\Vband{} in the blue channel, \Iband{} in the green channel, and \zband{} in
the red channel.

The COSMOS survey provides only \Iband{} HST imaging. For these galaxies, GZH
used ``pseudocolour'' images created by using the ACS \Iband{} data as an
illumination map and ground-based imaging from the \subaru{} telescope in
$B_J$, $r^+$, and $i^+$ filters to provide colour information (see
\citealt{gri12} for further details). This resulted in images with the angular
resolution of \hst{} ($\sim0.05$~\arcsec/pixel) for the overall intensity, but
colour gradients at ground-based resolution, with seeing between 0.95\arcsec{}
and 1.05\arcsec{} \citep{tan07}.

Stripe~82 single-epoch images were taken directly from the DR7 SDSS SkyServer,
which combined $g^{\prime}$, $r^{\prime}$, and $i^{\prime}$ exposures into the
RGB channels \citep{nie04}. The co-added Stripe~82 images were assembled from runs 106 and
206 in DR7 and processed into colour composites in the same manner as previous
iterations of Galaxy~Zoo.

In some cases, we found that attempting to emphasise faint features in the
images resulted in the sky noise taking the appearance of brightly-coloured
speckles. This impaired the aesthetics of the images and was considered a potential
distraction to visual classification.  To counteract this, a soft-edged object
mask was applied to the colour images and a desaturation operation performed.
This masking procedure was effective in preserving the colour balance for
galaxies and retained the visibility of faint features, while reducing the colour
contrast in the sky noise and greatly improving the appearance of the images.
This solution was applied to the coadded Stripe~82, COSMOS and 5-epoch GOODS
images.

In addition to the primary \hst{} legacy imaging sample with coloured images
described as above, GZH also includes samples of \hst{} galaxy images with
different colour prescriptions. In particular, there are two sets of 3,927~images
each, drawn from the COSMOS sample; the first has a dramatically reduced colour
saturation, and the second has reversed the colours so that the blue and red
filters have exchanged places in the RGB image. The ``faded'' galaxy set
facilitates measuring of possible variations in classification due to the
presence or absence of colour features. The ``recoloured'' set enables an
alternative test of potential colour biases in classifications.
% I guess this means we need to say somewhere later that we don't see an effect
% (we don't, right?).  Would also be good to cite here where we have done
% similar things in the past... - BDS

The simulated AGN host galaxy colour images (described in Section \ref{ssec:sim_agn})
were created using the same prescription as for the GOODS 2-epoch imaging. The
artificially redshifted colour images (Section \ref{ssec:ferengi}) were created using the
same prescription as for the AEGIS images.
% Edmond should verify the AEGIS bit above. - BDS

\subsection{Galaxy sample labels}\label{ssec:sample_labels}

The full GZH sample is composed of 8~different galaxy sub-samples. Throughout
this paper and in the published catalogues they are referred to with the
following labels:

\begin{itemize}

%\ref{tbl:catalog_hst}
\item \main: \hst{} imaging with RGB colours making use of all available
filters with typical saturation and in correct order, as described in
Section~\ref{ssec:images}. This sample includes AEGIS, COSMOS, GEMS, and
full-depth GOODS (North and South) images. (113,705~galaxies)

%\ref{tbl:catalog_faded}
\item \faded: a subset of COSMOS images with very low colour saturation.
(3,927~galaxies)

%\ref{tbl:catalog_recolored}
\item \recolored: a subset of COSMOS images with red and blue channels
reversed. Note: this sub-sample uses the same galaxies as the \texttt{faded}
sub-sample. (3,927~galaxies)

%\ref{tbl:catalog_goods_shallow}
\item \goods: Images from GOODS-North and GOODS-South, with
colour from 2~filters and imaged at 2-epoch depth. (6,144~galaxies)

%\ref{tbl:stripe82_single}
\item \stripe: Single-epoch images from SDSS Stripe~82.
(21,522~galaxies)

%\ref{tbl:stripe82_coadd}
\item \coadd: Co-added images from SDSS Stripe~82. Note: this
sub-sample includes all galaxies in the \texttt{stripe-82-single} sub-sample,
with additional sources detected in the deeper imaging. (30,339~galaxies)

%ferengi <-- no table?
\item \redshifted: Simulated \hst{} images constructed using SDSS
images and artificially processed to redshifts between $0.3 \leq z \leq 1$.
(288~original galaxies; 6,624~redshifted images)

%\ref{tbl:simulated_agn}
\item \simagn: Simulated AGN host galaxies constructed using \hst{} images and
PSFs. (96~original galaxies; 2,961~images of simulated AGN hosts)

\end{itemize}

\subsection{Galaxy metadata}

Photometric data for the bulk of the GZH \main, \faded, \recolored, and
\goods{} samples were largely drawn from the tables in \citet{gri12}. This
included photometric parameters such as the fluxes, magnitudes, radii,
ellipticities, position angles, and positions drawn from both \sextractor{} and
\galfit.  All photometric parameters were measured in both bands of the ACS
imaging, with the exception of the single-band COSMOS images. Photometric data
for the GOODS 5-epoch imaging, including \sextractor{} parameters, is from
\citet{gia12}.
%\galfit{} also provided the parametric \sersic{} index and effective
%half-light radius for the best-fit model. 

Redshifts for the GZH catalogue were compiled from a variety of sources. For each
galaxy, the primary redshift is in the $\tt Z\_BEST$ column of
Table~\ref{tbl:catalog_hst}. The redshift type (spectroscopic: $\tt SPEC\_Z$,
photometric: $\tt PHOTO\_Z$, or grism: $ \tt GRISM\_Z$) is listed in the column
$\tt Z\_BEST\_TYPE$, and the source catalogue of the redshift is included as $\tt
Z\_BEST\_SOURCE$. 

For galaxies which have published redshifts from multiple sources, the
following algorithm was used to select the $\tt Z\_BEST$ quantity. A
high-quality spectroscopic redshift in the ACS-GC is the primary option,
provided in the ACS-GC \citep{gri12}, 3DHST \citep{mom15}, and MUSYC
\citep{car10} catalogues and used in that order. For galaxies with multiple
spectroscopic redshifts, more than 98\% are consistent ($\Delta z<0.001$), and
so the order of selection made no practical difference. Galaxies with
inconsistent spectroscopic redshifts between any pair of catalogues are marked
with a flag in Table~\ref{tbl:catalog_hst}.  If no spectroscopic redshifts were
available, the 1-$\sigma$ errors of the photometric \citep[ACS-GC, 3DHST,
MUSYC, UltraVISTA;][]{ilb13} and UltraVISTA grism data were used. The
measurement with the smallest reported 1-$\sigma$ error was selected in each
case.
%Table~\ref{tbl:redshifts} shows the results of this selection. 
 
%\input{redshifts.tex}

Photometric and spectroscopic data for the \stripe{} and \coadd{} galaxies were taken
from the CasJobs DR7 tables. This included $ugriz$ Petrosian magnitudes and
fluxes, as well as the relative de~Vaucouleurs and exponential fits from the
model magnitudes. All redshifts used for SDSS galaxies were spectroscopic.
82.6\% of galaxies in the \stripe{} images and 65.1\% of galaxies in the
\coadd{} images had a measured DR7 spectroscopic redshift. 

The technique for redshift debiasing (Section~\ref{sec:debiasing}) requires
consistent measurements of the galaxy surface brightness. For both the
\redshifted{} calibration images and the \hst{} images which have their
morphologies corrected, we calculate the mean surface brightness $\mu$ within the
effective radius ($R_e$) as:

\begin{equation}
\mu = m + 2.5*\log_{10}{(2 \times (b/a) \times \pi R_e^2 )}.
\label{eqn:surface_brightness}
\end{equation}

All photometric parameters are taken from \sextractor. For the \ferengi{}
galaxies, $m$ is {\tt MAG\_AUTO} in the \Iband{} band, $(b/a)$ is the galaxy
ellipticity (the profile RMS along the semi-major and -minor axes), and $R_e$
is the 50\% {\tt FLUX\_RADIUS} converted into arcsec \citep{mel16}. For the \hst{} galaxies in the \main{} sample,
the parameters are identical except that we use {\tt MAG\_BEST} instead of
{\tt MAG\_AUTO} \citep{gri12} in either the \Iband, \iband, or \zband{} bands, depending 
on the available imaging (Table~\ref{tbl:gzh_numbers}).

\section{GZH interface and classifications}\label{sec:interface}

Below we describe the classification structure of GZH, including the software 
interface and the hierarchical structure of a classification. Section \ref{ssec:weighting}
describes the process of combining individual classifications into vote fractions for each
galaxy.

\subsection{Interface and decision tree}\label{ssec:interface}

\begin{figure*}
\center
\includegraphics[width=160mm]{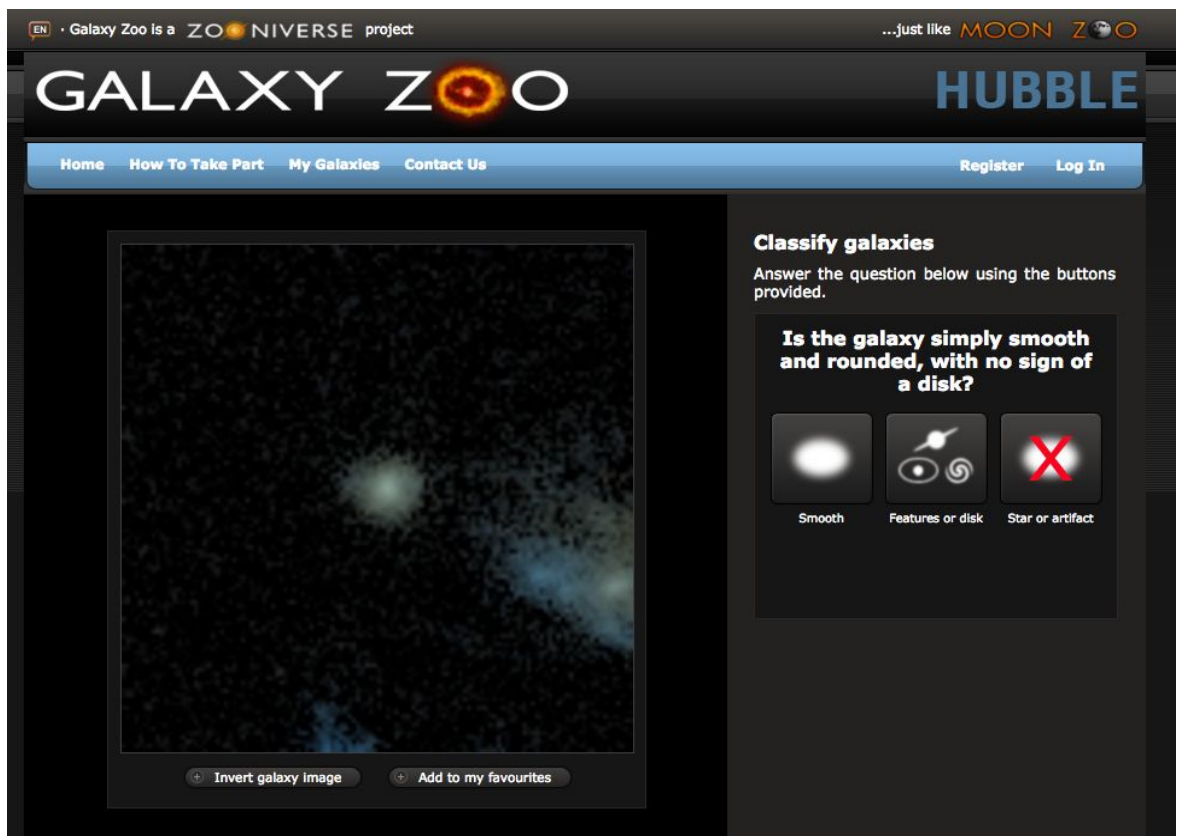}
\caption{Screenshot of the GZH interface at
the beginning of classifying a random galaxy, with the classifier ready
to select an answer for the first question in the decision
tree.\label{fig:interface}}
\end{figure*}

Classifications for GZH were made using a web-based interface
(Figure~\ref{fig:interface}), similar in design to Galaxy~Zoo and Galaxy~Zoo~2.
The front-end runs on a Ruby~on~Rails framework with classifications stored in
a MySQL backend. Classifiers were shown a randomly-selected colour composite image from
the GZH sample; the default showed the image with a black sky background,
although they had the option to invert the colour palette if desired. The
questions and responses for morphology appeared on the right side of the image as
a panel, including both text and icons. There was no tutorial required for
participation, although classifiers could access an extensive ``Help'' section
containing example images and descriptive text for all the morphological
labels. 

\begin{figure*}
\center
\includegraphics[width=\textwidth]{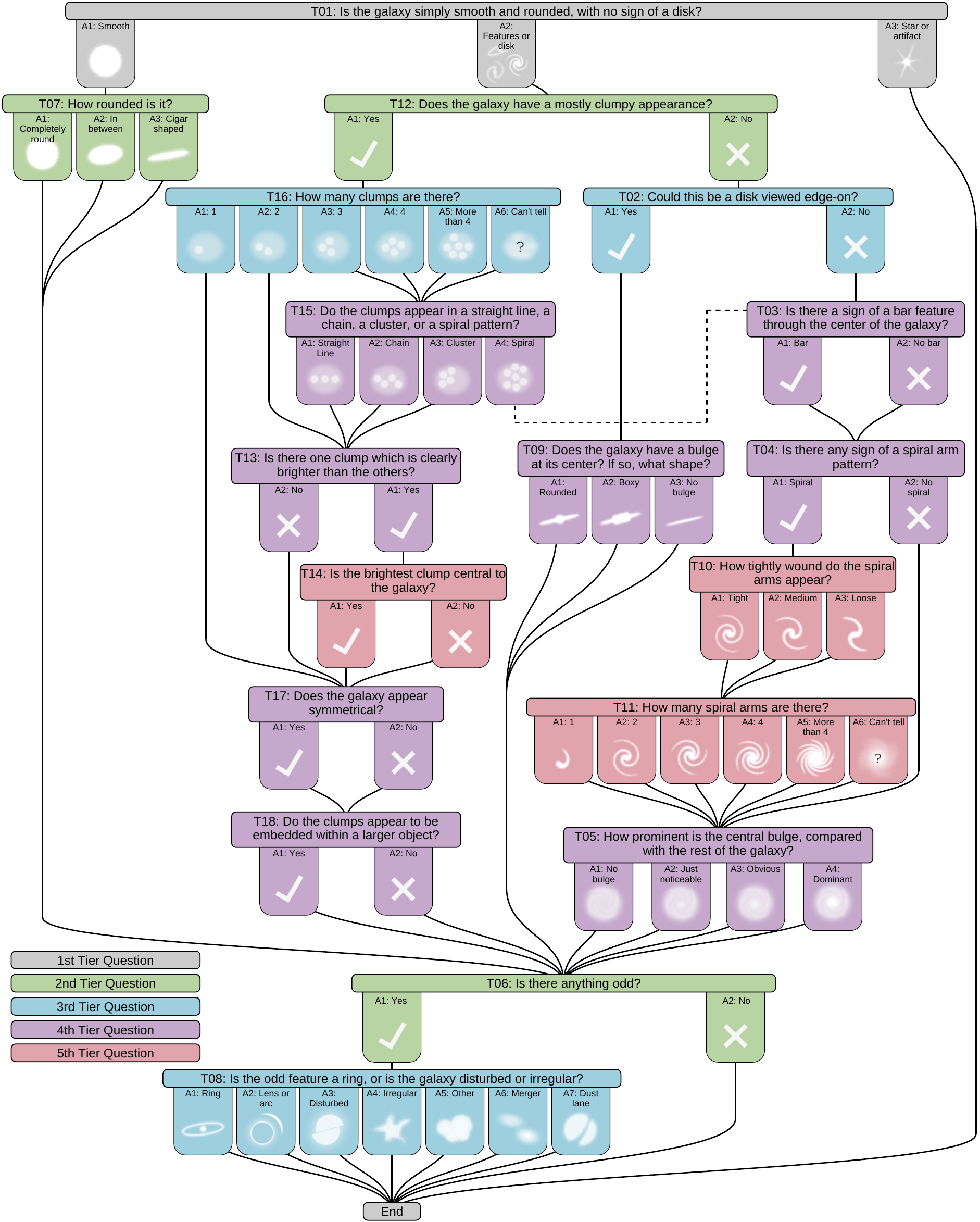}
\caption{Flowchart of the questions presented to GZH classifiers, labeled with
the corresponding task numbers. Tasks in the decision tree are colour-coded by
tier level: Grey-coloured tasks are 1$^\mathrm{st}$-tier questions which are
asked in every classification. Tasks coloured green, blue, purple, and pink
(respectively) are one, two, three, or four steps below branching points in the
decision tree. The dashed line between T15 and T03 indicates the unique case
where volunteers could label both clumpy and disk-like morphologies for
galaxies with clumps arranged in a spiral pattern.}
\label{fig:decisiontree}
\end{figure*}

The procedure for classifying an image in GZH followed a hierarchical decision
tree (Figure~\ref{fig:decisiontree}). Every classification began with the step of
identifying whether the object at the centre of the image was a ``smooth'' galaxy,
a galaxy with a disk or other features, or a star/artifact. Subsequent
questions in the tree depended on the previous answer(s) given by the classifier; the
decision tree was designed so that every question relevant to
the morphology in the process of being identified was answered. Questions that were not
answered were implicitly assumed to be absent in the image --- for example, if
the classifier identified a galaxy as being smooth, they were not asked to count the
number of spiral arms. For every task, the classifier chose a single answer before
continuing to the next question; they also had the option to restart any
classification in-progress. 

The GZH decision tree was designed to be similar to that used by GZ2 while
taking account of the likely differences in the morphologies of high redshift galaxies. There are four
broad sets of morphologies classified by the users in GZH.  The first set
identified stars or image artifacts (the result of either bad data or incorrect
identification of an object as a galaxy by the ACS pipeline); in this case, the
classification process ended and no further questions were asked. The second
set was for ``smooth'' galaxies, intended to select ellipticals/early-types;
volunteers also indicate the relative axial ratio (roundness) for these
galaxies. The third set was for disk/late-type galaxies, which labeled the
features necessary to place a galaxy on the standard Hubble tuning fork (bars,
spiral arm, strength of the central bulge). The final set, which was new in
this phase of Galaxy~Zoo and designed for high-redshift targets, identified
objects dominated by clumpy morphologies. Further annotations for clumpy
galaxies included assessing the number, arrangement, relative brightness, and
location of the clumps within the galaxy. Finally, every classification 
had the option of identifying ``odd'' features within the image; these
labels were for relatively rare $(\lesssim1\%)$ phenomena, including dust
lanes, gravitational lenses, and mergers. 

\begin{figure}
\center
\includegraphics[width=0.5\textwidth]{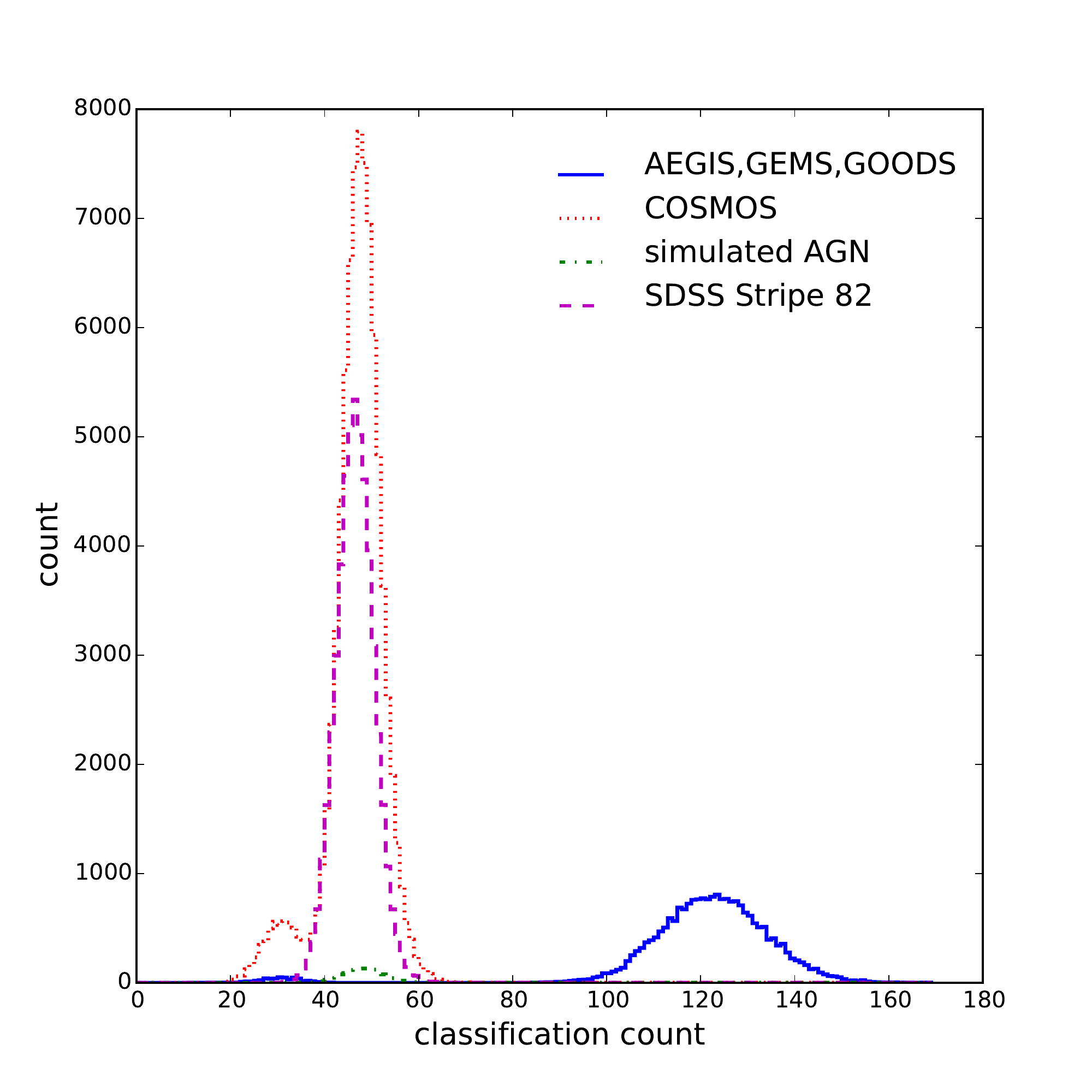}
\caption{Distribution of the total number of classifications per image for GZH,
split by survey.}
\label{fig:classification_hist}
\end{figure}

The number of independent classifications per subject collected by GZH was on
average higher than GZ1 or GZ2, due to both the increased complexity of the
decision tree and the relative difficulty of classifying images of small and
distant galaxies. Images from the \main{} AEGIS, GEMS, and GOODS data sets had
a median of 122~independent classifications per image. The remaining
images had fewer classifications either due to a later activation date (\main{} COSMOS, \simagn) or a lower
retirement limit (\stripe{} and \coadd). Images from these samples had a median
of 46--48~classifications per image (Figure~\ref{fig:classification_hist}).

The GZH project was launched on 23~Apr~2010 with the inclusion of the AEGIS,
GEMS, GOODS 2-epoch, and SDSS Stripe~82 images. Images from COSMOS and the
simulated AGN were activated in Dec 2010, as well as a small sample of images
from AEGIS, GEMS, and GOODS that were previously excluded from the original
sample due to cuts on blended and/or saturated objects and subsequently
confirmed as classifiable galaxies. The GZH site collected data until its
replacement, the fourth phase of
Galaxy~Zoo\footnote{\url{http://zoo4.galaxyzoo.org}} (including data from both
the \hst{} CANDELS survey and SDSS DR8) began on 10~Sep~2012. Classifications
for the GOODS 5-epoch images were separately obtained from Mar-Jun 2015 using
a new version of the Galaxy~Zoo site but the same GZH decision tree.
The GZH project finished with a total of 10,349,357~classifications from
93,898~registered participants. 
% Sum is the total of the GZH SQL table plus the GOODS full-depth imaging in GZ4.

\subsection{Classifier weighting}\label{ssec:weighting}

As a first step to producing a consensus measurement of each galaxy, the votes
of individual volunteers who classified galaxies in GZH were combined to make a
vote fraction for each response ($f_{response}$) to a question in the decision
tree. Votes were subsequently weighted and re-combined in an iterative method
similar to that in previous versions of Galaxy~Zoo \citep{lan08,wil13}, using a
method chosen to be as egalitarian as possible while also identifying and
downweighting classifiers who frequently disagreed with others. The weighting
factor $w$ was 1 for the top 95\% of classifiers as ranked by consistency.
For the bottom 5\% of classifiers, $w$ drops smoothly and is
effectively zero for the bottom 1\% of the distribution function. Since
downweighting only occurs for the bottom few percent of classifiers (and an even smaller
percentage of the classifications), the overall effect on the GZH dataset was
minimal. The method was effective, however, at filtering out contributions from
randomly classifying or malicious participants.

Classifications for GZH were weighted only if the classifier was logged
into the site under their username (this was encouraged, but not required for
participation). Classifications by participants who were not logged in were marked as
``Anonymous'' and receive the same $w=1$ weighting as the vast majority of
logged-in classifiers. Results from the Galaxy Zoo: CANDELS project (Simmons et~al., submitted)
show that the distributions of weights for anonymous and
logged-in users are similar, supporting the default technique of GZH.
% We should also try to estimate how many anonymous classifiers there were, if possible. 
% Not clear that I can do that given the SQL data

\section{Correcting for redshift-dependent classification bias}\label{sec:debiasing}

The previous versions of Galaxy Zoo morphology classifications
\citep{lin11,wil13} were based on observations of galaxies in the SDSS,
which have a median redshift of $z<0.2$. In these cases, it
was assumed that there was no cosmological evolution of the morphologies of
galaxies and therefore any observed changes in the morphological distribution
were due to a redshift-dependent bias that affects image quality and
classification accuracy (\ie, galaxies of a given mass/size will appear smaller
and dimmer at higher redshifts).  This bias is not unique to crowdsourcing
techniques; its dependence in data quality is a potential problem for
\emph{both} automated and visual classifications, and must be addressed in
order to accurately measure demographics over any significant redshift range. 

For both previous releases of Galaxy~Zoo morphologies, a correction for
redshift-dependent bias was applied based on matching the mean classification
fractions at the highest redshifts with those at the lowest redshift.
\citet{bam09} and \citet{wil13} provide complete descriptions of the process
for GZ1 and GZ2, respectively.

Instead, in GZH the redshift range is large enough that cosmological evolution of the
types and morphologies of galaxies is expected for the \hst{} sample. In
addition, the effects of band shifting will change the images even more across
these redshift ranges. As a result, the previous methods of correcting for
redshift-dependent bias do not work.  

In order to test and correct for the effects of redshift, GZH includes a set of
calibration images.  These are simulated images of a set of nearby galaxies as
they would appear observed at a variety of redshifts.  The input images are
from the SDSS \citep{yor00,str02}.  The sample consists of 288~well-resolved
galaxies at $z<0.013$. The galaxies spanned a variety of morphologies (as
selected by GZ2 classifications, including identifications of spiral structure,
ellipticals, mergers, edge-on disks, bulge prominence/shape, and bars) and
$r$-band surface brightnesses.  The selection of galaxies spanned the redshift
range of SDSS targets and maximised the number of \hst{} galaxies at the same
surface brightness and redshift. The \ferengi{} code was used to produce sets
of images corresponding to observations of these galaxies out to
$z_\mathrm{sim}=1.0$ (see Section~\ref{ssec:ferengi}). The resulting images
were classified in the GZH interface using the standard classification scheme. 

\begin{figure}
\begin{center}
\includegraphics[width=0.5\textwidth]{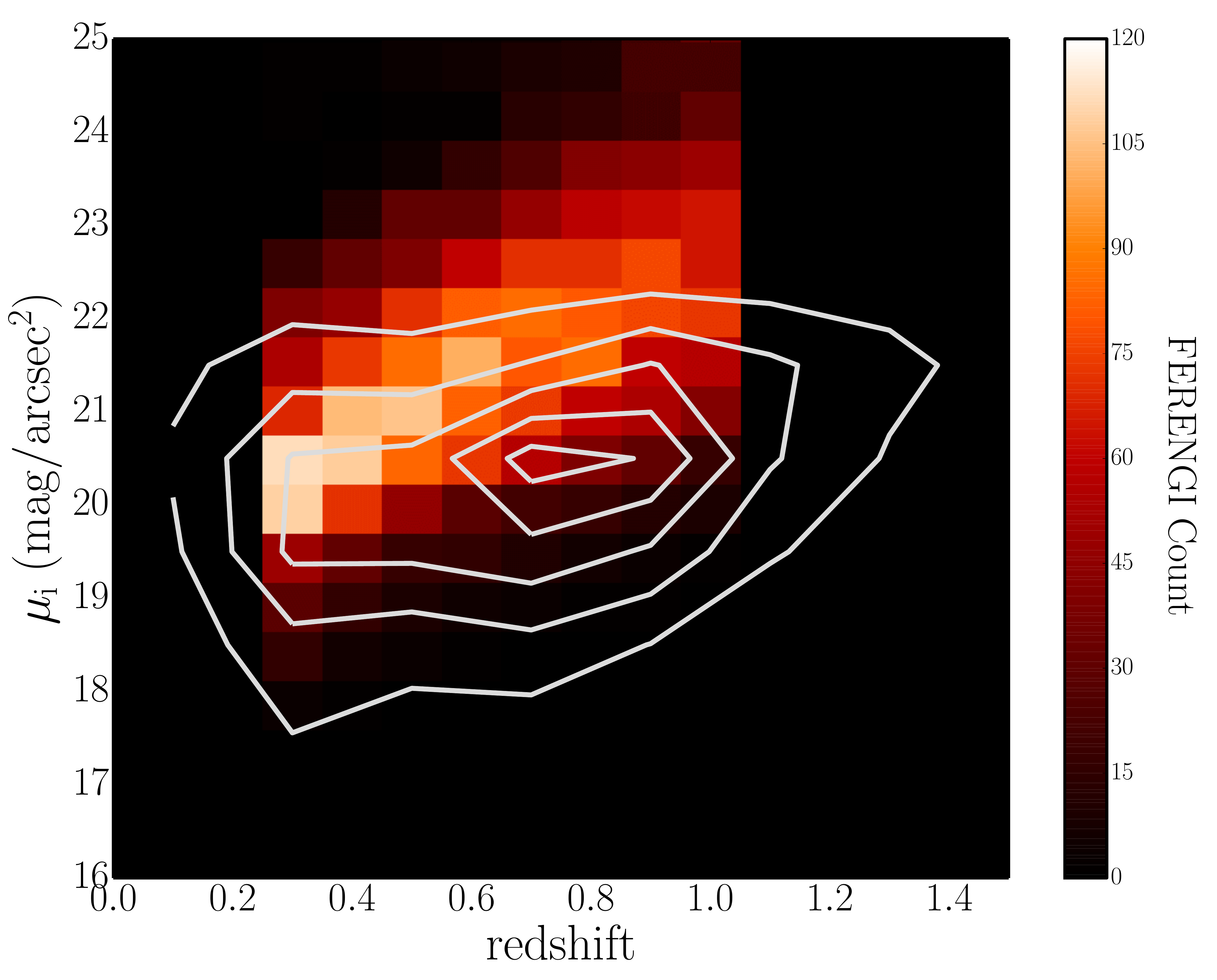}
\caption{Surface brightness as a function of redshift for 3,449~\ferengi{}
images and the 102,548~\main{} galaxies with measured $\mu$ and $z$ values. The
colour histogram shows the number of \ferengi{} images as a function of $\mu$
and $z_{\rm sim}$. White contours show counts for the galaxies in the \main{}
sample, with the outermost contour starting at $N=1500$ and separated by
intervals of 1500.} 
\label{fig:sb_redshift}
\end{center}
\end{figure}

Figure~\ref{fig:sb_redshift} shows the distribution in $\mu$ and $z$ of
the artificially-redshifted \textsc{FERENGI} images compared to the genuine
\hst{} images. The full bivariate distributions differ due to a combination of
the detection limits of the \hst{} surveys, intrinsic rareness of bright $\mu$
galaxies in the SDSS volume, and evolution of the stellar populations. Since
the dependence of the debiasing correction is evaluated in separate bins, the
main concern is the existence of an overlap between the two sets of images;
this is the case for almost all of the \hst{} images, which have $\mu >
19$~mag~arcsec$^{-2}$.  The brightest end of the distribution, which has fewer
examples of comparable \textsc{FERENGI} images, correspond to galaxies for
which the debiased correction is expected to be minor.

\subsection{Effects of morphological debiasing}\label{ssec:zeta_results}
The approach used in GZH for correcting the weighted classifications for
redshift bias rests on the assumption that the \emph{degree} of bias is a
function of the apparent size and brightness of the galaxy.
This is controlled by two types of parameters: \textbf{intrinsic} properties of
the galaxy itself, such as its physical diameter and luminosity, and
\textbf{extrinsic} properties, such as the distance (redshift) of the galaxy
and its relative orientation. The combination of all
such parameters forms a high-dimensional space, and there is no obvious technique
for separating these into individual effects. Instead, the method used here employs
only two parameters as a simple model of the effect of redshift brightness on classification: the
surface brightness ($\mu$; intrinsic) and redshift ($z$;
extrinsic).

\begin{figure*}
\centering
\includegraphics[width=\textwidth]{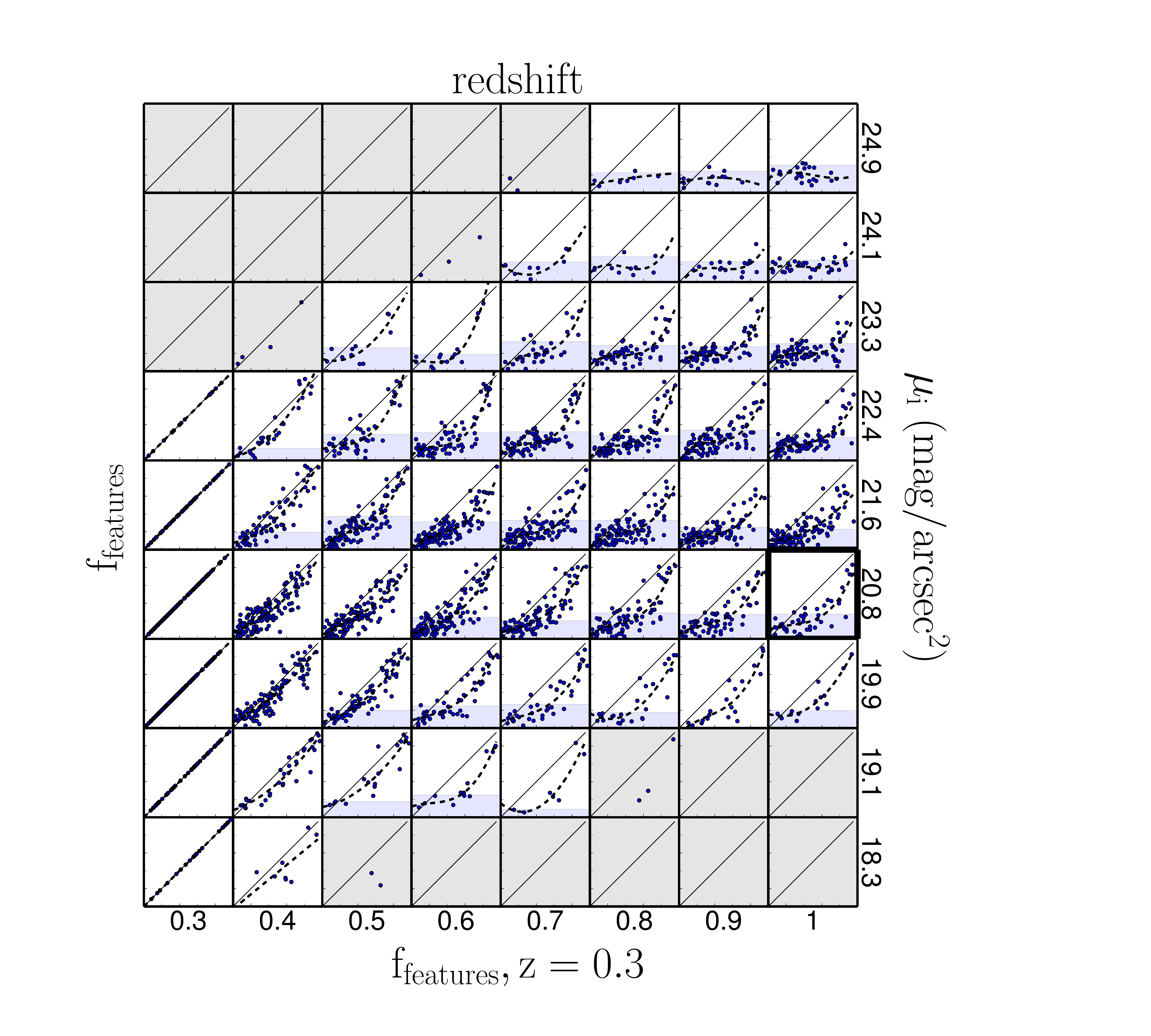}
\caption{Effects of redshift bias in 3,449~images in the \ferengi{} sample.
Each point \emph{in a given redshift and surface brightness bin} represents a
unique galaxy. On the $y$-axis in each bin is the \ffeatures{} value of the
image of that galaxy redshifted to the value corresponding to that redshift
bin. On the $x$-axis is the \ffeatures{} value of the image of the same galaxy
redshifted to $z=0.3$. The dashed black lines represent the best-fit
polynomials to the data in each square. The solid black line represents
\ffeaturesz=\ffeaturesrest. Regions in which there is a single-valued
relationship between \ffeatures{} at high redshift and at $z=0.3$ are white;
those in which there is not are blue, and those with not enough data ($N<5$)
are grey. A larger version of the bin outlined at $z=1.0$ and $20.3 < \mu <
21.0$ $\rm (mag/arcsec^2)$ is shown in Figure~\ref{fig:f_vs_f_zoom}.}
\label{fig:f_vs_f}
\end{figure*} 

\begin{figure}
\centering
\includegraphics[width=0.45\textwidth]{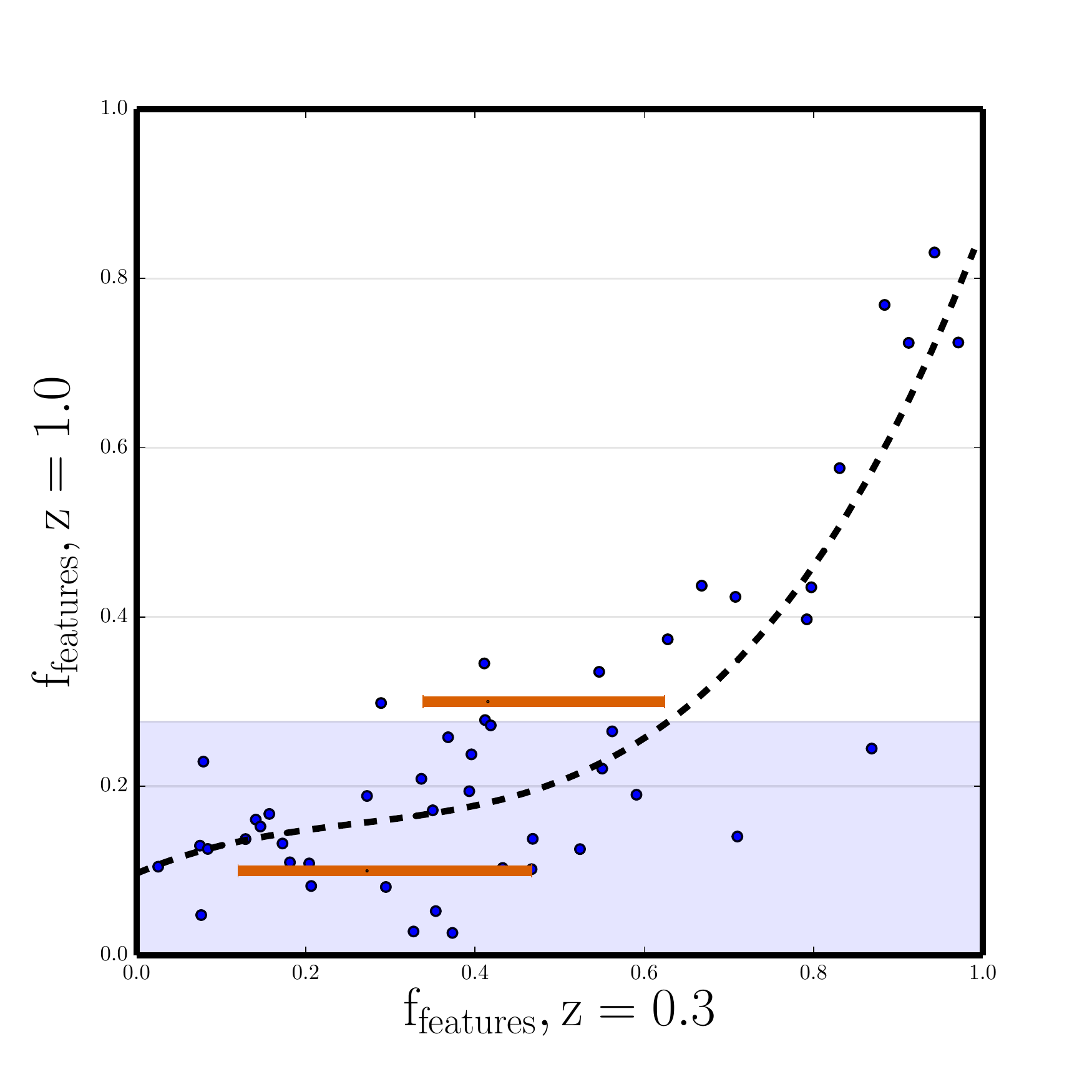}
\caption{A larger version of the dark-outlined square in Figure~\ref{fig:f_vs_f}, containing
\ferengi{} galaxies that have been artificially redshifted to $z=1.0$ and have
surface brightnesses between $20.3 < \mu < 21.0$ $\rm (mag/arcsec^2)$. The orange bars represent the inner
68\% (1$\sigma$) of the uncorrectable \ffeatures{} quantiles, which are used
to compute the limits on the range of debiased values.}
\label{fig:f_vs_f_zoom}
\end{figure}

Figure~\ref{fig:f_vs_f} shows the change in \ffeatures{} as a function of their
lowest simulated redshift for the 3,449~\ferengi{} images with robust photometric
measurements across the full range of redshifts.  For each simulated redshift
value \zsim{} at a fixed surface brightness, \ffeaturesz{} is the value
measured at that simulated redshift. We plot \ffeaturesz{} against \ffeaturesrest, the
value measured for the same galaxy at \zsim$=0.3$. 
 
The objective is to use these data to predict, for a galaxy with a measured
\ffeaturesz{} value, what its \ffeatures{} value \emph{would have been} if it
had been viewed at $z=0.3$. This predicted value is defined as the ``debiased''
vote fraction \ffeaturesdebiased, and is calculated by applying a correction to
the measured value of \ffeatures, determined by the $\zeta$ function described
in the following section (Equation~\ref{eqn:fzeta_mod}). A reliable predicted value can be obtained so long as
the relationship between \ffeaturesz{} and \ffeaturesrest{} is single-valued;
that is, for a given \ffeaturesz, there is exactly one corresponding value of
\ffeatures{} at $z=0.3$. 

Figure~\ref{fig:f_vs_f} shows that the relationship between \ffeaturesz{} and
\ffeaturesrest{} is \emph{not} always single-valued; hence, it is not
appropriate to correct galaxies that lie in certain regions of surface
brightness/redshift/\ffeatures{} space. Such regions tend to have low
\ffeatures{} values at high redshift, but a wide range of values at $z=0.3$.
These regions contain two morphological types of galaxies: the first set are
genuine ellipticals, which have low values of \ffeatures{} at both high and low
redshift. The second group are disks whose features become indistinct at high
redshift; hence their \ffeatures{} value at $z=0.3$ may be quite high, while
the value observed at high redshift is very low. This effect is strongest at
high $z$ and low $\mu$, where features become nearly impossible to discern in
the images (see top right panel of Figure~\ref{fig:f_vs_f}).

The criteria for determining whether a region of this space is single-valued,
and therefore correctable, is as follows: In each surface brightness and
redshift bin, the relationship between \ffeaturesz{} and \ffeaturesrest{} is
modelled by fitting the data with polynomials of degrees $n=3$, 2, and 1, and
using the best formal fit out of the three as measured by the sum of the
residuals. These fits are shown as the dashed black lines in
Figure~\ref{fig:f_vs_f}. Flat regions of the bins are areas in which there is
\emph{not} a clear single-valued relationship between \ffeaturesz{} and
\ffeaturesrest. We quantify this by measuring the slope of the best-fit
polynomial to the vote fractions. Regions within the bins with a slope less than 0.4
(a boundary selected through manual inspection and testing) are considered
\emph{not} one-to-one, and therefore \ffeaturesz{} cannot be boosted to its
\ffeaturesrest{} value.  Galaxies in this region are referred to as the
\emph{lower limit} sample, because the most stringent correction available is
that the weighted \ffeatures{} is a lower limit to the true value.  These
regions are highlighted in blue in Figure~\ref{fig:f_vs_f}. Uncoloured (white)
regions of the plot have sufficiently high slopes to consider the relationship
as single-valued; galaxies in these regions are considered ``correctable'', and
only these are used in measuring the parameters for the $\zeta$ function
(Section~\ref{ssec:zeta}). Only surface brightness/redshift bins with at least
5~galaxies were considered; regions with fewer than 5~galaxies are considered
to have ``not enough information'' to determine the \ffeaturesz{} and
\ffeaturesrest{} relationship, coloured grey in Figure~\ref{fig:f_vs_f}. These
galaxies are hereafter referred to as the ``NEI'' sample.

The unshaded regions in Figure~\ref{fig:f_vs_f} define discrete ranges of redshift,
surface brightness, and \ffeatures{} within which a galaxy must lie in order for the
debiased correction to be confidently applied.
While the appropriate correctable regions were defined as discrete bins, the true
correctable region is assumed to be a smooth function of $z$, $\mu$, and
\ffeatures{}. To define this smooth space, we calculate the shape of the convex
hull that encloses the correctable and lower-limit \ferengi{} galaxies in
$z$-$\mu$-\ffeatures{} space. The boundaries are then adjusted until the
contamination from both groups is minimised. The resulting hulls define the
correctable and lower-limit regions for categorising the \hst{} galaxies. The
results of this method and final categorisation of the \hst{} sample are in
Table~\ref{tbl:hubble_debiasable}. Of the galaxies at redshift higher than
$z=0.3$, 17\% can be debiased using the $\zeta$ method, 27\% cannot be debiased
since the relationship between \ffeaturesz{} and \ffeaturesrest{} is not
monotonic, and 56\% cannot be debiased since they either have an unknown
redshift or insufficient numbers of \ferengi{} images in their $z$-$\mu$ bin
to determine a reliable correction term ($\zeta$, described in Section~\ref{ssec:zeta}).

For the ``lower-limit'' galaxies for which a single debiased \ffeatures{} value
cannot be confidently assigned, the \emph{range} of debiased values is
estimated and included as a data product. This uses the
\ferengi{} simulated data to analyse the range of intrinsic \ffeaturesrest{}
values for any given observed \ffeatures{} value, again as a function of
surface brightness and redshift. In each unclassifiable (shaded blue) $z$,$\mu$ bin, 
we compute the spread of
intrinsic values of \ffeaturesrest{} for five quantiles of observed \ffeatures{}
(corresponding to the ``clean'' thresholds used in prior GZ publications).
The range of intrinsic values for GZH is defined by the upper and lower 1~$\sigma$ limits,
enclosing the inner 68\% of the data. This range is represented by the orange bars in 
Figure~\ref{fig:f_vs_f_zoom}. For any galaxy which
cannot be directly debiased by the $\zeta$ method, these ranges are used to
denote the upper and lower limits on the expected values \ffeaturesrest{} as a
function of the observed \ffeatures. 

\begin{table}
\center
\caption{Distribution of the ability of images to measure
morphological bias for
the \ferengi{} data (see Figure~\ref{fig:f_vs_f}).}
\label{tbl:ferengi_corrections}
\begin{tabular}{lrr}
\hline \hline
                                   & N       & \% \\
\hline 
Correctable                        & 1,690   &  49\% \\
Lower-limit                        & 1,678   &  49\% \\
NEI                                & 81      &   2\%\\
Total                              & 3,449   & 100\% \\
\hline \hline
\end{tabular}
\end{table}

\subsection{Correcting morphologies for classification bias}\label{ssec:zeta}

For the ``correctable'' sample of galaxies, we observe a decline in the vote fraction
\ffeatures{} with increasing simulated redshift for each unique galaxy. 
We model this relationship for artificially-redshifted images with a simple exponential
function which bounds $f_{\mu, z>0.3}$  between $f_{\mu, z=0.3}$ and 0:

\begin{equation}
f_{\mu,z} = 1 - (1 - f_{\mu,z=0.3})e^{\frac{z-z_0}{\hat\zeta}}
\label{eqn:fzeta}
\end{equation}
%old equation
%We model the debiased vote fraction $f_{\mu,z}$ for an
%artificially-redshifted galaxy as:

%\begin{equation}
%f_{\mu,z} = \left(f_{\mu,z=0.3}\right) \times e^{{\frac{z-z_0}{\zeta}}},
%\label{eqn:fzeta}
%\end{equation}

\noindent where $f_{\mu,z=0.3}$ is the vote fraction at the lowest redshift in
the artificially-redshifted sample ($z_0=0.3$). Here, $\zeta$ is a parameter that
controls the rate at which \ffeatures{} decreases with increasing redshift (and
that may depend on other galaxy properties). This function bounds the observed
vote fractions between $f_{\mu,z=0.3}$ and zero based on two assumptions: 1)
the vote fractions for featured galaxies decrease monotonically with increasing
redshift, therefore restricting vote fractions for a given galaxy to be less
than $f_{\mu,z=0.3}$ (which is almost always true of the data), and 2) the
vote fractions cannot be less than zero (which is always true).

\begin{figure*}
\center
\includegraphics[width=\textwidth]{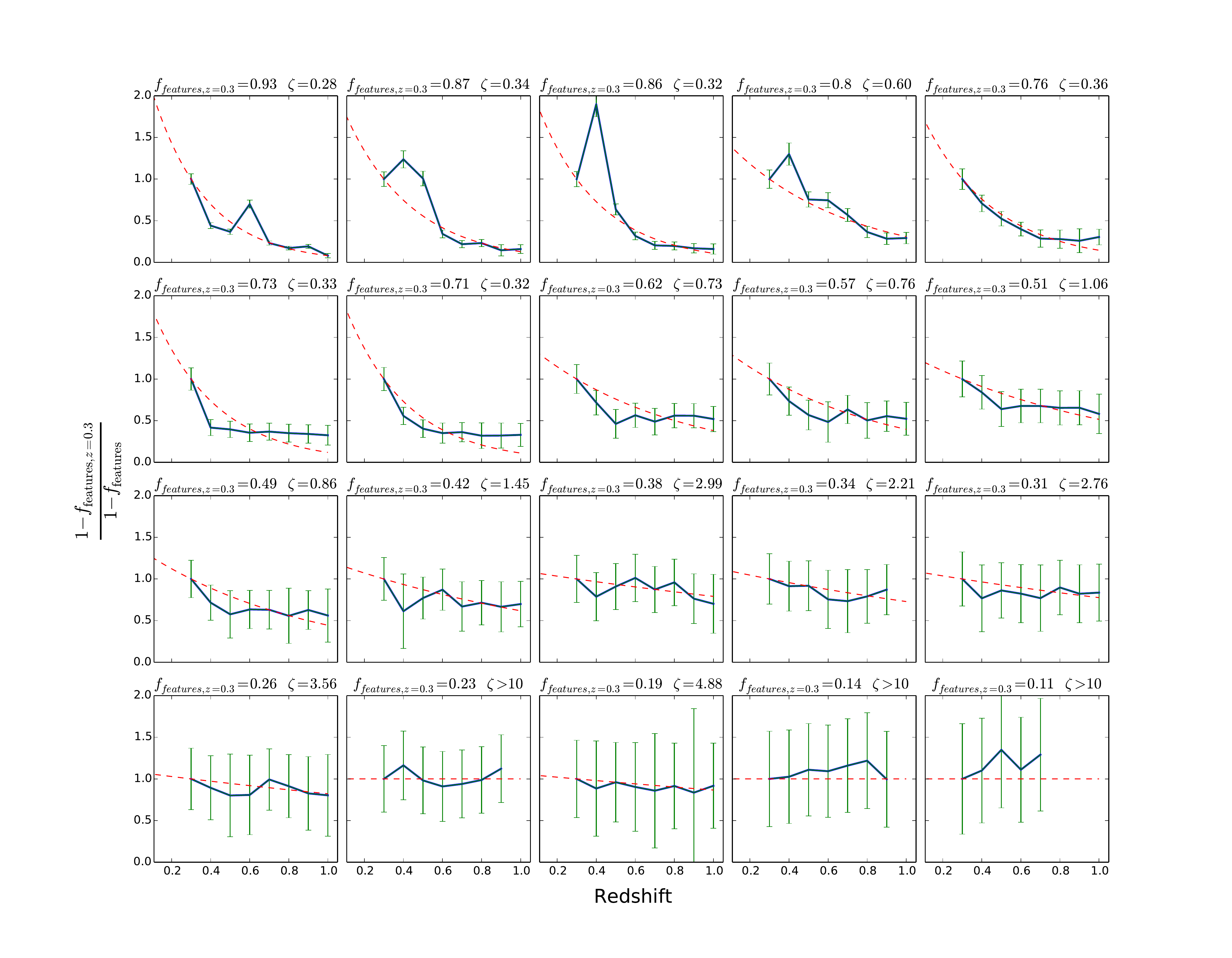}
\caption{Behaviour of the normalised, weighted vote fractions of features
visible in a galaxy ($f_\textrm{features}$) as a function of redshift in the
artificial \ferengi{} images. These plots demonstrate overall trends
in the sensitivity to feature detection as a function of redshift.
Galaxies in this plot were randomly selected from
a distribution with evolutionary correction $e=0$ and at least three detectable images 
in redshift bins of $z\ge0.3$. The displayed bins are sorted by \ffeaturesrest, 
labeled above each plot. Measured vote fractions (blue
solid line) are fit with an exponential function (red dashed line;
Equation~\ref{eqn:fzeta}); the best-fit parameter for $\zeta$ is given above
each plot.}
% Error bars are binomial for each f, symmetrised, then added in quadrature.
\label{fig:zeta_examples}
\end{figure*}

Figure~\ref{fig:zeta_examples} shows the change in vote fraction and the
best-fit model for a random set of galaxies in the \ferengi{} sample. The
results show that there is a clear dependence on redshift for the observed
changes in \ffeatures. To examine whether a better fit should involve
additional parameters, we tested the global dependence of $\zeta$ on the
galaxy surface brightness. This is motivated by the fact that brighter galaxies
would presumably be easier to identify; \citet{bam09} and \citet{wil13} found
changes in \ffeatures{} depend on a combination of physical size and
absolute magnitude. To simplify our model, we subsume these into a single
parameter of surface brightness, $\mu$.

\begin{figure}
\center
\includegraphics[width=0.5\textwidth]{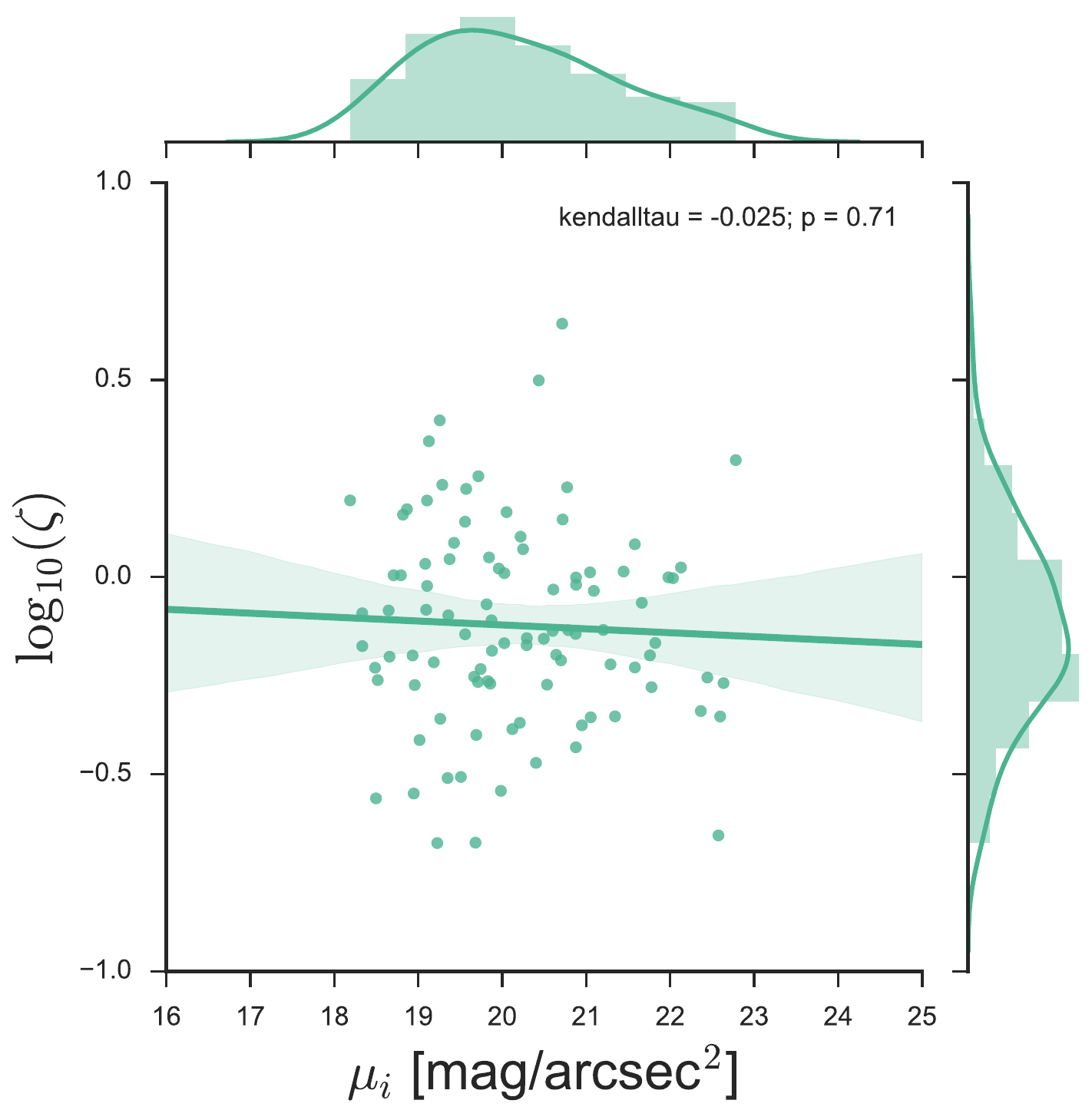}
\caption{All fits for the \ferengi{} galaxies of the vote fraction dropoff
parameter $\zeta$ for \ffeatures{} as a function of surface brightness. This
includes only the simulated galaxies with a bounded range on the dropoff
($-10<\zeta<10$) and sufficient points to fit each function (28~original
galaxies, each with varying images artificially redshifted in one to eight bins over a range from $0.3\lesssim z_\mathrm{sim}\lesssim1.0$). 
}
\label{fig:zeta_mu}
\end{figure}

Figure~\ref{fig:zeta_mu} examines the results of fitting the \ferengi{} images
with Equation~\ref{eqn:fzeta}. This fit only includes galaxies where
\sextractor{} robustly measured the photometry in all images between
$0.3<z_\mathrm{sim}<1.0$ (such that $\sigma_{M_I}<1.0$~mag; \citealt{mel16}). Interestingly, the
derived correction is only a very weak function of surface brightness.
Higher-surface brightness galaxies have on average slightly stronger
corrections, possibly because these galaxies have larger $f_\textrm{features}$
values at high redshifts. Low surface brightness galaxies are more likely to
begin low and remain low; the bounded nature of the dropoff (and variance among
the individual voters) means that the average magnitude of $\zeta$ will be
lower. 

Since there is little evidence for any strong systematic dependence of $\zeta$
on $\mu$, we do not include any additional parameters in fitting to our
calibration model.  We fit the data in Figure~\ref{fig:zeta_mu} with a linear
function:

\begin{equation}
\log_{10}(\hat\zeta) = \zeta_0 + (\zeta_1 \times \mu),
\label{eqn:zetafit}
\end{equation}

\noindent where $\hat\zeta$ is the correction factor applied to each galaxy as
a function of surface brightness. The best-fit parameters to the linear fit
from least-squares optimization are $\zeta_0=0.50$, $\zeta_1=-0.03$. To make the
final debiased correction for the genuine \hst{} data, we apply a 
correction similar to Equation~\ref{eqn:fzeta}:

\begin{equation}
f_\textrm{features,debiased} = 1 - (1 - f_\mathrm{features,weighted})e^{\frac{-(z-z_0)}{\hat\zeta}}
\label{eqn:fzeta_mod}
\end{equation}

\noindent where $f_\mathrm{features,weighted}$ is the weighted vote fraction
described in Section~\ref{ssec:weighting}, and $f_{\rm features,debiased}$ is bounded 
between $f_{\rm features,weighted}$ and 1. 

\begin{figure}
\center
\includegraphics[width=0.5\textwidth]{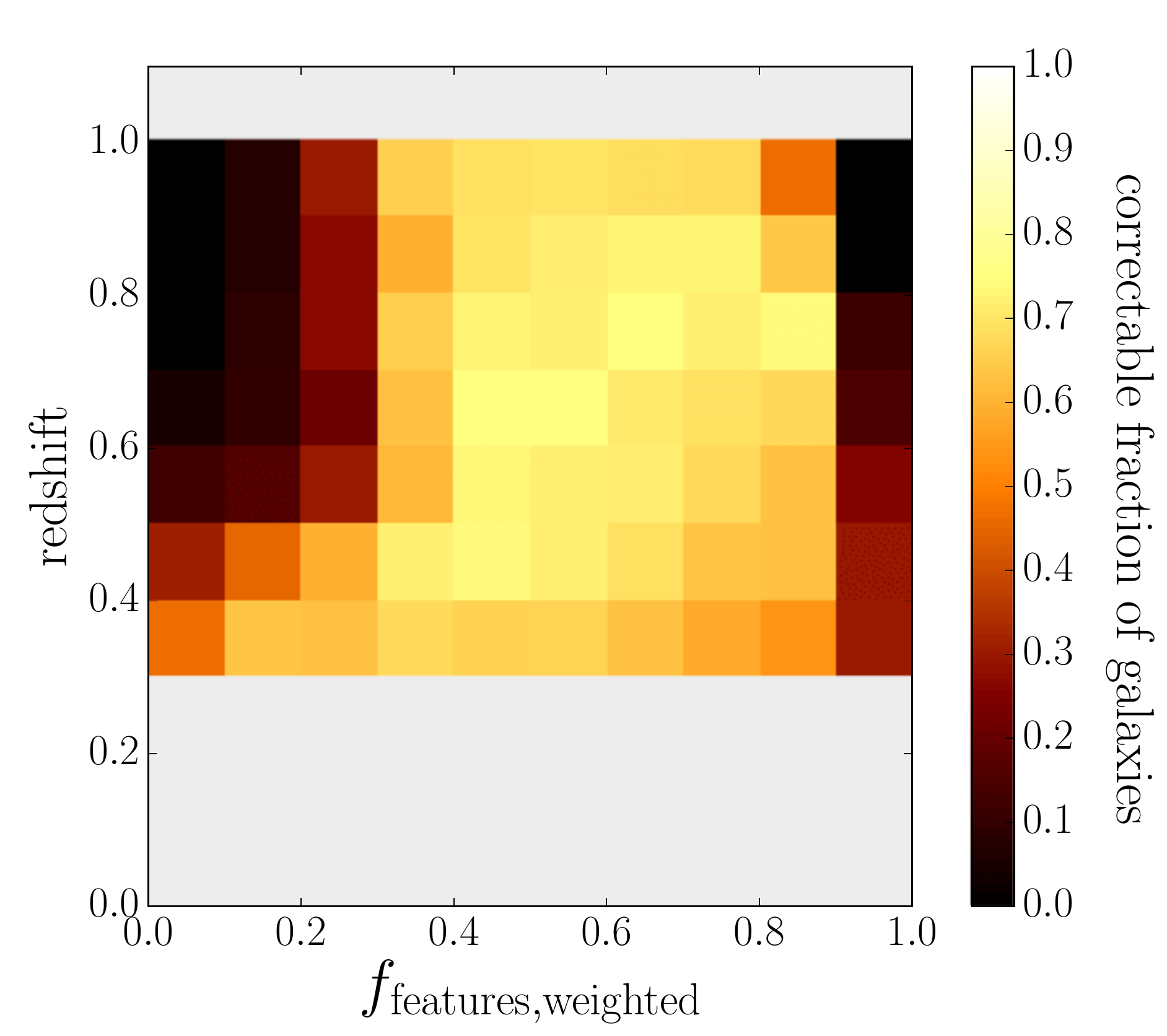}
\caption{Histogram showing the fraction of galaxies that have a finite correction
for the debiased vote fractions \ffeaturesdebiased{} as a function of \ffeatures{}
and redshift. The parameter space for corrections is limited to $0.3 \leq z \leq 1.0$
due to the sampling of the parent SDSS galaxies and detectability in the \ferengi{} images.
}
\label{fig:correctable_fraction}
\end{figure}

Figure~\ref{fig:correctable_fraction} shows the
distribution of the fraction of galaxies that are ultimately correctable as a function
of \ffeatures{} and redshift. The distributions for individual samples (AEGIS, COSMOS,
GEMS, GOODS) are individually very similar. We emphasize that a correction is more likely to be derived
and applied for galaxies at higher redshift and with higher weighted values of \ffeatures.
This has important consequences for selection of physically meaningful samples from GZH
(see Section~\ref{sec:cookbook}), meaning that comparative studies of galaxies
should use a threshold on the likelihood of features that evolves as a function of redshift.

\begin{figure*}
\center
\includegraphics[width=\textwidth]{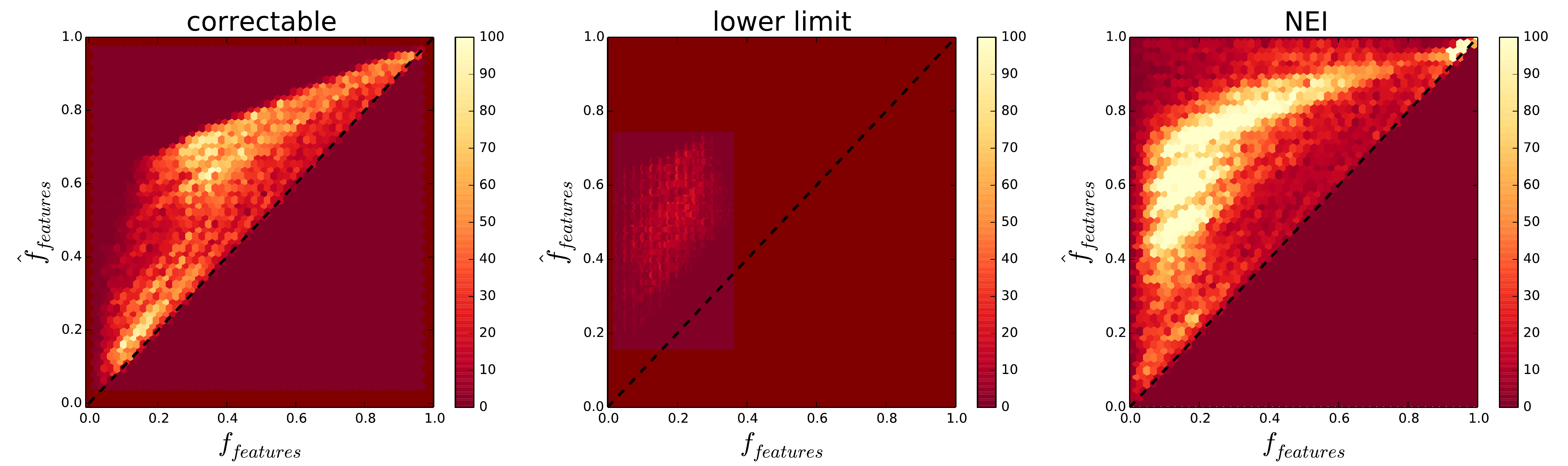}
\caption{Debiased \ffeatures{} corrected to $z=0.3$ vs weighted \ffeatures{}
for the correctable (left), lower-limit (middle), and ``not enough
information'' (NEI; right) galaxies in the GZH sample.}
\label{fig:debiased_corrections}
\end{figure*}

Few galaxies in the sample have sufficiently high corrections to completely
change them from being confidently ``smooth'' to ``featured'' following the
correction for redshift bias (Figure~\ref{fig:debiased_corrections}). As a
check, we compare morphologies of highly boosted galaxies to the expert visual
classifications in CANDELS \citep{kar15}.  Only nine galaxies that
were strongly boosted (\ffeatures$<0.25$ and \ffeaturesdebiased$>0.50$) in GZH also
appear in the CANDELS expert sample. Of those, the CANDELS expert sample
classifies 4 as spheroids/disky, 2 as disks, 1 as a spheroid, 1 as
irregular/disky and 1 as unclassifiable. The \fbest{} values for GZH are all
between 0.5 and 0.7, making them all intermediate disk candidates. Visual
inspection of the GZH images shows extended disk structure in at least
5~images; the remaining 4 are either extremely faint or have imaging artifacts
in the ACS data.  Since the surveys use different rest-frame filters and the
overlap sample is tiny, though, detailed comparisons between the overall
morphologies are highly difficult (see
Sections~\ref{ssec:gzh_gzc}--\ref{ssec:comparisons}).

\subsection{Challenges of debiasing questions beyond ``smooth or features''}\label{ssec:higher_order_tasks}

As with the HST images, each \ferengi{} subject had a varying number of
classifiers answering the various questions in the hierarchical decision tree.
Every classifier answers the first question, {\it ``Is the galaxy smooth and
rounded, with no sign of a disk?''}; as such, the vote fractions \fsmooth,
\ffeatures, and \fartifact{} all have the lowest statistical error for any
question in the tree, based only on the total number of responses (between
40 and 120; see Section~\ref{sec:interface}).  The number of participants
answering subsequent questions, however, is always equal to or less than the
number who answered the preceding question. The average number of responses per
task for fourth- or fifth-tier questions (such as spiral arm structure; Tasks
12--14) is only $4\pm4$ for the \ferengi{} sample. While this distribution is
strongly bimodal (reflecting the true morphologies of selected galaxies), the
very low absolute numbers of votes introduce very high variance when attempting
to calculate a statistical correction.

In the \ferengi{} data, these numbers severely limit the amount of information
that can be extracted for the higher-tier questions. The debiasing technique
used (Section~\ref{ssec:zeta}) requires that at least 10~classifiers answer each
question for a galaxy with $z_\mathrm{sim}=0.3$ \emph{and} the corresponding
image at higher redshift. This requirement is (by default) met by all galaxies
for the smooth/features question. However, this is often \emph{not} met for
questions beyond Task~01. On average, $60\%\pm24\%$ of the galaxies do not have
sufficient data to measure a correction, as compared to 2.0\% achieved for
Task~01 (Table~\ref{tbl:ferengi_corrections}). This leaves the average
surface brightness/redshift bin 
with insufficient data points to confidently measure the change in vote fraction ($\lesssim10$~galaxies per bin). 
For these reasons, debiased vote fractions are only provided in the GZH catalogue for Task~01
(smooth/features). We suggest that use of morphological data for higher order tasks should instead use
the weighted vote fractions (see Section~\ref{sec:cookbook}).

\begin{table*}
\caption{Number of correctable galaxies for the top-level task in GZH, split by \hst{} survey.}\label{tbl:hubble_debiasable}
\begin{tabular}{lcrrrrr|r}
\hline\hline
                                   & Correction type & AEGIS   & COSMOS & GEMS  & GOODS-N & GOODS-S  &  Total  \\
                                   &                 &         &        &       & 5-epoch & 5-epoch  &         \\
\hline
correctable                        & 0               & 2,908   & 21,169 & 2,802 & 1,459   & 1,189    &  29,527 \\
lower-limit                        & 1               &   833   &  5,169 & 1,021 & 1,377   & 1,267    &   9,667 \\
no correction needed ($z \le 0.3$) & 2               &   955   & 10,870 & 1,175 &   415   &   400    &  13,815 \\ 
not enough information (NEI)       & 3               & 2,677   & 43,058 & 3,559 & 2,077   & 2,184    &  53,555 \\
no redshift information            & 4               & 1,134   &  4,688 &   530 &   687   &   102    &   7,141 \\
\hline
total                              &                 & 8,507   & 84,954 & 9,087 & 6,015   & 5,142    & 113,705 \\
\hline\hline
\end{tabular}
\end{table*}

\section{The Galaxy Zoo: Hubble catalogue}\label{sec:results}

The catalogue for GZH includes morphological data for 189,149~images
(generated from a total of 145,741~unique galaxies). The full table can be
accessed at \url{http://data.galaxyzoo.org}. The online data also includes a
secondary metadata table, which is drawn from a variety of sources detailed in
Section \ref{sec:data}. 
% 113705+6144+30339-4447 = 145,741
% main + goods_shallow + stripe82_coadd - goods_union

Each image is listed under a unique project ID (eg,~AHZ000001); the actual
galaxy in the image is identified by the combination of the OBJNO and original
survey. For each of the 55~responses in the GZH decision tree, the following
classification data is provided: for each question, $\tt N_{votes}$ is the
number of classifiers who answered that question. For each unique answer, $\tt fraction$
is the fraction of classifiers who selected that answer ($\rm N_{answer}/N_{votes}$), and
$\tt weighted$ is the weighted fraction, which takes into account overall
consistency (Section~\ref{ssec:weighting}). 

The GZH vote fractions can be largely dependent on the resolution of the image.
Two otherwise morphologically identical galaxies which differ significantly in
redshift, brightness, or size may result in very different vote fractions for
any given question, given that many features of a galaxy are difficult to
discern in less-resolved images (bars, spiral arms, disk structure, etc). For
this reason, caution must be used when taking vote fractions as cut-offs to
determine morphological structure; guidelines for careful classification are
given in Section~\ref{sec:cookbook}. 

The GZH catalogue is corrected for redshift bias only for the first
question of the GZH decision tree (Section~\ref{sec:debiasing}), which asks
{\it ``Is the galaxy smooth and round, with no sign of a disk?''} For this
question, the catalogue provides the additional parameters $\tt debiased$, $\tt
lower~limit$, $\tt upper~limit$, and $\tt best$ vote fractions. The $\tt best$
fraction for \ffeatures{} is chosen based on the categorization of the galaxy:
if it is ``correctable'', $\tt best = debiased$; if it is a lower limit, $\tt
best = lower~limit$; if neither condition applies, then $\tt best = weighted$.

The debiased and best vote fractions for \fsmooth{} are calculated on the
criteria that vote fractions for all answers must sum to unity:

\begin{equation}
f_\mathrm{smooth} \equiv 1 - f_\mathrm{features,best} - f_\mathrm{artifact}.
\label{eqn:constraint}
\end{equation}

\noindent In rare cases ($1.2\%$ of the \main{} sample), this requirement resulted in
negative vote fractions for \fsmooth; these were cases in which the
\ffeatures{} vote fraction was boosted to a high value relative to \fartifact.
In these cases, the constraint of Equation~\ref{eqn:constraint} is met by
setting \fsmooth$=0.0$ and decreasing \fbest{} accordingly. This correction is
typically small, with a median change of
\ffeatures{} or \fsmooth{} of $\Delta f = 0.04$.

Tables~\ref{tbl:catalog_hst}-\ref{tbl:simulated_agn} also contain flags for
Task 01 that identify clean (but not complete) samples of galaxies with high
likelihoods of being either smooth, featured, or a star/artifact.  These are
set as $\tt smooth/featured/artifact\_flag=1$ if $\rm
f_{smooth/features/artifact\_best}>0.8$. For a galaxy to be flagged as
``smooth'', an additional criterion of $\tt correction\_type =~$0, 1, or 2 is
applied. This accounts for the uncertainty in distinguishing between genuine
ellipticals and disks whose features have been washed out due to surface
brightness and redshift effects (Section~\ref{sec:debiasing}).

Data products for GZH are split by the type of image being classified. Each
sample in Section~\ref{ssec:sample_labels} corresponds to the data in
Tables~\ref{tbl:catalog_hst}-\ref{tbl:simulated_agn}.

\clearpage
\tabletypesize{\scriptsize}
\begin{deluxetable}{lllcrlllllll}
\centering
\tablecolumns{12}
\tablewidth{0pc}
\tablecaption{GZH morphological classifications for \hst{} images from AEGIS, COSMOS, GEMS, and GOODS\label{tbl:catalog_hst}} 
\tabletypesize{\scriptsize}
\tablehead{
 & & & 
\multicolumn{2}{c}{\underline{t01\_smooth\_or\_features\_}} &
\multicolumn{6}{c}{\underline{t01\_smooth\_or\_features\_a01\_smooth\_}} &
\colhead{$\ldots$}
\\
\colhead{Zooniverse ID} & 
\colhead{Survey ID} & 
\colhead{Imaging} & 
\colhead{Correction$^{1}$} & 
\colhead{$N_\mathrm{votes}$} & 
\colhead{fraction} & 
\colhead{weighted} & 
\colhead{debiased} & 
\colhead{best} &
\colhead{lower~limit} & 
\colhead{upper~limit} &
\colhead{}
}
\small
\startdata
AHZ100002g  & 10010842  & AEGIS             & 0        & 127       & 0.118    & 0.128     & 0.089    & 0.089    & 0.032     & 0.106 \\
AHZ100002h  & 10010870  & AEGIS             & 4        & 127       & 0.567    & 0.592     & -0.126    & 0.592    & $-$       & $-$   \\
$\ldots$    \\
AHZ20004kd  & 20014731  & COSMOS            & 0        &  44       & 0.682    & 0.675     & 0.243    & 0.330    & 0.615       & 0.243   \\
AHZ20004ke  & 20014732  & COSMOS            & 2        &  45       & 0.689    & 0.756     & 0.844    & 0.756    & $-$       & $-$   \\
$\ldots$    \\
AHZ400043g  & 90022729  & GEMS              & 3        & 121       & 0.702    & 0.733     & 0.486    & 0.733    & $-$       & $-$ \\
AHZ4000416  & 90022735  & GEMS              & 0        & 127       & 0.646    & 0.698     & 0.509    & 0.509    & 0.347     & 0.394 \\
$\ldots$    \\
AGZ0007z47  & 10014     & GOODS-N-FULLDEPTH & 1        & 40        & 0.475    & 0.475     & 0.203    & 0.475    & 0.126     & 0.475 \\
AGZ0007z48  & 10017     & GOODS-N-FULLDEPTH & 3        & 40        & 0.675    & 0.675     & 0.094    & 0.675    & $-$       & $-$ \\
$\ldots$    \\
AGZ00083jb  & 8869      & GOODS-S-FULLDEPTH & 0        & 40        & 0.425    & 0.425     & 0.135    & 0.135    & 0.185     & 0.481 \\
AGZ00083jc  & 8878      & GOODS-S-FULLDEPTH & 0        & 40        & 0.205    & 0.205     & 0.050    & 0.050    & $-$0.028  & 0.181 \\
$\ldots$    \\
\enddata
\tablenotetext{1}{Flag indicating how the vote fractions for this galaxy were
corrected through debiasing (Section~\ref{ssec:zeta_results}), if possible.
0~=~correctable, 1~=~lower-limit ($f_\mathrm{raw}$ --- $f_\mathrm{adj}$ is not
single-valued), 2~=~uncorrected ($z < 0.3$), 3~=~uncorrected
(insufficient \ferengi{} galaxies in this $z$-$\mu$ bin), 4~=~uncorrected (no
galaxy redshift available).}
\tablecomments{The full version of this table is available in electronic form,
as well as at \url{http://data.galaxyzoo.org}. The complete version includes
data for 113,705~galaxies and morphological information for all tasks in the
tree. A subset of the information is shown here to illustrate form and
content.}
\end{deluxetable}

\tabletypesize{\scriptsize}
\begin{deluxetable}{lllcrlllllll|c}
\centering
\tablecolumns{13}
\tablewidth{0pc}
\tablecaption{GZH morphological classifications for color-faded \hst{} images\label{tbl:catalog_faded}} 
\tabletypesize{\scriptsize}
\tablehead{
 & & & 
\multicolumn{2}{c}{\underline{t01\_smooth\_or\_features\_}} &
\multicolumn{7}{c}{\underline{t01\_smooth\_or\_features\_a01\_smooth\_}} &
\colhead{$\ldots$}
\\
\colhead{Zooniverse ID} & 
\colhead{Survey ID} & 
\colhead{Imaging} & 
\colhead{Correction} & 
\colhead{$N_\mathrm{votes}$} & 
\colhead{fraction} & 
\colhead{weighted} & 
\colhead{debiased} & 
\colhead{best} &
\colhead{lower~limit} & 
\colhead{upper~limit} &
\colhead{flag} &
\colhead{}
}
\small
\startdata
AHZF000001  &   20000002    &   COSMOS  &   1   &   48 &  0.708   &     0.755   &   0.294    &    0.755 &   0.450   &   0.831 & 0 \\
AHZF000003  &   20000004    &   COSMOS  &   0   &   49 &  0.367   &     0.379   &   0.136    &    0.136 &   0.161   &   0.289 & 0 \\
AHZF000004  &   20000006    &   COSMOS  &   3   &   49 &  0.265   &     0.271   &   0.038    &    0.271 &   $-$     &   $-$   & 0 \\
AHZF00000z  &   20000102    &   COSMOS  &   1   &   44 &  0.727   &     0.780   &   0.296    &    0.780 &   0.421   &   0.797 & 0 \\
AHZF000010  &   20000104    &   COSMOS  &   2   &   53 &  0.811   &     0.849   &   0.894    &    0.849 &   $-$     &   $-$   & 0 \\
$\ldots$    \\
\enddata
\tablecomments{The full version of this table is available in electronic form,
as well as at \url{http://data.galaxyzoo.org}. The complete version includes
data for 3,927~galaxies and morphological information for all tasks in the
tree. A subset of the information is shown here to illustrate form and
content.}
\end{deluxetable}

\tabletypesize{\scriptsize}
\begin{deluxetable}{lllcrllllllll|c}
\centering
\tablecolumns{13}
\tablewidth{0pc}
\tablecaption{GZH morphological classifications for color-inverted \hst{} images\label{tbl:catalog_recolored}} 
\tabletypesize{\scriptsize}
\tablehead{
 & & & 
\multicolumn{2}{c}{\underline{t01\_smooth\_or\_features\_}} &
\multicolumn{7}{c}{\underline{t01\_smooth\_or\_features\_a01\_smooth\_}} &
\colhead{$\ldots$}
\\
\colhead{Zooniverse ID} & 
\colhead{Survey ID} & 
\colhead{Imaging} & 
\colhead{Correction} & 
\colhead{$N_\mathrm{votes}$} & 
\colhead{fraction} & 
\colhead{weighted} & 
\colhead{debiased} & 
\colhead{best} &
\colhead{lower~limit} & 
\colhead{upper~limit} &
\colhead{flag} &
\colhead{}
}
\small
\startdata
AHZC000001  &   20000002      &   COSMOS      &   1  &  52  &   0.615   &   0.664   &   0.223   &  0.664   &  0.396   &  0.777  & 0  \\
AHZC000003  &   20000004      &   COSMOS      &   0  &  48  &   0.333   &   0.364   &   0.049   &  0.049   &  0.026   &  0.154  & 0  \\
AHZC000004  &   20000006      &   COSMOS      &   3  &  51  &   0.235   &   0.252   &   0.016   &  0.252   &  $-$     &  $-$    & 0  \\
AHZC00000z  &   20000102      &   COSMOS      &   1  &  49  &   0.755   &   0.757   &   0.327   &  0.620   &  0.279   &  0.620  & 0  \\
AHZC000010  &   20000104      &   COSMOS      &   2  &  51  &   0.843   &   0.882   &   0.926   &  0.881   &  $-$     &  $-$    & 1  \\
$\ldots$    \\
\enddata
\tablecomments{The full version of this table is available in electronic form,
as well as at \url{http://data.galaxyzoo.org}. The complete version includes
data for 3,927~galaxies and morphological information for all tasks in the
tree. A subset of the information is shown here to illustrate form and
content.}
\end{deluxetable}

\tabletypesize{\scriptsize}
\begin{deluxetable}{lllcrlllllll|c}
\centering
\tablecolumns{13}
\tablewidth{0pc}
\tablecaption{GZH morphological classifications for GOODS 2-epoch images\label{tbl:catalog_goods_shallow}} 
\tabletypesize{\scriptsize}
\tablehead{
 & & &  
\multicolumn{2}{c}{\underline{t01\_smooth\_or\_features\_}} &
\multicolumn{7}{c}{\underline{t01\_smooth\_or\_features\_a01\_smooth\_}} &
\colhead{$\ldots$}
\\
\colhead{Zooniverse ID} & 
\colhead{Survey ID} & 
\colhead{Imaging} & 
\colhead{Correction} & 
\colhead{$N_\mathrm{votes}$} & 
\colhead{fraction} & 
\colhead{weighted} & 
\colhead{debiased} & 
\colhead{best} &
\colhead{lower~limit} & 
\colhead{upper~limit} &
\colhead{flag} &
\colhead{}
}
\small
\startdata
AHZ3000001  &   50000000    &     GOODS-N   &   3   &   123 &   0.390   &   0.415   &   0.079 &   0.415 &   $-$     &   $-$   & 0 \\
AHZ3000002  &   50000001    &     GOODS-N   &   2   &   126 &   0.341   &   0.355   &   0.356 &   0.356 &   $-$     &   $-$   & 0 \\
AHZ3000003  &   50000005    &     GOODS-N   &   0   &   129 &   0.760   &   0.826   &   0.641 &   0.641 &   0.622   &   0.835 & 0 \\
AHZ3000004  &   50000008    &     GOODS-N   &   3   &   120 &   0.758   &   0.787   &   0.629 &   0.787 &   $-$     &   $-$   & 0 \\
AHZ3000005  &   50000010    &     GOODS-N   &   3   &   123 &   0.854   &   0.890   &   0.605 &   0.890 &   $-$     &   $-$   & 0 \\
$\ldots$    \\
\enddata
\tablecomments{The full version of this table is available in electronic form,
as well as at \url{http://data.galaxyzoo.org}. The complete version includes
data for 6,144~galaxies and morphological information for all tasks in the
tree. A subset of the information is shown here to illustrate form and
content.}
\end{deluxetable}

\clearpage
\tabletypesize{\scriptsize}
\begin{deluxetable}{lllcllc|c}
\centering
\tablecolumns{10}
\tablewidth{0pc}
\tablecaption{GZH morphological classifications for SDSS Stripe 82 single-epoch images\label{tbl:stripe82_single}} 
\tabletypesize{\scriptsize}
\tablehead{
 & & & 
\multicolumn{1}{c}{\underline{t01\_smooth\_or\_features\_}} &
\multicolumn{3}{c}{\underline{t01\_smooth\_or\_features\_a01\_smooth\_}} &
\colhead{$\ldots$}
\\
\colhead{Zooniverse ID} & 
\colhead{Survey ID} & 
\colhead{Imaging} & 
\colhead{$N_\mathrm{votes}$} & 
\colhead{fraction} &
\colhead{weighted} & 
\colhead{flag} &
\colhead{}
}
\small
\startdata
AHZ5000001  &   587730845812064684  &   SDSS    &        41  &   0.585   &    0.595  &     0  \\
AHZ5000002  &   587730845812065247  &   SDSS    &        46  &   0.609   &    0.651  &     0  \\
AHZ5000003  &   587730845812196092  &   SDSS    &        51  &   0.039   &    0.044  &     0  \\
AHZ5000004  &   587730845812196825  &   SDSS    &        35  &   0.514   &    0.605  &     0  \\
AHZ5000005  &   587730845812524122  &   SDSS    &        47  &   0.766   &    0.812  &     1  \\
AHZ5000006  &   587730845812654984  &   SDSS    &        42  &   0.500   &    0.542  &     0  \\
AHZ5000007  &   587730845812655541  &   SDSS    &        41  &   0.488   &    0.526  &     0  \\
AHZ5000008  &   587730845812720365  &   SDSS    &        53  &   0.792   &    0.84   &     1  \\
AHZ5000009  &   587730845812720640  &   SDSS    &        43  &   0.000   &    0.0    &     0  \\
AHZ500000a  &   587730845812720699  &   SDSS    &        40  &   0.425   &    0.478  &     0  \\
$\ldots$    \\
\enddata
\tablecomments{The full version of this table is available in electronic form,
as well as at \url{http://data.galaxyzoo.org}. The complete version includes
data for 21,522~galaxies and morphological information for all tasks in the
tree. A subset of the information is shown here to illustrate form and
content.}
\end{deluxetable}

\tabletypesize{\scriptsize}
\begin{deluxetable}{lllcllc|c}
\centering
\tablecolumns{10}
\tablewidth{0pc}
\tablecaption{GZH morphological classifications for SDSS Stripe 82 co-added images\label{tbl:stripe82_coadd}} 
\tabletypesize{\scriptsize}
\tablehead{
 & & & 
\multicolumn{1}{c}{\underline{t01\_smooth\_or\_features\_}} &
\multicolumn{3}{c}{\underline{t01\_smooth\_or\_features\_a01\_smooth\_}} &
\colhead{$\ldots$}
\\
\colhead{Zooniverse ID} & 
\colhead{Survey ID} & 
\colhead{Imaging} & 
\colhead{$N_\mathrm{votes}$} & 
\colhead{fraction} & 
\colhead{weighted} & 
\colhead{flag} &
\colhead{}
}
\small
\startdata
AHZ6000001  & 8647474690312306978   &   SDSS    &     40  & 0.275   &    0.289  &   0 \\
AHZ6000002  & 8647474690312307154   &   SDSS    &     43  & 0.605   &    0.634  &   0 \\
AHZ6000003  & 8647474690312307877   &   SDSS    &     51  & 0.608   &    0.627  &   0 \\
AHZ6000004  & 8647474690312308301   &   SDSS    &     52  & 0.038   &    0.038  &   0 \\
AHZ6000005  & 8647474690312308318   &   SDSS    &     44  & 0.614   &    0.632  &   0 \\
AHZ6000006  & 8647474690312308880   &   SDSS    &     36  & 0.667   &    0.683  &   0 \\
AHZ6000007  & 8647474690312372644   &   SDSS    &     48  & 0.646   &    0.674  &   0 \\
AHZ6000008  & 8647474690312372789   &   SDSS    &     45  & 0.489   &    0.571  &   0 \\
AHZ6000009  & 8647474690312372931   &   SDSS    &     47  & 0.553   &    0.587  &   0 \\
AHZ600000a  & 8647474690312373190   &   SDSS    &     47  & 0.574   &    0.559  &   0 \\
$\ldots$    \\
\enddata
\tablecomments{The full version of this table is available in electronic form,
as well as at \url{http://data.galaxyzoo.org}. The complete version includes
data for 30,339~galaxies and morphological information for all tasks in the
tree. A subset of the information is shown here to illustrate form and
content.}
\end{deluxetable}

\tabletypesize{\scriptsize}
\begin{deluxetable}{lllccccrccc}
\centering
\tablecolumns{11}
\tablewidth{0pc}
\tablecaption{GZH morphological classifications for \hst{} images with simulated AGN\label{tbl:simulated_agn}} 
\tabletypesize{\scriptsize}
\tablehead{
 & & & & & & 
\colhead{\underline{t01\_smooth\_or\_features\_}} &
\multicolumn{3}{r}{\underline{t01\_smooth\_or\_features\_a01\_smooth\_}} &
\colhead{$\ldots$}
\\
\colhead{Zooniverse ID} & 
\colhead{Survey ID} & 
\colhead{Imaging} & 
\colhead{Version} & 
\colhead{$L_\mathrm{ratio}$} & 
\colhead{AGN color$^{1}$} & 
\colhead{$N_\mathrm{votes}$} & 
\colhead{fraction} & 
\colhead{weighted} &
\colhead{flag} & 
\colhead{}
}
\small
\startdata
AHZ7000001  &   9002470011  &   GEMS       &   1   &   0.2   &   1   &   42  &   0.238   &   0.239  & 0  \\
AHZ7000002  &   9002470012  &   GEMS       &   1   &   1.0   &   1   &   51  &   0.255   &   0.265  & 0  \\
AHZ7000003  &   9002470013  &   GEMS       &   1   &   5.0   &   1   &   47  &   0.170   &   0.167  & 0  \\
AHZ7000004  &   9002470014  &   GEMS       &   1   &   10.0  &   1   &   41  &   0.195   &   0.195  & 0 \\
AHZ7000005  &   9002470015  &   GEMS       &   1   &   50.0  &   1   &   47  &   0.170   &   0.178  & 0  \\
$\ldots$    \\
AHZ70002a4  &   9010781875  &   GOODS-S       &   2   &   50.0   &   2   &   33  &   0.242   &   0.285  & 0 \\
AHZ70002a5  &   9010781881  &   GOODS-S       &   2   &   0.2    &   3   &   32  &   0.312   &   0.323  & 0 \\
AHZ70002a6  &   9010781882  &   GOODS-S       &   2   &   1.0    &   3   &   35  &   0.543   &   0.559  & 0 \\
AHZ70002a7  &   9010781883  &   GOODS-S       &   2   &   5.0    &   3   &   28  &   0.429   &   0.460  & 0 \\
AHZ70002a8  &   9010781884  &   GOODS-S       &   2   &   10.0   &   3   &   25  &   0.200   &   0.200  & 0 \\
AHZ70002a9  &   9010781885  &   GOODS-S       &   2   &   50.0   &   3   &   30  &   0.167   &   0.167  & 0 \\
$\ldots$    \\
\enddata
\tablenotetext{1}{Flag indicating the color of the PSF in the simulated AGN. 0~=~no simulated AGN, 1~=~blue, 2~=~flat, 3~=~red.}
\tablecomments{The full version of this table is available in electronic form,
as well as at \url{http://data.galaxyzoo.org}. The complete version includes
data for 2,961~galaxies and morphological information for all tasks in the
tree. A subset of the information is shown here to illustrate form and
content.}
\end{deluxetable}

\clearpage

\section{Using the GZH catalogue}\label{sec:cookbook}

The primary purpose of the GZH catalogue is to provide a reliable method of
selecting galaxies of a desired morphological type.  This section provides
instructions for creating such pure samples using the vote fractions
corresponding to the tasks shown in Figure~\ref{fig:decisiontree}
\citep[eg,][]{mas11c,che15,gal15}. Increasing the levels of the thresholds can
create purer, but not necessarily complete, morphologically-selected samples.
These are useful for selecting galaxies of rare or unique types that merit
individual study.  Looser cuts can be applied to obtain samples with a higher
level of completeness, although the rate of false positives must be closely
monitored.  We stress that the details of any selection process will vary based
on the particular science case; for example, \citet{bam09,ski09,sme15} all
demonstrate the advantages of using the vote fractions directly as weights
rather than applying discrete thresholds.

To select galaxies of a morphological type identified with a particular task, a
cut is placed on the vote fraction for that task ($f_{\rm task}$), \emph{as
well as} the vote fractions for the tasks preceding it, because of the
dependency induced by the decision-tree structure. For example, to select
barred galaxies, a cut may be placed on \fbar{} such that only galaxies where a
high fraction of votes for this task voted for the $bar,yes$ answer. This is
not the only necessary cut, however, since not all classifiers answer this question;
only those who have previously selected ``features'' in Task~01, ``not clumpy''
in Task~12, and ``not edge-on'' in Task~02 will have the opportunity to vote on
the bar question, Task~03. To ensure that \fbar{} is well-sampled, cuts on all
previous tasks must be applied. 

The flexibility of this catalogue allows users to set their own selection
criteria for vote fraction thresholds to create a morphologically pure sample.
Table~\ref{tbl:thresholds} provides suggested cuts for selecting galaxies with a
variety of morphologies. These thresholds were determined by visual inspection
of various subsamples of GZH data. The thresholds employ a combination of 20~votes for the
task being considered, as well as limits on the vote fractions for previous
response(s) in the decision tree. 

We visually analysed subsamples of 50~galaxies meeting both
criteria, as well as a control sample of galaxies which had 20~classifiers vote on
the task, but did not meet the threshold cut set for the previous task. The
threshold cut was adjusted and new subsamples were inspected until both the
original and control samples achieved $>80\%$ purity.

To use data from Table~\ref{tbl:thresholds} to create, for example, a sample of  
3-armed spiral galaxies, we suggest selecting objects with $N_{arm~number} \ge
20$, \ffeatures$>0.23, f_{\rm clumpy,no} > 0.30, f_{edgeon,no}>0.25$, and
$f_{spiral,yes}>0.25$. These cuts define a sample of galaxies of ``arm number
candidates''; \ie, galaxies for which answering the arm number question makes
physical sense and the vote fraction $f_{\rm arm~number}$ is well-sampled. 
Such galaxies will be featured, non-clumpy, non-edge~on spirals. At
this point a final cut can be made on $f_{\rm arm~number=3}$ to select spirals with
three arms. 

Tasks 03, 04 and 05 have an additional possible pathway; as shown in
Figure~\ref{fig:decisiontree}, a classifier might also be shown this question
if they select ``featured/disk'' in Task~01, ``clumpy'' in Task~12, two or more
clumps in Task~16, and ``spiral arrangement'' in Task~15. After applying
the appropriate thresholds for this path, $< 0.5\%$ of the galaxies which
have $\ge 20$ answers to these questions used this pathway to arrive at
these Tasks. None of these images show obvious disk
structure, although the clumps within are arranged in a spiral pattern. 

This section described how to use Table~\ref{tbl:thresholds} to select
galaxies for which a particular task is reliably sampled. The
following two examples extend this and show how to use the vote fractions to
obtain samples of galaxies with a specific morphological type.

\subsection{Example 1: Selecting barred galaxies} 

Bars in galaxies are a trace of the dynamical state of the disk
\citep[\eg,][]{com09a,ath12}. Disks which have significant vertical motions
typically do not form bars (with the exception of tidally triggered bars;
\citealt{bar91}), but once a disk settles to a thin, dynamically cool state,
the formation of bars proceeds quickly. Most theoretical predictions show that
bars are long-lasting in the absence of significant galaxy interactions
\citep{ath05}, so the fraction of bars in disk galaxies can measure the
dynamical maturity of that population. For this reason, tracking the fraction
of bars in disk galaxies as a function of redshift has attracted significant
interest since the first resolved images of high redshift galaxies were
obtained \citep[\eg,][]{abr99,elm04,jog04,she08a, mel14,sim14}.

We create a sample of barred disk galaxies in GZH by applying cuts on the previous
tasks in Table~\ref{tbl:thresholds}. We first identify 11,049 ``bar
candidates,'' which are disk galaxies that are sufficiently face-on to attempt
visual identification of a bar. These galaxies were selected by applying the cuts $N_{bar} \ge 20$,
$f_{\rm features}>0.23, f_{\rm clumpy,no} > 0.30,$ and $f_{edgeon,no}>0.25$.
These galaxies are featured, non-clumpy, non-edge~on galaxies. Of these, a pure
sample of 730~barred disks was identified by applying a cut of \fbar$>0.7$. A
subsample of 50~galaxies were visually inspected and 94\% were found to contain
strong bars. A complete sample of strong and weak bars was created by applying
a cut of \fbar$>0.3$.  This sample contained 3,218~galaxies, 86\% of which were
found to contain weak or strong bars through visual inspection.

The resulting bar sample can be used to estimate the redshift evolution of bar
fraction; we find a steady decrease of \fbar$\sim 0.32$ at $z=0.4$ to
\fbar$\sim 0.24$ at $z=1.0$. This decrease agrees with
\citet{mel14}, although they report a lower overall bar fraction going from
\fbar$=0.22$ at $z=0.4$ to \fbar$=0.11$ at $z=1.0$. The difference in total bar
fraction is expected, as this analysis used a looser cut on \fbar, there is no
luminosity cut, and the use of debiased values for \ffeatures{} increases the
total amount of disks in the sample compared to \citet{mel14}. However, the
results from GZH indicate that if features are reliably identified, the bar is
always visible; for galaxies at increasing redshifts, the disks fade first but
the galactic bar remains visible \citep{mel16}. As a result, we do \emph{not}
recommend using an evolving cut on \fbar{} as a function of redshift for the
selection of barred disks. 

Another existing study using GZH to select barred galaxies was presented in
\citet{che15}. This study used a slightly different bar sample selection to
demonstrate that AGN hosts show no statistically significant enhancement in bar
fraction or average bar likelihood compared to closely-matched inactive
galaxies. We note that their technique matched test and control galaxies over
the same redshift ranges, which minimises the possible impact of redshift bias.

\subsection{Example 2: Identifying clump multiplicity} Clumps are known to be a
characteristic feature of galaxies at high redshift, and there is
evidence they play a crucial role in the evolution of modern spirals,
particularly in the formation of central bulges
\citep{elm05,elm14,guo15,beh16}. Simulations show clumps migrate from the outer
disk to the galactic centre in only a few orbital periods \citep{man15} and
observations show increasing bulge to clump mass and density ratios as the
Universe evolves since $z\sim 1.5$ \citep{elm09}, suggesting that clumps
coalesce over time to form the modern bulges of disk galaxies. GZH includes a
``clumpy'' path in the decision tree for the purposes of identifying clumps
and investigating their evolution with redshift. 

For galaxies identified as ``clumpy'' in GZH, the number of clumps can be
determined using Task~16. Table~\ref{tbl:thresholds} can be used to select
8,444~galaxies measured as ``clumpy'' using \ffeatures$> 0.23$ and $f_{\rm
clumpy,yes}>0.80$ to ensure the vote fractions for Task~16 are well-sampled.
The clump number can be reasonably identified for 1,112 of the clumpy galaxies;
for the remainder, the unique clumps were less distinguishable from each other
and the exact number of clumps could not be deduced without careful visual
inspection. In the 1,112 which did have distinguishable clumps, there are
61 one-clump galaxies using $f_{\rm 1~clump}>0.50$, 442 two-clump galaxies
using $f_{\rm 2~clumps}>0.80$, 275 three-clump galaxies using $f_{\rm
3~clumps}>0.75$, 71 four-clump galaxies using $f_{\rm 4~clumps} > 0.70$,
and 263~galaxies with more than four clumps using $f_{\rm >4~clumps}>0.70$.
Alternatively, these data may be used to create more general samples of
clumpy galaxies with few clumps and many clumps. A sample of 989 ``few
clumps'' galaxies can be made using $(f_{\rm 1~clump} + f_{\rm 2~clumps}) >
0.5$ and 2,910 ``many clumps'' galaxies using $(f_{\rm 3~clumps}+f_{\rm
4~clumps}+f_{\rm >4~clumps})>0.5$.

\begin{table}
\caption{Suggested thresholds for selecting morphological samples from GZH. \label{tbl:thresholds}}
\begin{tabular}{llll}
\hline\hline
No.      &  Task                & Previous         & Vote fraction threshold            \\
         &                      & task(s)          & $N_{\rm task} \ge 20$              \\
\hline
01       & smooth or features   & $-$              & $-$                                \\
02       & edge on              & 01,12            & $f_{\rm clumpy,no}>0.30$           \\
03       & bar                  & 01,12,02         & $f_{\rm edgeon,no}>0.25$           \\
         &                      & 01,12,16,15      & $f_{\rm clumpy~spiral}>0.65$       \\
04       & spiral arms          & 01,12,02         & $f_{\rm edgeon,no}>0.25$           \\
         &                      & 01,12,16,15      & $f_{\rm clumpy~spiral}>0.65$       \\
05       & bulge prominence     & 01,12,02         & $f_{\rm edgeon,no}>0.25$           \\
         &                      & 01,12,16,15      & $f_{\rm clumpy~spiral}>0.65$       \\
06       & odd yes/no           & $-$              & $-$                                \\
07       & rounded              & 01               & $f_{\rm smooth}>0.70$              \\
08       & odd feature          & 06               & $f_{\rm odd,yes}> 0.50$            \\
09       & bulge shape          & 01,12,02         & $f_{\rm edgeon,yes}>0.40$          \\
10       & arms winding         & 01,12,02,04      & $f_{\rm spiral,yes}>0.25$          \\
11       & arms number          & 01,12,02,04      & $f_{\rm spiral,yes}>0.25$          \\
12       & clumpy               & 01               & $f_{\rm features}>0.23$            \\
13       & bright clump         & 01,12,16         & $f_{\rm one~clump}< 0.40$          \\
14       & bright central clump & 01,12,16,13      & $f_{\rm bright~clump,yes}>0.50$    \\
15       & clump arrangement    & 01,12,16         & $f_{\rm multiple~clumps}>0.45$     \\
16       & clump count          & 01,12            & $f_{\rm clumpy,yes}>0.80$          \\
17       & clumps symmetrical   & 01,12            & $f_{\rm clumpy,yes}>0.80$          \\
18       & clumps embedded      & 01,12            & $f_{\rm clumpy,yes}>0.80$          \\
\hline\hline
\end{tabular}
\end{table}

\section{Analysis}\label{sec:analysis}

%\subsection{Effect of changing depth (GOODS, Stripe 82)}

\subsection{Demographics of morphology}

Any analysis of the morphological distribution of galaxies must properly
consider morphology with respect to other physical properties of the sample,
such as colour, mass, size, environment, and redshift. We defer such analyses to
later papers, and comment here only on a few broad characteristics of the
overall GZH sample. 

\begin{figure*}
\center
\includegraphics[width=0.90\textwidth]{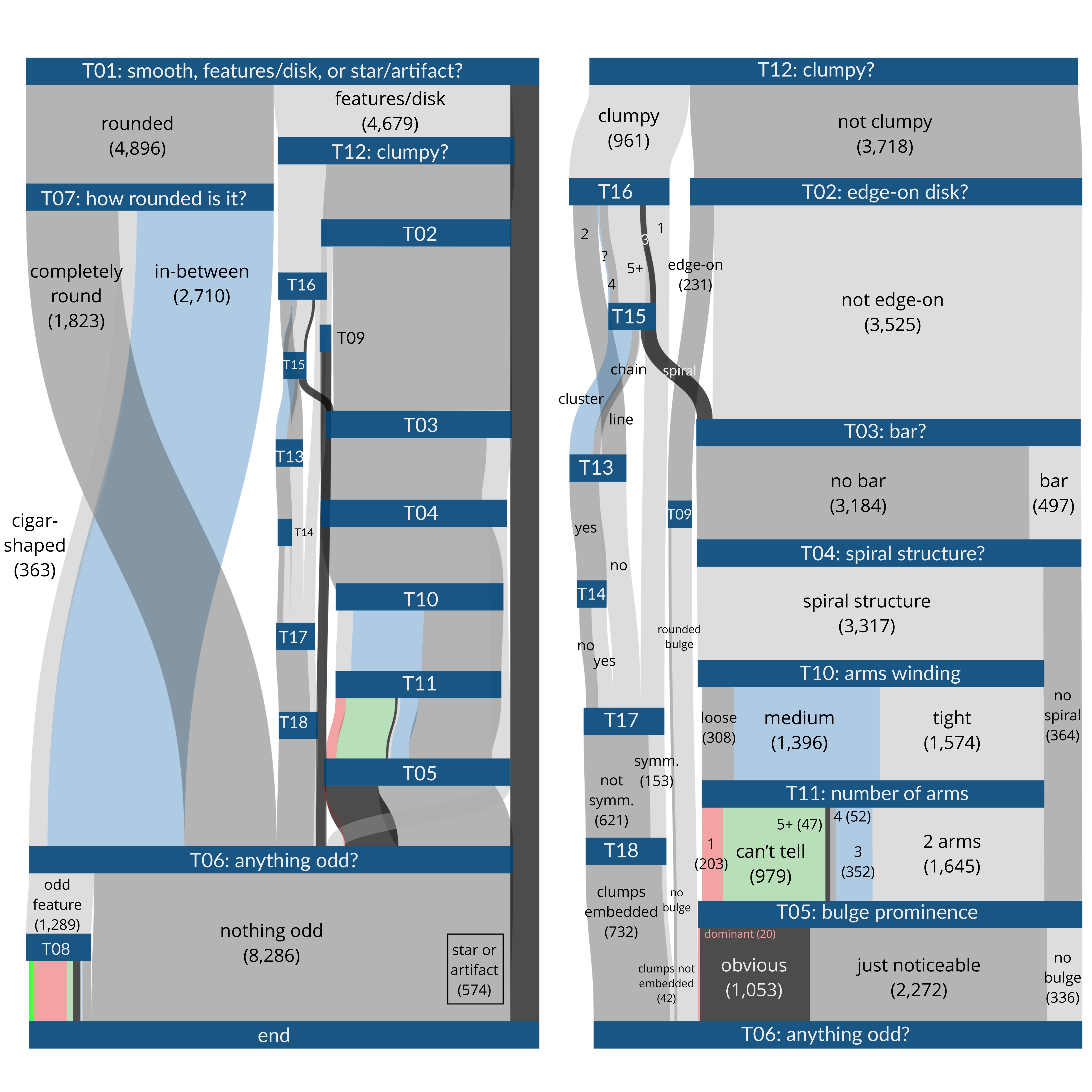}
\caption{Demographics of the morphologies for a volume-limited
sample ($z<1.0, m_\mathrm{I}<22.5$) of the COSMOS galaxies in GZH. Each node in the diagram (dark blue horizontal
bars) represents a task in the tree (Figure~\ref{fig:decisiontree}). The left diagram shows
the full decision tree. The right diagram zooms in on the features/clumpy
tasks only. Paths between tasks represent
each possible answer to the task, flowing from top to bottom between their
origin and the subsequent task in the tree. Labels are assigned to
galaxies based on the plurality answer for each task, with each galaxy assigned
only one label per node. Widths of the paths are proportional to the number
of galaxies assigned to that path. The widths of the nodes are proportional to
the number of galaxies for which the question was reliably answered.}
\label{fig:sankey}
\end{figure*}

As an example of visualising the overall morphological distribution,
Figure~\ref{fig:sankey} shows the breakdown of GZH morphologies as a flow
diagram. In order to show a physically meaningful sample,
Figure~\ref{fig:sankey} only includes images from a volume-limited sample
($z<1.0, m<22.5$ in $I$-band) in a single \hst{} survey (COSMOS, which has the
largest number of total galaxies).  We use a simple plurality vote for the
responses to each task to characterise the morphologies.  This emphasises one
of the basic results of the GZH project; namely, that there are comparable
numbers of spheroid and disk galaxies in the Universe at $z\sim1$, with
significant diversity in the arrangement of spiral arms and internal clumps.

The overall distribution of galaxy types is significantly different than the
low-redshift sample classified in SDSS imaging from GZ1 and GZ2. \citet{lin11}
found that elliptical galaxies exceeded spiral galaxies by a factor of
$\sim2:1$ in their spectroscopic sample when using a plurality vote criterion
(although \citealt{bam09} show that this strongly depends on the selection
method; spirals are the dominant population in a volume-limited sample at
$z<0.088$). The data for GZH show that smooth galaxies have a comparable total
population to ``featured'' galaxies of which roughly 20\% are dominated by
clumps rather than well-organized disks.  The fraction of objects identified as
stars or artifacts is also much higher in the \hst{} imaging. By plurality
votes, these encompass only $\sim0.1\%$ of images in SDSS \citep{wil13}, but
6\% of images in GZH. 

Within the sample of galaxies identified as ``not smooth'', it is clear that
the addition of the clumpy branch is necessary to describe a large fraction of
the sample, since disk-dominated galaxies outnumber clumpy morphologies by less than
a factor of 2. Disk galaxies are mostly unbarred \citep{mel14} and
possess two visible spiral arms over a flat distribution of pitch angles and
bulge prominence. Clumpy galaxies are identified across the full range of clump
multiplicities, with the exception of 1-clump galaxies (which would be
difficult to differentiate from compact spheroids). Roughly half of the
galaxies have at least 1~clump identified as the brightest. Clumps are most
commonly asymmetrically arranged in clusters and usually embedded within
larger objects. 

\subsection{Galaxy Zoo: Hubble and Galaxy Zoo: CANDELS}\label{ssec:gzh_gzc}

There are 7,681~galaxies in the GOODS-S field with morphological
classifications in both GZH and the Galaxy~Zoo:~CANDELS project (Simmons
et~al., submitted). Since both the sensitivity and filters for the two sets of
images are significantly different (and there is no correction for redshift
bias applied to GZC), there is no prior reason to expect a perfect correlation
between the separate vote fractions for the projects. Briefly, we note that the
\ffeatures{} value for GZH is on average higher than GZC; the effect is
strongest at $f_\mathrm{features,GZC}<0.3$, for which roughly half the galaxies
have $f_\mathrm{features,GZH}>0.5$. However, the correlation between vote
fractions is single-valued (although not linear, with a Pearson correlation
coefficient of $r=0.6$), and should be possible to calibrate using a similar
approach to that described in Section~\ref{sec:debiasing}; the correlation
between other tasks, such as edge-on galaxies, is significantly stronger
($r=0.9$). While the raw vote fractions are not directly comparable, the
initial analysis indicates that the broad morphologies are at least consistent.

\subsection{Comparing GZH morphologies to other catalogues}\label{ssec:comparisons}

All of the Legacy surveys included in the GZH imaging have had morphological
catalogues previously published; these have significant differences in
the number of galaxies, size and magnitude limits, classification scheme, and
the methods used for measuring morphology. These catalogues have been
cross-matched to GZH to compare results. 
%This is not presented as an
%endorsement of any particular method, but as an exploration of the strengths
%and weaknesses of the Galaxy~Zoo crowdsourced catalogues compared to products
%made with machine learning, automatic fits, and expert visual classification. 

The types and accuracy of morphological classification strongly depend on the
sample and methods being used. In an attempt to make a consistent comparison
between different techniques, galaxies are broadly grouped into three
categories: bulge-dominated/elliptical/smooth, disk-dominated/spiral, and
irregular/clumpy. These categories are matched to two GZH parameters:
\fbest, which identifies disk-dominated and clumpy galaxies, and \fodd, which
identifies deviations from well-formed spirals or S0s and which constitutes a
``catch-all'' for the variety of asymmetric morphologies that can constitute an
irregular galaxy. 

The comparison and analysis of the GZH morphologies is done on a set of matched,
volume-limited samples at $z<1.0$ and $m_{I|i|z} < 22.5$~mag.  The redshift and
magnitude limits for the sample are chosen to match the shared constraints for
the shallowest depths \citep[GEMS;][]{bun05} and the limits on morphological
reliability \citep[COSMOS/ZEST;][]{sca07} in the comparison catalogues. We match
all catalogues against GZH using a positional radius of 0.5\arcsec.

\begin{figure*}
\center
\includegraphics[width=1.0\textwidth]{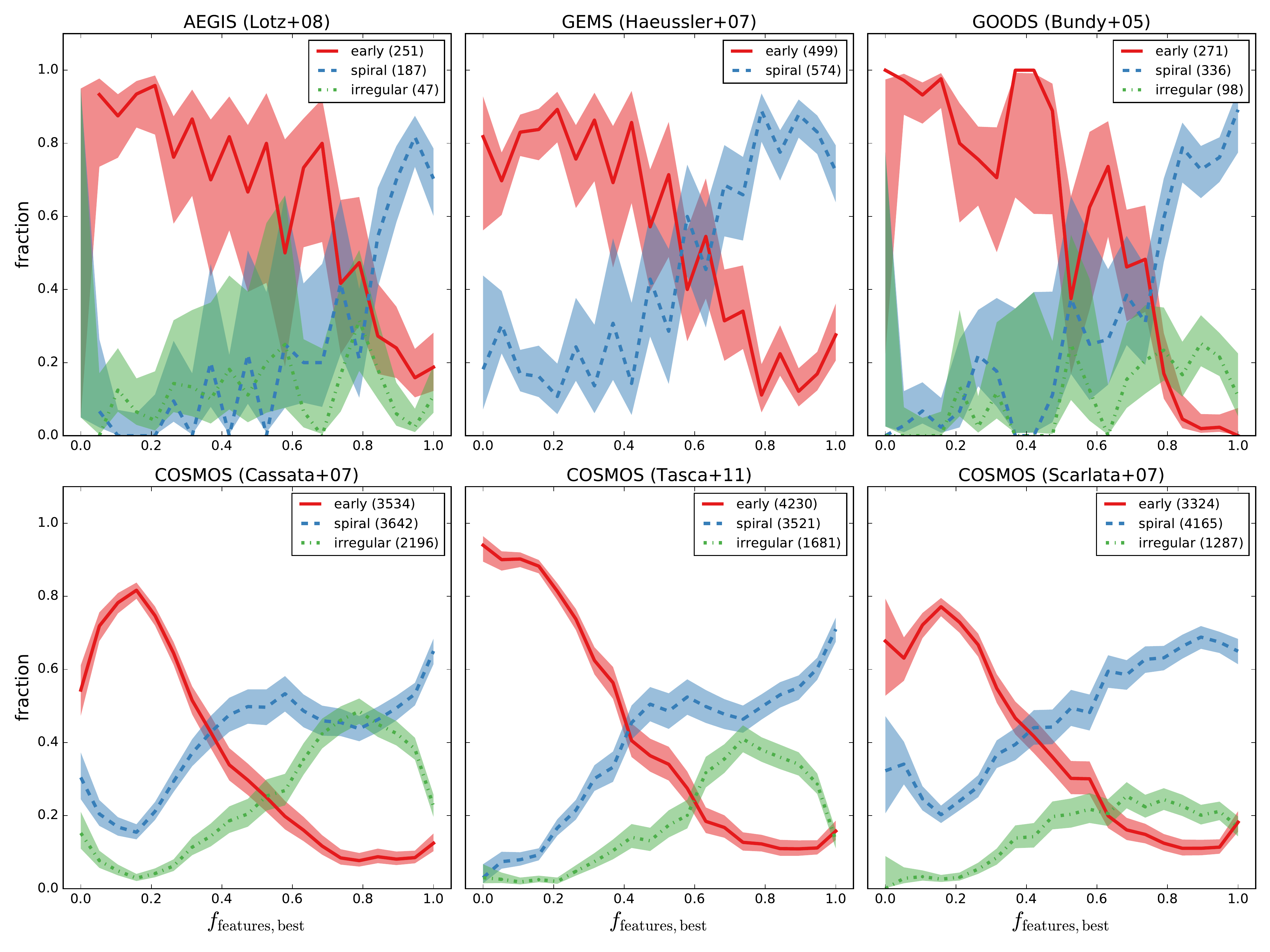}
\caption{Distributions of morphological parameters for a volume-limited sample
($z<1.0$, $m<22.5$) of galaxies matched between GZH and six published
morphological catalogues, split by survey (AEGIS, COSMOS, GEMS, and GOODS). This
plot shows the fraction of overall galaxies for each of the external
morphological categories as a function of \fbest{} in GZH. The shaded regions
around the binned fractions are confidence intervals calculated for a binomial
population \citep{cam11}.}
\label{fig:comparisons_features}
\end{figure*}

Figure~\ref{fig:comparisons_features} shows the proportion of galaxies as split
by their automated/expert visual morphologies for each of the six catalogues
matched to GZH. Galaxies from every catalogue show a strong correlation between
\fbest{} and the fraction of galaxies identified as spirals, with a
corresponding anti-correlation between \fbest{} and the fraction of
ellipticals.

AEGIS galaxies were morphologically classified using non-parametric
measurements by \citet{lot08}. Their method used a combination of the Gini
coefficient ($G$), which measures the relative inequality in pixel brightness,
and $M_{20}$, the second-order moment of the brightest 20\% of the light
\citep{lot04}. A linear combination of $G$ and $M_{20}$ delineates three broad
categories of galaxy morphology: E/S0/Sa (``elliptical''), Sb/Sc/Ir
(``spiral''), and mergers (``irregular''). We limit the AEGIS galaxies to those
with reliably-measured morphologies and $S/N>3$ in both $V$- and $I$-bands in
\citet{lot08}. 

The AEGIS sample had only 485~total galaxies matched with GZH, and so the
statistical uncertainties in the analysis are higher. However, at \fbest$>0.8$
there is a clear separation of featured and unfeatured galaxies in
Figure~\ref{fig:comparisons_features}, though this is not seen at lower \fbest.

Morphologies for GEMS galaxies were measured by \citet{hau07}, who used
single-component \sersic{} fits to the F850LP imaging. There are 8,846~galaxies
with measurements in both \citet{hau07} and GZH. The primary morphological
parameter in the automated catalogue is the \sersic{} index $n$ defining the
radial surface brightness profile.  ``Elliptical'' galaxies are selected by
$n>2.5$ and ``spirals''  by $n<2.5$.  There is no automatic measurement of
irregular or clumpy structure. 

Galaxies from GEMS also have a strong correlation with \fbest{}
(Figure~\ref{fig:comparisons_features}), which is somewhat surprising (but
encouraging) considering the known limits of assigning galaxy morphology based
only on a single \sersic{} parameter.  Visual inspection of images of the
exceptions --- where both \fbest{} and $n$ are high --- show that most are
obvious spirals but with prominent bulges. This means that the single-component
\sersic{} fit is likely choosing too small of an effective radius and missing
the extended disk structure for a large population of galaxies. This is a known
issue with using $n$ as a standalone measure of galaxy morphology -- it is
relatively common, for example, to measure high $n$ for an object with an
exponential disk but a central cusp in the light profile (eg, due to an AGN,
cluster, foreground star, noise, etc).

Galaxies in both of the GOODS fields down to a limit of $z_{AB}=22.5$ were
visually classified by a single expert (R.S.~Ellis), inspecting both $z$-band
and composite $Viz$ colour images \citep{bun05}. These morphologies are assigned
a numerical value based on categories in \citet{bri98a}: the corresponding
morphologies used are ``elliptical'' (classes 0,1,2), ``spirals'' (classes
3,4,5), and ``irregular'' (classes 6,7,8).

The GOODS galaxies show a very similar trend to the AEGIS images in
Figure~\ref{fig:comparisons_features}, with nearly 100\% pure samples at the
lowest and highest values of \fbest. We conjecture that the strong agreement
is at least partially due to the shared method of visual classification,
and reinforces the findings from several phases of Galaxy~Zoo that crowdsourced
classifications can be competitive with dedicated experts.

COSMOS galaxies have multiple published datasets automatically classifying
morphology, all using a variation of non-parametric measurements. \citet{cas07}
used a combination of concentration ($C$), asymmetry ($A$), $G$, and $M_{20}$
\citep{cas05} make empirical divisions based on several hundred training images
to assign galaxy morphology. A similar method was employed by \citet{tas11},
using the same non-parametric indices but with a different method of
calculating the Petrosian radius and total light profile, and using a
nearest-neighbors method to label morphologies. \citet[][ZEST]{sca07} used $C$,
$A$, $G$, $M_{20}$, the galaxy ellipticity ($\epsilon$), and \sersic{} index
($n$) to quantify morphology; a principal component analysis is used to assign
galaxies to categories. All three COSMOS catalogs use discrete morphological categories
of ellipticals, spirals, and peculiar/irregulars. We use the same categories
with the exception of assigning S0/bulge-dominated disks (type~2.0) from ZEST
into the unfeatured category.

Galaxies from the COSMOS catalogues all have morphologies that are
reasonably well-predicted by the GZH based on \fbest. The relation is strongest
for early-type galaxies, since the comparable fractions of irregulars and
spirals shows that clumpy/asymmetric morphologies are clearly also being
captured by the \fbest{} parameter. The small dropoff in the fraction of
early-types for the Cassata and ZEST samples at \fbest$<0.1$ in
Figure~\ref{fig:comparisons_features} is somewhat puzzling, but could
potentially be due to the presence of a small population of extremely
compact objects (stars or quasars) that would be classified as artifacts in
GZH.

A detailed comparison of all external catalogues is outside the scope of this
paper, but we emphasise only that it is impossible for any catalogue to be fully
consistent with \emph{all} of the previously published datasets (for example, only
64\% of the galaxies in \citealt{tas11} have the same morphological class as
the same galaxies in \citealt{cas07}). We interpret the broad agreements
between early-types, spirals, and irregulars between GZH and all three methods
of automated classification as a validation of both the crowdsourced visual
classifications provided by GZH and those automated methods.

It has been proposed that automated methods are preferable to visual
classification due to being quantified and more easily reproducible
\citep[\eg,][]{con14,paw16}. We argue that this distinction applies only to
visual classification done by individual and/or small numbers of experts. The
technique of crowdsourcing using a large number of independent classifiers
provides visual classification which is both quantified, and reproducible, and
(as demonstrated in this section), broadly agrees with all the automated
methods and expert visual classifications.

We repeated all analyses comparing morphologies between GZH and external
publications on the entire GZH dataset, rather than the volume-limited sample
($z<1.0,m<22.5$) discussed above. If galaxies at all redshifts and at the
original magnitude limit of $m<23.5$ are used, the correlations between all of
the automatic morphological populations and GZH decrease significantly.  The
exceptions are the GEMS galaxies, which are the only sample that was visually
classified and had a shallower magnitude cut on their catalogue.  Based on
these comparisons, we suggest that use of morphological catalogues with these
\hst{} images be limited to galaxies with $m_\mathrm{I|i|z}<22.5$; fainter
targets have been demonstrated to be less reliable, regardless of the method of
morphological classification being used.

\section{Summary}\label{sec:summary}

%Now people go and do science with these awesome GZH classifications. 

This paper presents the catalogue release for the Galaxy Zoo Hubble project,
which uses crowdsourced visual classifications to measure galaxy morphologies.
The first two phases of Galaxy Zoo \citep{lin11,wil13} used images of
low-redshift galaxies from SDSS; this is the first result of the project with
space-based images of high-redshift targets (in addition to the Galaxy Zoo:
CANDELS collaboration; Simmons et al., submitted). The final sample includes
classifications for 189,149~images generated from 145,741~unique galaxies
(of which 115,402 are in \hst{} Legacy imaging).

Galaxies were selected from a brightness-limited sample from multiple Legacy
surveys using the Advanced Camera for Surveys on the \hubble,
including AEGIS, GEMS, GOODS-N, GOODS-S, and COSMOS. The catalogue also includes
classifications for SDSS images from Stripe~82 at $z<0.25$; these images
serve both as a low-redshift anchor for
cosmological studies and a potential comparison for the different epochs of
classification between GZH and Galaxy Zoo 2 \citep{wil13}. 

The data in the GZH catalogues have been extensively reduced and tested. A
dominant effect is the known bias against identifying disky and asymmetric
sub-structures at either low resolution or surface brightness. This can be the
result either of genuinely small or dim galaxies, or a perceived effect from
observing galaxies at further distances (higher redshift). To calibrate this
\emph{without} potentially overcorrecting for the genuine morphological
evolution of galaxies over cosmic time, the GZH project uses SDSS images of
low-redshift galaxies, processes them to appear as if they were at higher
redshift, and classifies them through the GZH interface in an identical
fashion. The resulting change in \ffeatures{} as a function of $z$ and $\mu$ is
applied as a multiplicative correction to the top-level vote fractions for
$\sim50\%$ of the GZH galaxies. However, any population studies using
GZH morphological data must use a combination of debiased vote fractions
and upper limits from the ``uncorrectable'' galaxies, due to the evolving nature
of the thresholds as a function of redshift.

Galaxies in GZH show significant changes in the disk/elliptical fraction as a
function of redshift, along with an increasing number of galaxies dominated by
smaller clumps and presumed to be in the process of building up their baryonic
mass through a combination of hierarchical merging and {\it in-situ} star formation.
While the majority of scientific interpretation is left to future work, this
paper confirms the decrease in observed bar fraction with increasing redshift
and identifies a new way of selecting clumpy galaxies as a
function of clump multiplicity.

The full data tables for the catalogues can be accessed in machine-readable form
from both the journal and at \url{http://data.galaxyzoo.org}. All the code and
data tables used to generate this manuscript can be found at
\url{https://github.com/willettk/gzhubble}.
 
\section*{Acknowledgements}
KW, MG, CS, MB, and LF gratefully acknowledge support from the US National
Science Foundation Grant AST1413610.
Support for BDS was provided by the National Aeronautics and Space
Administration through Einstein Postdoctoral Fellowship Award Number PF5-160143
issued by the Chandra X-ray Observatory Center, which is operated by the
Smithsonian Astrophysical Observatory for and on behalf of the National
Aeronautics Space Administration under contract NAS8-03060.
KS gratefully acknowledges support from
Swiss National Science Foundation Grant PP00P2\_138979/1. TM acknowledges
funding from the Science and Technology Facilities Council ST/J500665/1. 
RJS is supported by the STFC grant code ST/K502236/1.
The development and hosting of Galaxy Zoo Hubble was supported by a grant from
the Alfred P. Sloan Foundation. The Zooniverse acknowledges support from a
Google Global Impact Award. 

We thank Meg Schwamb and the ASIAA for hosting the ``Citizen Science in
Astronomy'' workshop, 3-7 Mar 2014 in Taipei, Taiwan, at which some of this
analysis was initiated. We thank Jennifer Lotz for sharing her $G$-$M_{20}$
measurements for the AEGIS sample. We thank Coleman Krawczyk for his assistance
in producing Figure~\ref{fig:decisiontree}. We thank Nathan Cloutier and Brent
Hilgart for useful discussions. We also thank the referee for thoughtful comments
which improved the quality of this paper. 

This project made heavy use of the Astropy packages in Python \citep{ast13},
the \texttt{seaborn} plotting package \citep{was15}, astroML \citep{van12}, and
TOPCAT \citep{tay05,tay11}. Modified code from Nick~Wherry and David~Schlegel
was used to create the JPG images. Figure~\ref{fig:sankey} was generated with
\url{http://sankeymatic.com/}. \citet{hol05} provided valuable assistance
in interpreting \sextractor{} output.

This work is based on (GO-10134, GO-09822, GO-09425.01, GO-09583.01, GO-9500)
program observations with the NASA/ESA Hubble Space Telescope, obtained at the
Space Telescope Science Institute, which is operated by the Association of
Universities for Research in Astronomy, Inc., under NASA contract NAS 5-26555. 

Funding for the SDSS and SDSS-II has been provided by the Alfred P. Sloan
Foundation, the Participating Institutions, the National Science Foundation,
the U.S. Department of Energy, the National Aeronautics and Space
Administration, the Japanese Monbukagakusho, the Max Planck Society, and the
Higher Education Funding Council for England. The SDSS website is
\url{http://www.sdss.org/}. 

The SDSS is managed by the Astrophysical Research Consortium for the
Participating Institutions. The Participating Institutions are the American
Museum of Natural History, Astrophysical Institute Potsdam, University of
Basel, University of Cambridge, Case Western Reserve University, University of
Chicago, Drexel University, Fermilab, the Institute for Advanced Study, the
Japan Participation Group, Johns Hopkins University, the Joint Institute for
Nuclear Astrophysics, the Kavli Institute for Particle Astrophysics and
Cosmology, the Korean Scientist Group, the Chinese Academy of Sciences
(LAMOST), Los Alamos National Laboratory, the Max-Planck-Institute for
Astronomy (MPIA), the Max-Planck-Institute for Astrophysics (MPA), New Mexico
State University, Ohio State University, University of Pittsburgh, University
of Portsmouth, Princeton University, the United States Naval Observatory and
the University of Washington. 

\bibliographystyle{mnras}
\bibliography{gz_hubble_data}

\begin{thebibliography}{}
\makeatletter
\relax
\def\mn@urlcharsother{\let\do\@makeother \do\$\do\&\do\#\do\^\do\_\do\%\do\~}
\def\mn@doi{\begingroup\mn@urlcharsother \@ifnextchar [ {\mn@doi@}
  {\mn@doi@[]}}
\def\mn@doi@[#1]#2{\def\@tempa{#1}\ifx\@tempa\@empty \href
  {http://dx.doi.org/#2} {doi:#2}\else \href {http://dx.doi.org/#2} {#1}\fi
  \endgroup}
\def\mn@eprint#1#2{\mn@eprint@#1:#2::\@nil}
\def\mn@eprint@arXiv#1{\href {http://arxiv.org/abs/#1} {{\tt arXiv:#1}}}
\def\mn@eprint@dblp#1{\href {http://dblp.uni-trier.de/rec/bibtex/#1.xml}
  {dblp:#1}}
\def\mn@eprint@#1:#2:#3:#4\@nil{\def\@tempa {#1}\def\@tempb {#2}\def\@tempc
  {#3}\ifx \@tempc \@empty \let \@tempc \@tempb \let \@tempb \@tempa \fi \ifx
  \@tempb \@empty \def\@tempb {arXiv}\fi \@ifundefined
  {mn@eprint@\@tempb}{\@tempb:\@tempc}{\expandafter \expandafter \csname
  mn@eprint@\@tempb\endcsname \expandafter{\@tempc}}}

\bibitem[\protect\citeauthoryear{{Abazajian} et~al.,}{{Abazajian}
  et~al.}{2009}]{aba09}
{Abazajian} K.~N.,  et~al., 2009, \mn@doi [\apjs]
  {10.1088/0067-0049/182/2/543}, \href
  {http://adsabs.harvard.edu/abs/2009ApJS..182..543A} {182, 543}

\bibitem[\protect\citeauthoryear{{Abraham}, {Merrifield}, {Ellis}, {Tanvir}  \&
  {Brinchmann}}{{Abraham} et~al.}{1999}]{abr99}
{Abraham} R.~G.,  {Merrifield} M.~R.,  {Ellis} R.~S.,  {Tanvir} N.~R.,
  {Brinchmann} J.,  1999, \mn@doi [\mnras] {10.1046/j.1365-8711.1999.02766.x},
  \href {http://adsabs.harvard.edu/abs/1999MNRAS.308..569A} {308, 569}

\bibitem[\protect\citeauthoryear{{Abraham}, {van den Bergh}  \&
  {Nair}}{{Abraham} et~al.}{2003}]{abr03}
{Abraham} R.~G.,  {van den Bergh} S.,   {Nair} P.,  2003, \mn@doi [\apj]
  {10.1086/373919}, \href {http://adsabs.harvard.edu/abs/2003ApJ...588..218A}
  {588, 218}

\bibitem[\protect\citeauthoryear{{Astropy Collaboration} et~al.,}{{Astropy
  Collaboration} et~al.}{2013}]{ast13}
{Astropy Collaboration} et~al., 2013, \mn@doi [\aap]
  {10.1051/0004-6361/201322068}, \href
  {http://adsabs.harvard.edu/abs/2013A%26A...558A..33A} {558, A33}

\bibitem[\protect\citeauthoryear{{Athanassoula}}{{Athanassoula}}{2012}]{ath12}
{Athanassoula} E.,  2012, \mn@doi [\mnras] {10.1111/j.1745-3933.2012.01320.x},
  \href {http://adsabs.harvard.edu/abs/2012MNRAS.426L..46A} {426, L46}

\bibitem[\protect\citeauthoryear{{Athanassoula}, {Lambert}  \&
  {Dehnen}}{{Athanassoula} et~al.}{2005}]{ath05}
{Athanassoula} E.,  {Lambert} J.~C.,   {Dehnen} W.,  2005, \mn@doi [\mnras]
  {10.1111/j.1365-2966.2005.09445.x}, \href
  {http://adsabs.harvard.edu/abs/2005MNRAS.363..496A} {363, 496}

\bibitem[\protect\citeauthoryear{{Baillard} et~al.,}{{Baillard}
  et~al.}{2011}]{bai11}
{Baillard} A.,  et~al., 2011, \mn@doi [\aap] {10.1051/0004-6361/201016423},
  \href {http://adsabs.harvard.edu/abs/2011A%26A...532A..74B} {532, A74}

\bibitem[\protect\citeauthoryear{{Bamford}, {Rojas}, {Nichol}, {Miller},
  {Wasserman}, {Genovese}  \& {Freeman}}{{Bamford} et~al.}{2008}]{bam08}
{Bamford} S.~P.,  {Rojas} A.~L.,  {Nichol} R.~C.,  {Miller} C.~J.,  {Wasserman}
  L.,  {Genovese} C.~R.,   {Freeman} P.~E.,  2008, \mn@doi [\mnras]
  {10.1111/j.1365-2966.2008.13963.x}, \href
  {http://adsabs.harvard.edu/abs/2008MNRAS.391..607B} {391, 607}

\bibitem[\protect\citeauthoryear{{Bamford} et~al.,}{{Bamford}
  et~al.}{2009}]{bam09}
{Bamford} S.~P.,  et~al., 2009, \mn@doi [\mnras]
  {10.1111/j.1365-2966.2008.14252.x}, \href
  {http://adsabs.harvard.edu/abs/2009MNRAS.393.1324B} {393, 1324}

\bibitem[\protect\citeauthoryear{{Barden}, {Jahnke}  \&
  {H{\"a}u{\ss}ler}}{{Barden} et~al.}{2008}]{bar08a}
{Barden} M.,  {Jahnke} K.,   {H{\"a}u{\ss}ler} B.,  2008, \mn@doi [\apjs]
  {10.1086/524039}, \href {http://adsabs.harvard.edu/abs/2008ApJS..175..105B}
  {175, 105}

\bibitem[\protect\citeauthoryear{{Barden}, {H{\"a}u{\ss}ler}, {Peng},
  {McIntosh}  \& {Guo}}{{Barden} et~al.}{2012}]{bar12}
{Barden} M.,  {H{\"a}u{\ss}ler} B.,  {Peng} C.~Y.,  {McIntosh} D.~H.,   {Guo}
  Y.,  2012, \mn@doi [\mnras] {10.1111/j.1365-2966.2012.20619.x}, \href
  {http://adsabs.harvard.edu/abs/2012MNRAS.422..449B} {422, 449}

\bibitem[\protect\citeauthoryear{{Barnes} \& {Hernquist}}{{Barnes} \&
  {Hernquist}}{1991}]{bar91}
{Barnes} J.~E.,  {Hernquist} L.~E.,  1991, \mn@doi [\apjl] {10.1086/185978},
  \href {http://adsabs.harvard.edu/abs/1991ApJ...370L..65B} {370, L65}

\bibitem[\protect\citeauthoryear{{Beckwith} et~al.,}{{Beckwith}
  et~al.}{2006}]{bec06}
{Beckwith} S.~V.~W.,  et~al., 2006, \mn@doi [\aj] {10.1086/507302}, \href
  {http://adsabs.harvard.edu/abs/2006AJ....132.1729B} {132, 1729}

\bibitem[\protect\citeauthoryear{{Behrendt}, {Burkert}  \&
  {Schartmann}}{{Behrendt} et~al.}{2016}]{beh16}
{Behrendt} M.,  {Burkert} A.,   {Schartmann} M.,  2016, \mn@doi [\apjl]
  {10.3847/2041-8205/819/1/L2}, \href
  {http://adsabs.harvard.edu/abs/2016ApJ...819L...2B} {819, L2}

\bibitem[\protect\citeauthoryear{{Bell} et~al.,}{{Bell} et~al.}{2012}]{bel12}
{Bell} E.~F.,  et~al., 2012, \mn@doi [\apj] {10.1088/0004-637X/753/2/167},
  \href {http://adsabs.harvard.edu/abs/2012ApJ...753..167B} {753, 167}

\bibitem[\protect\citeauthoryear{{Bertin} \& {Arnouts}}{{Bertin} \&
  {Arnouts}}{1996}]{ber96}
{Bertin} E.,  {Arnouts} S.,  1996, \mn@doi [\aaps] {10.1051/aas:1996164}, \href
  {http://adsabs.harvard.edu/abs/1996A%26AS..117..393B} {117, 393}

\bibitem[\protect\citeauthoryear{{Brinchmann} et~al.,}{{Brinchmann}
  et~al.}{1998}]{bri98a}
{Brinchmann} J.,  et~al., 1998, \mn@doi [\apj] {10.1086/305621}, \href
  {http://adsabs.harvard.edu/abs/1998ApJ...499..112B} {499, 112}

\bibitem[\protect\citeauthoryear{{Bundy}, {Ellis}  \& {Conselice}}{{Bundy}
  et~al.}{2005}]{bun05}
{Bundy} K.,  {Ellis} R.~S.,   {Conselice} C.~J.,  2005, \mn@doi [\apj]
  {10.1086/429549}, \href {http://adsabs.harvard.edu/abs/2005ApJ...625..621B}
  {625, 621}

\bibitem[\protect\citeauthoryear{{Buta}}{{Buta}}{2013}]{but13}
{Buta} R.~J.,  2013, {Galaxy Morphology}.
Springer, pp 1--89, \mn@doi{10.1007/978-94-007-5609-0_1}

\bibitem[\protect\citeauthoryear{{Caldwell} et~al.,}{{Caldwell}
  et~al.}{2008}]{cal08}
{Caldwell} J.~A.~R.,  et~al., 2008, \mn@doi [\apjs] {10.1086/521080}, \href
  {http://adsabs.harvard.edu/abs/2008ApJS..174..136C} {174, 136}

\bibitem[\protect\citeauthoryear{{Cameron}}{{Cameron}}{2011}]{cam11}
{Cameron} E.,  2011, \mn@doi [\pasa] {10.1071/AS10046}, \href
  {http://adsabs.harvard.edu/abs/2011PASA...28..128C} {28, 128}

\bibitem[\protect\citeauthoryear{{Cameron}, {Carollo}, {Oesch}, {Bouwens},
  {Illingworth}, {Trenti}, {Labb{\'e}}  \& {Magee}}{{Cameron}
  et~al.}{2011}]{cam11a}
{Cameron} E.,  {Carollo} C.~M.,  {Oesch} P.~A.,  {Bouwens} R.~J.,
  {Illingworth} G.~D.,  {Trenti} M.,  {Labb{\'e}} I.,   {Magee} D.,  2011,
  \mn@doi [\apj] {10.1088/0004-637X/743/2/146}, \href
  {http://adsabs.harvard.edu/abs/2011ApJ...743..146C} {743, 146}

\bibitem[\protect\citeauthoryear{{Cardamone} et~al.,}{{Cardamone}
  et~al.}{2010}]{car10}
{Cardamone} C.~N.,  et~al., 2010, \mn@doi [\apjs]
  {10.1088/0067-0049/189/2/270}, \href
  {http://adsabs.harvard.edu/abs/2010ApJS..189..270C} {189, 270}

\bibitem[\protect\citeauthoryear{{Cassata} et~al.,}{{Cassata}
  et~al.}{2005}]{cas05}
{Cassata} P.,  et~al., 2005, \mn@doi [\mnras]
  {10.1111/j.1365-2966.2005.08657.x}, \href
  {http://adsabs.harvard.edu/abs/2005MNRAS.357..903C} {357, 903}

\bibitem[\protect\citeauthoryear{{Cassata} et~al.,}{{Cassata}
  et~al.}{2007}]{cas07}
{Cassata} P.,  et~al., 2007, \mn@doi [\apjs] {10.1086/516591}, \href
  {http://adsabs.harvard.edu/abs/2007ApJS..172..270C} {172, 270}

\bibitem[\protect\citeauthoryear{Cheung et~al.,}{Cheung et~al.}{2015}]{che15}
Cheung E.,  et~al., 2015, \mn@doi [\mnras] {10.1093/mnras/stu2462}, \href
  {http://labs.adsabs.harvard.edu/adsabs/abs/2015MNRAS.447..506C/} {447, 510}

\bibitem[\protect\citeauthoryear{{Chevance}, {Weijmans}, {Damjanov}, {Abraham},
  {Simard}, {van den Bergh}, {Caris}  \& {Glazebrook}}{{Chevance}
  et~al.}{2012}]{che12}
{Chevance} M.,  {Weijmans} A.-M.,  {Damjanov} I.,  {Abraham} R.~G.,  {Simard}
  L.,  {van den Bergh} S.,  {Caris} E.,   {Glazebrook} K.,  2012, \mn@doi
  [\apjl] {10.1088/2041-8205/754/2/L24}, \href
  {http://adsabs.harvard.edu/abs/2012ApJ...754L..24C} {754, L24}

\bibitem[\protect\citeauthoryear{{Combes}}{{Combes}}{2009}]{com09a}
{Combes} F.,  2009, in {Jogee} S.,  {Marinova} I.,  {Hao} L.,   {Blanc} G.~A.,
  eds,  Astronomical Society of the Pacific Conference Series Vol. 419, Galaxy
  Evolution: Emerging Insights and Future Challenges. p.~31 (\mn@eprint {arXiv}
  {0901.0178})

\bibitem[\protect\citeauthoryear{{Conselice}}{{Conselice}}{2003}]{con03}
{Conselice} C.~J.,  2003, \mn@doi [\apjs] {10.1086/375001}, \href
  {http://adsabs.harvard.edu/abs/2003ApJS..147....1C} {147, 1}

\bibitem[\protect\citeauthoryear{{Conselice}}{{Conselice}}{2014}]{con14}
{Conselice} C.~J.,  2014, \mn@doi [\araa]
  {10.1146/annurev-astro-081913-040037}, \href
  {http://adsabs.harvard.edu/abs/2014ARA%26A..52..291C} {52, 291}

\bibitem[\protect\citeauthoryear{{Darg} et~al.,}{{Darg} et~al.}{2010}]{dar10}
{Darg} D.~W.,  et~al., 2010, \mn@doi [\mnras]
  {10.1111/j.1365-2966.2009.15786.x}, \href
  {http://adsabs.harvard.edu/abs/2010MNRAS.401.1552D} {401, 1552}

\bibitem[\protect\citeauthoryear{{Davis} et~al.,}{{Davis} et~al.}{2007}]{dav07}
{Davis} M.,  et~al., 2007, \mn@doi [\apjl] {10.1086/517931}, \href
  {http://adsabs.harvard.edu/abs/2007ApJ...660L...1D} {660, L1}

\bibitem[\protect\citeauthoryear{{Dressler}}{{Dressler}}{1980}]{dre80}
{Dressler} A.,  1980, \mn@doi [\apj] {10.1086/157753}, \href
  {http://adsabs.harvard.edu/abs/1980ApJ...236..351D} {236, 351}

\bibitem[\protect\citeauthoryear{{Elmegreen} \& {Elmegreen}}{{Elmegreen} \&
  {Elmegreen}}{2014}]{elm14}
{Elmegreen} D.~M.,  {Elmegreen} B.~G.,  2014, \mn@doi [\apj]
  {10.1088/0004-637X/781/1/11}, \href
  {http://adsabs.harvard.edu/abs/2014ApJ...781...11E} {781, 11}

\bibitem[\protect\citeauthoryear{{Elmegreen}, {Elmegreen}  \&
  {Hirst}}{{Elmegreen} et~al.}{2004}]{elm04}
{Elmegreen} B.~G.,  {Elmegreen} D.~M.,   {Hirst} A.~C.,  2004, \mn@doi [\apj]
  {10.1086/422407}, \href {http://adsabs.harvard.edu/abs/2004ApJ...612..191E}
  {612, 191}

\bibitem[\protect\citeauthoryear{{Elmegreen}, {Elmegreen}, {Rubin}  \&
  {Schaffer}}{{Elmegreen} et~al.}{2005}]{elm05}
{Elmegreen} D.~M.,  {Elmegreen} B.~G.,  {Rubin} D.~S.,   {Schaffer} M.~A.,
  2005, \mn@doi [\apj] {10.1086/432502}, \href
  {http://adsabs.harvard.edu/abs/2005ApJ...631...85E} {631, 85}

\bibitem[\protect\citeauthoryear{{Elmegreen}, {Elmegreen}, {Ferguson}  \&
  {Mullan}}{{Elmegreen} et~al.}{2007}]{elm07}
{Elmegreen} D.~M.,  {Elmegreen} B.~G.,  {Ferguson} T.,   {Mullan} B.,  2007,
  \mn@doi [\apj] {10.1086/518715}, \href
  {http://adsabs.harvard.edu/abs/2007ApJ...663..734E} {663, 734}

\bibitem[\protect\citeauthoryear{{Elmegreen}, {Elmegreen}, {Marcus},
  {Shahinyan}, {Yau}  \& {Petersen}}{{Elmegreen} et~al.}{2009}]{elm09}
{Elmegreen} D.~M.,  {Elmegreen} B.~G.,  {Marcus} M.~T.,  {Shahinyan} K.,  {Yau}
  A.,   {Petersen} M.,  2009, \mn@doi [\apj] {10.1088/0004-637X/701/1/306},
  \href {http://adsabs.harvard.edu/abs/2009ApJ...701..306E} {701, 306}

\bibitem[\protect\citeauthoryear{{F{\"o}rster Schreiber}, {Shapley}, {Erb},
  {Genzel}, {Steidel}, {Bouch{\'e}}, {Cresci}  \& {Davies}}{{F{\"o}rster
  Schreiber} et~al.}{2011}]{for11a}
{F{\"o}rster Schreiber} N.~M.,  {Shapley} A.~E.,  {Erb} D.~K.,  {Genzel} R.,
  {Steidel} C.~C.,  {Bouch{\'e}} N.,  {Cresci} G.,   {Davies} R.,  2011,
  \mn@doi [\apj] {10.1088/0004-637X/731/1/65}, \href
  {http://adsabs.harvard.edu/abs/2011ApJ...731...65F} {731, 65}

\bibitem[\protect\citeauthoryear{{Fortson} et~al.,}{{Fortson}
  et~al.}{2012}]{for12}
{Fortson} L.,  et~al., 2012, {Galaxy Zoo: Morphological Classification and
  Citizen Science}.
CRC Press, Taylor \& Francis Group, pp 213--236

\bibitem[\protect\citeauthoryear{{Freeman}, {Izbicki}, {Lee}, {Newman},
  {Conselice}, {Koekemoer}, {Lotz}  \& {Mozena}}{{Freeman}
  et~al.}{2013}]{fre13}
{Freeman} P.~E.,  {Izbicki} R.,  {Lee} A.~B.,  {Newman} J.~A.,  {Conselice}
  C.~J.,  {Koekemoer} A.~M.,  {Lotz} J.~M.,   {Mozena} M.,  2013, \mn@doi
  [\mnras] {10.1093/mnras/stt1016}, \href
  {http://adsabs.harvard.edu/abs/2013MNRAS.434..282F} {434, 282}

\bibitem[\protect\citeauthoryear{{Galloway} et~al.,}{{Galloway}
  et~al.}{2015}]{gal15}
{Galloway} M.~A.,  et~al., 2015, \mn@doi [\mnras] {10.1093/mnras/stv235}, \href
  {http://adsabs.harvard.edu/abs/2015MNRAS.448.3442G} {448, 3442}

\bibitem[\protect\citeauthoryear{{Genel} et~al.,}{{Genel} et~al.}{2014}]{gen14}
{Genel} S.,  et~al., 2014, \mn@doi [\mnras] {10.1093/mnras/stu1654}, \href
  {http://adsabs.harvard.edu/abs/2014MNRAS.445..175G} {445, 175}

\bibitem[\protect\citeauthoryear{{Giavalisco}}{{Giavalisco}}{2012}]{gia12}
{Giavalisco} M.,  2012, VizieR Online Data Catalog, \href
  {http://adsabs.harvard.edu/abs/2012yCat.2318....0G} {2318}

\bibitem[\protect\citeauthoryear{{Giavalisco} et~al.,}{{Giavalisco}
  et~al.}{2004}]{gia04}
{Giavalisco} M.,  et~al., 2004, \mn@doi [\apjl] {10.1086/379232}, \href
  {http://adsabs.harvard.edu/abs/2004ApJ...600L..93G} {600, L93}

\bibitem[\protect\citeauthoryear{{Griffith} et~al.,}{{Griffith}
  et~al.}{2012}]{gri12}
{Griffith} R.~L.,  et~al., 2012, \mn@doi [\apjs] {10.1088/0067-0049/200/1/9},
  \href {http://adsabs.harvard.edu/abs/2012ApJS..200....9G} {200, 9}

\bibitem[\protect\citeauthoryear{{Grogin} et~al.,}{{Grogin}
  et~al.}{2011}]{gro11}
{Grogin} N.~A.,  et~al., 2011, \mn@doi [\apjs] {10.1088/0067-0049/197/2/35},
  \href {http://adsabs.harvard.edu/abs/2011ApJS..197...35G} {197, 35}

\bibitem[\protect\citeauthoryear{{Guo} et~al.,}{{Guo} et~al.}{2015}]{guo15}
{Guo} Y.,  et~al., 2015, \mn@doi [\apj] {10.1088/0004-637X/800/1/39}, \href
  {http://adsabs.harvard.edu/abs/2015ApJ...800...39G} {800, 39}

\bibitem[\protect\citeauthoryear{{H{\"a}u{\ss}ler} et~al.,}{{H{\"a}u{\ss}ler}
  et~al.}{2007}]{hau07}
{H{\"a}u{\ss}ler} B.,  et~al., 2007, \mn@doi [\apjs] {10.1086/518836}, \href
  {http://adsabs.harvard.edu/abs/2007ApJS..172..615H} {172, 615}

\bibitem[\protect\citeauthoryear{{Hinshaw} et~al.,}{{Hinshaw}
  et~al.}{2013}]{hin13}
{Hinshaw} G.,  et~al., 2013, \mn@doi [\apjs] {10.1088/0067-0049/208/2/19},
  \href {http://adsabs.harvard.edu/abs/2013ApJS..208...19H} {208, 19}

\bibitem[\protect\citeauthoryear{{Holwerda}}{{Holwerda}}{2005}]{hol05}
{Holwerda} B.~W.,  2005, ArXiv Astrophysics e-prints, \href
  {http://adsabs.harvard.edu/abs/2005astro.ph.12139H} {}

\bibitem[\protect\citeauthoryear{{Hopkins} et~al.,}{{Hopkins}
  et~al.}{2010}]{hop10}
{Hopkins} P.~F.,  et~al., 2010, \mn@doi [\apj] {10.1088/0004-637X/715/1/202},
  \href {http://adsabs.harvard.edu/abs/2010ApJ...715..202H} {715, 202}

\bibitem[\protect\citeauthoryear{{Hubble}}{{Hubble}}{1926}]{hub26}
{Hubble} E.~P.,  1926, \mn@doi [\apj] {10.1086/143018}, \href
  {http://adsabs.harvard.edu/abs/1926ApJ....64..321H} {64, 321}

\bibitem[\protect\citeauthoryear{{Hubble}}{{Hubble}}{1936}]{hub36}
{Hubble} E.~P.,  1936, {Realm of the Nebulae}.
Yale University Press

\bibitem[\protect\citeauthoryear{{Ilbert} et~al.,}{{Ilbert}
  et~al.}{2013}]{ilb13}
{Ilbert} O.,  et~al., 2013, \mn@doi [\aap] {10.1051/0004-6361/201321100}, \href
  {http://adsabs.harvard.edu/abs/2013A%26A...556A..55I} {556, A55}

\bibitem[\protect\citeauthoryear{{Jogee} et~al.,}{{Jogee} et~al.}{2004}]{jog04}
{Jogee} S.,  et~al., 2004, \mn@doi [\apjl] {10.1086/426138}, \href
  {http://adsabs.harvard.edu/abs/2004ApJ...615L.105J} {615, L105}

\bibitem[\protect\citeauthoryear{{Kartaltepe} et~al.,}{{Kartaltepe}
  et~al.}{2015}]{kar15}
{Kartaltepe} J.~S.,  et~al., 2015, \mn@doi [\apjs]
  {10.1088/0067-0049/221/1/11}, \href
  {http://adsabs.harvard.edu/abs/2015ApJS..221...11K} {221, 11}

\bibitem[\protect\citeauthoryear{{Kaviraj}}{{Kaviraj}}{2014a}]{kav14a}
{Kaviraj} S.,  2014a, \mn@doi [\mnras] {10.1093/mnrasl/slt136}, \href
  {http://adsabs.harvard.edu/abs/2014MNRAS.437L..41K} {437, L41}

\bibitem[\protect\citeauthoryear{{Kaviraj}}{{Kaviraj}}{2014b}]{kav14}
{Kaviraj} S.,  2014b, \mn@doi [\mnras] {10.1093/mnras/stu338}, \href
  {http://adsabs.harvard.edu/abs/2014MNRAS.440.2944K} {440, 2944}

\bibitem[\protect\citeauthoryear{{Koekemoer}, {Fruchter}, {Hook}  \&
  {Hack}}{{Koekemoer} et~al.}{2002}]{koe02}
{Koekemoer} A.~M.,  {Fruchter} A.~S.,  {Hook} R.~N.,   {Hack} W.,  2002, in The
  2002 HST Calibration Workshop : Hubble after the Installation of the ACS and
  the NICMOS Cooling System, Proceedings of a Workshop held at the Space
  Telescope Science Institute, Baltimore, Maryland, October 17 and 18, 2002.
  Edited by Santiago Arribas, Anton Koekemoer, and Brad Whitmore. Baltimore,
  MD: Space Telescope Science Institute, 2002., p.339. pp 339--+

\bibitem[\protect\citeauthoryear{{Koekemoer}, {Fruchter}  \&
  {Hack}}{{Koekemoer} et~al.}{2003}]{koe03}
{Koekemoer} A.,  {Fruchter} A.,   {Hack} W.,  2003, Space Telescope European
  Coordinating Facility Newsletter, Volume 33, p.10, \href
  {http://adsabs.harvard.edu/abs/2003STECF..33...10K} {33, 10}

\bibitem[\protect\citeauthoryear{{Koekemoer} et~al.,}{{Koekemoer}
  et~al.}{2007}]{koe07}
{Koekemoer} A.~M.,  et~al., 2007, \mn@doi [\apjs] {10.1086/520086}, \href
  {http://adsabs.harvard.edu/abs/2007ApJS..172..196K} {172, 196}

\bibitem[\protect\citeauthoryear{{Koekemoer} et~al.,}{{Koekemoer}
  et~al.}{2011}]{koe11}
{Koekemoer} A.~M.,  et~al., 2011, \mn@doi [\apjs] {10.1088/0067-0049/197/2/36},
  \href {http://adsabs.harvard.edu/abs/2011ApJS..197...36K} {197, 36}

\bibitem[\protect\citeauthoryear{{Krist}}{{Krist}}{1993}]{kri93}
{Krist} J.,  1993, in {R.~J.~Hanisch, R.~J.~V.~Brissenden, \& J.~Barnes} ed.,
  Astronomical Society of the Pacific Conference Series Vol. 52, Astronomical
  Data Analysis Software and Systems II. pp 536--+

\bibitem[\protect\citeauthoryear{{Lackner} \& {Gunn}}{{Lackner} \&
  {Gunn}}{2012}]{lac12}
{Lackner} C.~N.,  {Gunn} J.~E.,  2012, \mn@doi [\mnras]
  {10.1111/j.1365-2966.2012.20450.x}, \href
  {http://adsabs.harvard.edu/abs/2012MNRAS.421.2277L} {421, 2277}

\bibitem[\protect\citeauthoryear{{Lahav} et~al.,}{{Lahav} et~al.}{1995}]{lah95}
{Lahav} O.,  et~al., 1995, \mn@doi [Science] {10.1126/science.267.5199.859},
  \href {http://adsabs.harvard.edu/abs/1995Sci...267..859L} {267, 859}

\bibitem[\protect\citeauthoryear{{Land} et~al.,}{{Land} et~al.}{2008}]{lan08}
{Land} K.,  et~al., 2008, \mn@doi [\mnras] {10.1111/j.1365-2966.2008.13490.x},
  \href {http://adsabs.harvard.edu/abs/2008MNRAS.388.1686L} {388, 1686}

\bibitem[\protect\citeauthoryear{{Law}, {Shapley}, {Steidel}, {Reddy},
  {Christensen}  \& {Erb}}{{Law} et~al.}{2012a}]{law12}
{Law} D.~R.,  {Shapley} A.~E.,  {Steidel} C.~C.,  {Reddy} N.~A.,  {Christensen}
  C.~R.,   {Erb} D.~K.,  2012a, \mn@doi [\nat] {10.1038/nature11256}, \href
  {http://adsabs.harvard.edu/abs/2012Natur.487..338L} {487, 338}

\bibitem[\protect\citeauthoryear{{Law}, {Steidel}, {Shapley}, {Nagy}, {Reddy}
  \& {Erb}}{{Law} et~al.}{2012b}]{law12a}
{Law} D.~R.,  {Steidel} C.~C.,  {Shapley} A.~E.,  {Nagy} S.~R.,  {Reddy} N.~A.,
    {Erb} D.~K.,  2012b, \mn@doi [\apj] {10.1088/0004-637X/745/1/85}, \href
  {http://adsabs.harvard.edu/abs/2012ApJ...745...85L} {745, 85}

\bibitem[\protect\citeauthoryear{{Lilly} et~al.,}{{Lilly} et~al.}{1998}]{lil98}
{Lilly} S.,  et~al., 1998, \mn@doi [\apj] {10.1086/305713}, \href
  {http://adsabs.harvard.edu/abs/1998ApJ...500...75L} {500, 75}

\bibitem[\protect\citeauthoryear{{Lintott} et~al.,}{{Lintott}
  et~al.}{2008}]{lin08}
{Lintott} C.~J.,  et~al., 2008, \mn@doi [\mnras]
  {10.1111/j.1365-2966.2008.13689.x}, \href
  {http://adsabs.harvard.edu/abs/2008MNRAS.389.1179L} {389, 1179}

\bibitem[\protect\citeauthoryear{{Lintott} et~al.,}{{Lintott}
  et~al.}{2011}]{lin11}
{Lintott} C.,  et~al., 2011, \mn@doi [\mnras]
  {10.1111/j.1365-2966.2010.17432.x}, \href
  {http://adsabs.harvard.edu/abs/2011MNRAS.410..166L} {410, 166}

\bibitem[\protect\citeauthoryear{{Lotz}, {Primack}  \& {Madau}}{{Lotz}
  et~al.}{2004}]{lot04}
{Lotz} J.~M.,  {Primack} J.,   {Madau} P.,  2004, \mn@doi [\aj]
  {10.1086/421849}, \href {http://adsabs.harvard.edu/abs/2004AJ....128..163L}
  {128, 163}

\bibitem[\protect\citeauthoryear{{Lotz} et~al.,}{{Lotz} et~al.}{2008}]{lot08}
{Lotz} J.~M.,  et~al., 2008, \mn@doi [\apj] {10.1086/523659}, \href
  {http://adsabs.harvard.edu/abs/2008ApJ...672..177L} {672, 177}

\bibitem[\protect\citeauthoryear{{Loveday} et~al.,}{{Loveday}
  et~al.}{2012}]{lov12}
{Loveday} J.,  et~al., 2012, \mn@doi [\mnras]
  {10.1111/j.1365-2966.2011.20111.x}, \href
  {http://adsabs.harvard.edu/abs/2012MNRAS.420.1239L} {420, 1239}

\bibitem[\protect\citeauthoryear{{Lupton}, {Blanton}, {Fekete}, {Hogg},
  {O'Mullane}, {Szalay}  \& {Wherry}}{{Lupton} et~al.}{2004}]{lup04}
{Lupton} R.,  {Blanton} M.~R.,  {Fekete} G.,  {Hogg} D.~W.,  {O'Mullane} W.,
  {Szalay} A.,   {Wherry} N.,  2004, \mn@doi [\pasp] {10.1086/382245}, \href
  {http://adsabs.harvard.edu/abs/2004PASP..116..133L} {116, 133}

\bibitem[\protect\citeauthoryear{{Mandelker}, {Dekel}, {Ceverino}, {DeGraf},
  {Guo}  \& {Primack}}{{Mandelker} et~al.}{2015}]{man15}
{Mandelker} N.,  {Dekel} A.,  {Ceverino} D.,  {DeGraf} C.,  {Guo} Y.,
  {Primack} J.,  2015, preprint, \href
  {http://adsabs.harvard.edu/abs/2015arXiv151208791M} {} (\mn@eprint {arXiv}
  {1512.08791})

\bibitem[\protect\citeauthoryear{{Mao}, {Mo}  \& {White}}{{Mao}
  et~al.}{1998}]{mao98}
{Mao} S.,  {Mo} H.~J.,   {White} S.~D.~M.,  1998, \mn@doi [\mnras]
  {10.1046/j.1365-8711.1998.01766.x}, \href
  {http://adsabs.harvard.edu/abs/1998MNRAS.297L..71M} {297, L71}

\bibitem[\protect\citeauthoryear{{Masters} et~al.,}{{Masters}
  et~al.}{2011}]{mas11c}
{Masters} K.~L.,  et~al., 2011, \mn@doi [\mnras]
  {10.1111/j.1365-2966.2010.17834.x}, \href
  {http://adsabs.harvard.edu/abs/2011MNRAS.411.2026M} {411, 2026}

\bibitem[\protect\citeauthoryear{{Melvin}}{{Melvin}}{2016}]{mel16}
{Melvin} T.,  2016, PhD thesis, University of Portsmouth

\bibitem[\protect\citeauthoryear{{Melvin} et~al.,}{{Melvin}
  et~al.}{2014}]{mel14}
{Melvin} T.,  et~al., 2014, \mn@doi [\mnras] {10.1093/mnras/stt2397}, \href
  {http://adsabs.harvard.edu/abs/2014MNRAS.tmp...97M} {}

\bibitem[\protect\citeauthoryear{{Momcheva} et~al.,}{{Momcheva}
  et~al.}{2015}]{mom15}
{Momcheva} I.~G.,  et~al., 2015, preprint, \href
  {http://adsabs.harvard.edu/abs/2015arXiv151002106M} {} (\mn@eprint {arXiv}
  {1510.02106})

\bibitem[\protect\citeauthoryear{{Mortlock} et~al.,}{{Mortlock}
  et~al.}{2013}]{mor13}
{Mortlock} A.,  et~al., 2013, \mn@doi [\mnras] {10.1093/mnras/stt793}, \href
  {http://adsabs.harvard.edu/abs/2013MNRAS.433.1185M} {433, 1185}

\bibitem[\protect\citeauthoryear{{Nair} \& {Abraham}}{{Nair} \&
  {Abraham}}{2010}]{nai10}
{Nair} P.~B.,  {Abraham} R.~G.,  2010, \mn@doi [\apjs]
  {10.1088/0067-0049/186/2/427}, \href
  {http://adsabs.harvard.edu/abs/2010ApJS..186..427N} {186, 427}

\bibitem[\protect\citeauthoryear{{Nieto-Santisteban}, {Szalay}  \&
  {Gray}}{{Nieto-Santisteban} et~al.}{2004}]{nie04}
{Nieto-Santisteban} M.~A.,  {Szalay} A.~S.,   {Gray} J.,  2004, in {Ochsenbein}
  F.,  {Allen} M.~G.,   {Egret} D.,  eds,  Astronomical Society of the Pacific
  Conference Series Vol. 314, Astronomical Data Analysis Software and Systems
  (ADASS) XIII. p.~666

\bibitem[\protect\citeauthoryear{{Pawlik}, {Wild}, {Walcher}, {Johansson},
  {Villforth}, {Rowlands}, {Mendez-Abreu}  \& {Hewlett}}{{Pawlik}
  et~al.}{2016}]{paw16}
{Pawlik} M.~M.,  {Wild} V.,  {Walcher} C.~J.,  {Johansson} P.~H.,  {Villforth}
  C.,  {Rowlands} K.,  {Mendez-Abreu} J.,   {Hewlett} T.,  2016, \mn@doi
  [\mnras] {10.1093/mnras/stv2878}, \href
  {http://adsabs.harvard.edu/abs/2016MNRAS.456.3032P} {456, 3032}

\bibitem[\protect\citeauthoryear{{Peng}, {Ho}, {Impey}  \& {Rix}}{{Peng}
  et~al.}{2002}]{pen02a}
{Peng} C.~Y.,  {Ho} L.~C.,  {Impey} C.~D.,   {Rix} H.-W.,  2002, \mn@doi [\aj]
  {10.1086/340952}, \href {http://adsabs.harvard.edu/abs/2002AJ....124..266P}
  {124, 266}

\bibitem[\protect\citeauthoryear{{Pierce} et~al.,}{{Pierce}
  et~al.}{2010}]{pie10a}
{Pierce} C.~M.,  et~al., 2010, \mn@doi [\mnras]
  {10.1111/j.1365-2966.2010.16502.x}, \href
  {http://adsabs.harvard.edu/abs/2010MNRAS.405..718P} {405, 718}

\bibitem[\protect\citeauthoryear{{Rix} et~al.,}{{Rix} et~al.}{2004}]{rix04}
{Rix} H.-W.,  et~al., 2004, \mn@doi [\apjs] {10.1086/420885}, \href
  {http://adsabs.harvard.edu/abs/2004ApJS..152..163R} {152, 163}

\bibitem[\protect\citeauthoryear{{S{\'a}nchez} et~al.,}{{S{\'a}nchez}
  et~al.}{2004}]{san04}
{S{\'a}nchez} S.~F.,  et~al., 2004, \mn@doi [\apj] {10.1086/423234}, \href
  {http://adsabs.harvard.edu/cgi-bin/nph-bib_query?bibcode=2004ApJ...614..586S&db_key=AST}
  {614, 586}

\bibitem[\protect\citeauthoryear{{Sandage}}{{Sandage}}{1961}]{san61}
{Sandage} A.,  1961, {The Hubble atlas of galaxies}.
{Carnegie Institute of Washington}

\bibitem[\protect\citeauthoryear{{Scarlata} et~al.,}{{Scarlata}
  et~al.}{2007}]{sca07}
{Scarlata} C.,  et~al., 2007, \mn@doi [\apjs] {10.1086/516582}, \href
  {http://adsabs.harvard.edu/abs/2007ApJS..172..406S} {172, 406}

\bibitem[\protect\citeauthoryear{{Schawinski} et~al.,}{{Schawinski}
  et~al.}{2014}]{sch14}
{Schawinski} K.,  et~al., 2014, \mn@doi [\mnras] {10.1093/mnras/stu327}, \href
  {http://adsabs.harvard.edu/abs/2014MNRAS.440..889S} {440, 889}

\bibitem[\protect\citeauthoryear{{Schaye} et~al.,}{{Schaye}
  et~al.}{2015}]{sch15}
{Schaye} J.,  et~al., 2015, \mn@doi [\mnras] {10.1093/mnras/stu2058}, \href
  {http://adsabs.harvard.edu/abs/2015MNRAS.446..521S} {446, 521}

\bibitem[\protect\citeauthoryear{{Scoville} et~al.,}{{Scoville}
  et~al.}{2007}]{sco07}
{Scoville} N.,  et~al., 2007, \mn@doi [\apjs] {10.1086/516585}, \href
  {http://adsabs.harvard.edu/abs/2007ApJS..172....1S} {172, 1}

\bibitem[\protect\citeauthoryear{{Sheth} et~al.,}{{Sheth}
  et~al.}{2008}]{she08a}
{Sheth} K.,  et~al., 2008, \mn@doi [\apj] {10.1086/524980}, \href
  {http://adsabs.harvard.edu/abs/2008ApJ...675.1141S} {675, 1141}

\bibitem[\protect\citeauthoryear{{Silk} \& {Mamon}}{{Silk} \&
  {Mamon}}{2012}]{sil12}
{Silk} J.,  {Mamon} G.~A.,  2012, \mn@doi [Research in Astronomy and
  Astrophysics] {10.1088/1674-4527/12/8/004}, \href
  {http://adsabs.harvard.edu/abs/2012RAA....12..917S} {12, 917}

\bibitem[\protect\citeauthoryear{{Simard}, {Mendel}, {Patton}, {Ellison}  \&
  {McConnachie}}{{Simard} et~al.}{2011}]{sim11}
{Simard} L.,  {Mendel} J.~T.,  {Patton} D.~R.,  {Ellison} S.~L.,
  {McConnachie} A.~W.,  2011, \mn@doi [\apjs] {10.1088/0067-0049/196/1/11},
  \href {http://adsabs.harvard.edu/abs/2011ApJS..196...11S} {196, 11}

\bibitem[\protect\citeauthoryear{{Simmons} \& {Urry}}{{Simmons} \&
  {Urry}}{2008}]{sim08}
{Simmons} B.~D.,  {Urry} C.~M.,  2008, \mn@doi [\apj] {10.1086/589827}, \href
  {http://adsabs.harvard.edu/abs/2008ApJ...683..644S} {683, 644}

\bibitem[\protect\citeauthoryear{{Simmons}, {Van Duyne}, {Urry}, {Treister},
  {Koekemoer}, {Grogin}  \& {The GOODS Team}}{{Simmons} et~al.}{2011}]{simm11}
{Simmons} B.~D.,  {Van Duyne} J.,  {Urry} C.~M.,  {Treister} E.,  {Koekemoer}
  A.~M.,  {Grogin} N.~A.,   {The GOODS Team} 2011, \mn@doi [\apj]
  {10.1088/0004-637X/734/2/121}, \href
  {http://adsabs.harvard.edu/abs/2011ApJ...734..121S} {734, 121}

\bibitem[\protect\citeauthoryear{{Simmons} et~al.,}{{Simmons}
  et~al.}{2013}]{sim13}
{Simmons} B.~D.,  et~al., 2013, \mn@doi [\mnras] {10.1093/mnras/sts491}, \href
  {http://adsabs.harvard.edu/abs/2013MNRAS.429.2199S} {429, 2199}

\bibitem[\protect\citeauthoryear{{Simmons} et~al.,}{{Simmons}
  et~al.}{2014}]{sim14}
{Simmons} B.~D.,  et~al., 2014, \mn@doi [\mnras] {10.1093/mnras/stu1817}, \href
  {http://adsabs.harvard.edu/abs/2014MNRAS.445.3466S} {445, 3466}

\bibitem[\protect\citeauthoryear{{Skibba} et~al.,}{{Skibba}
  et~al.}{2009}]{ski09}
{Skibba} R.~A.,  et~al., 2009, \mn@doi [\mnras]
  {10.1111/j.1365-2966.2009.15334.x}, \href
  {http://adsabs.harvard.edu/abs/2009MNRAS.399..966S} {399, 966}

\bibitem[\protect\citeauthoryear{{Skibba} et~al.,}{{Skibba}
  et~al.}{2012}]{ski12}
{Skibba} R.~A.,  et~al., 2012, \mn@doi [\mnras]
  {10.1111/j.1365-2966.2012.20972.x}, \href
  {http://adsabs.harvard.edu/abs/2012MNRAS.423.1485S} {423, 1485}

\bibitem[\protect\citeauthoryear{{Smethurst} et~al.,}{{Smethurst}
  et~al.}{2015}]{sme15}
{Smethurst} R.~J.,  et~al., 2015, \mn@doi [\mnras] {10.1093/mnras/stv161},
  \href {http://adsabs.harvard.edu/abs/2015MNRAS.450..435S} {450, 435}

\bibitem[\protect\citeauthoryear{{Smethurst} et~al.,}{{Smethurst}
  et~al.}{2016}]{sme16}
{Smethurst} R.~J.,  et~al., 2016, preprint, \href
  {http://adsabs.harvard.edu/abs/2016arXiv160900023S} {} (\mn@eprint {arXiv}
  {1609.00023})

\bibitem[\protect\citeauthoryear{{Steinmetz} \& {Navarro}}{{Steinmetz} \&
  {Navarro}}{2002}]{ste02}
{Steinmetz} M.,  {Navarro} J.~F.,  2002, \mn@doi [\na]
  {10.1016/S1384-1076(02)00102-1}, \href
  {http://adsabs.harvard.edu/abs/2002NewA....7..155S} {7, 155}

\bibitem[\protect\citeauthoryear{{Stetson}}{{Stetson}}{1987}]{ste87}
{Stetson} P.~B.,  1987, \mn@doi [\pasp] {10.1086/131977}, \href
  {http://adsabs.harvard.edu/abs/1987PASP...99..191S} {99, 191}

\bibitem[\protect\citeauthoryear{{Strauss} et~al.,}{{Strauss}
  et~al.}{2002}]{str02}
{Strauss} M.~A.,  et~al., 2002, \mn@doi [\aj] {10.1086/342343}, \href
  {http://adsabs.harvard.edu/abs/2002AJ....124.1810S} {124, 1810}

\bibitem[\protect\citeauthoryear{{Taniguchi} et~al.,}{{Taniguchi}
  et~al.}{2007}]{tan07}
{Taniguchi} Y.,  et~al., 2007, \mn@doi [\apjs] {10.1086/516596}, \href
  {http://adsabs.harvard.edu/abs/2007ApJS..172....9T} {172, 9}

\bibitem[\protect\citeauthoryear{{Tasca}}{{Tasca}}{2011}]{tas11}
{Tasca} L.~A.~M.,  2011, VizieR Online Data Catalog, \href
  {http://adsabs.harvard.edu/abs/2011yCat.7265....0T} {7265}

\bibitem[\protect\citeauthoryear{{Taylor}}{{Taylor}}{2005}]{tay05}
{Taylor} M.~B.,  2005, in {Shopbell} P.,  {Britton} M.,   {Ebert} R.,  eds,
  Astronomical Society of the Pacific Conference Series Vol. 347, Astronomical
  Data Analysis Software and Systems XIV. p.~29

\bibitem[\protect\citeauthoryear{{Taylor}}{{Taylor}}{2011}]{tay11}
{Taylor} M.,  2011, {TOPCAT: Tool for OPerations on Catalogues And Tables}
  (\mn@eprint {ascl} {1101.010})

\bibitem[\protect\citeauthoryear{{Toomre} \& {Toomre}}{{Toomre} \&
  {Toomre}}{1972}]{too72}
{Toomre} A.,  {Toomre} J.,  1972, \mn@doi [\apj] {10.1086/151823}, \href
  {http://adsabs.harvard.edu/abs/1972ApJ...178..623T} {178, 623}

\bibitem[\protect\citeauthoryear{{Vanderplas}, {Connolly}, {Ivezi{\'c}}  \&
  {Gray}}{{Vanderplas} et~al.}{2012}]{van12}
{Vanderplas} J.,  {Connolly} A.,  {Ivezi{\'c}} {\v Z}.,   {Gray} A.,  2012, in
  Conference on Intelligent Data Understanding (CIDU). pp 47 --54,
  \mn@doi{10.1109/CIDU.2012.6382200}

\bibitem[\protect\citeauthoryear{{Vogelsberger} et~al.,}{{Vogelsberger}
  et~al.}{2014}]{vog14a}
{Vogelsberger} M.,  et~al., 2014, \mn@doi [\mnras] {10.1093/mnras/stu1536},
  \href {http://adsabs.harvard.edu/abs/2014MNRAS.444.1518V} {444, 1518}

\bibitem[\protect\citeauthoryear{Waskom et~al.,}{Waskom et~al.}{2015}]{was15}
Waskom M.,  et~al., 2015, seaborn: v0.6.0 (June 2015),
  \mn@doi{10.5281/zenodo.19108}, \url {http://dx.doi.org/10.5281/zenodo.19108}

\bibitem[\protect\citeauthoryear{{Willett} et~al.,}{{Willett}
  et~al.}{2013}]{wil13}
{Willett} K.~W.,  et~al., 2013, \mn@doi [\mnras] {10.1093/mnras/stt1458}, \href
  {http://adsabs.harvard.edu/abs/2013MNRAS.435.2835W} {435, 2835}

\bibitem[\protect\citeauthoryear{{Willett} et~al.,}{{Willett}
  et~al.}{2015}]{wil15}
{Willett} K.~W.,  et~al., 2015, \mn@doi [\mnras] {10.1093/mnras/stv307}, \href
  {http://adsabs.harvard.edu/abs/2015MNRAS.449..820W} {449, 820}

\bibitem[\protect\citeauthoryear{{Williams} et~al.,}{{Williams}
  et~al.}{1996}]{wil96}
{Williams} R.~E.,  et~al., 1996, \mn@doi [\aj] {10.1086/118105}, \href
  {http://adsabs.harvard.edu/abs/1996AJ....112.1335W} {112, 1335}

\bibitem[\protect\citeauthoryear{{Wright}}{{Wright}}{2006}]{wri06}
{Wright} E.~L.,  2006, \mn@doi [\pasp] {10.1086/510102}, \href
  {http://adsabs.harvard.edu/abs/2006PASP..118.1711W} {118, 1711}

\bibitem[\protect\citeauthoryear{{York} et~al.,}{{York} et~al.}{2000}]{yor00}
{York} D.~G.,  et~al., 2000, \mn@doi [\aj] {10.1086/301513}, \href
  {http://adsabs.harvard.edu/abs/2000AJ....120.1579Y} {120, 1579}

\bibitem[\protect\citeauthoryear{{de~Vaucouleurs}}{{de~Vaucouleurs}}{1959}]{dev59}
{de~Vaucouleurs} G.,  1959, Handbuch der Physik, \href
  {http://adsabs.harvard.edu/abs/1959HDP....53..275D} {53, 275}

\bibitem[\protect\citeauthoryear{{van den Bergh}}{{van den
  Bergh}}{1976}]{van76}
{van den Bergh} S.,  1976, \mn@doi [\apj] {10.1086/154452}, \href
  {http://adsabs.harvard.edu/abs/1976ApJ...206..883V} {206, 883}

\makeatother
\end{thebibliography}

\newpage
\clearpage

\appendix

\section{GOODS Shallow Depth data}

\begin{table}
\caption{Correctable fractions for the top-level task in GZH in the GOODS
shallow-depth (2-epoch) images. \label{tbl:goods_shallow_categories}}
\begin{tabular}{lrr|r}
\hline\hline
                                   & GOODS-N & GOODS-S & Total \\
\hline
correctable                        & 1,051   &   730   & 1,781 \\
lower-limit                        &   131   &   334   &   465 \\
no correction needed ($z \le 0.3$) &   267   &   267   &   534 \\ 
NEI                                &   943   & 2,078   & 3,021 \\
no redshift information            &   159   &   184   &   343 \\
\hline
total                              & 2,551   & 3,593   & 6,144 \\
\hline\hline
\end{tabular}
\end{table}

GZH used both 5-epoch and 2-epoch sets of data to construct the GOODS set of
images. The 11,157 full depth 5-epoch images are used in the \main{} catalogue; the
classifications for the 6,144 \goods{} images are provided as a
supplementary table. This section analyses the effect of image depth on the
ability of the Galaxy~Zoo classifiers to identify features or disk structure in the images. 

\subsection{Comparing shallow and full depth morphologies}\label{ssec:depth_comparison}

\begin{figure*}
\begin{center}
\includegraphics[width=0.50\textwidth]{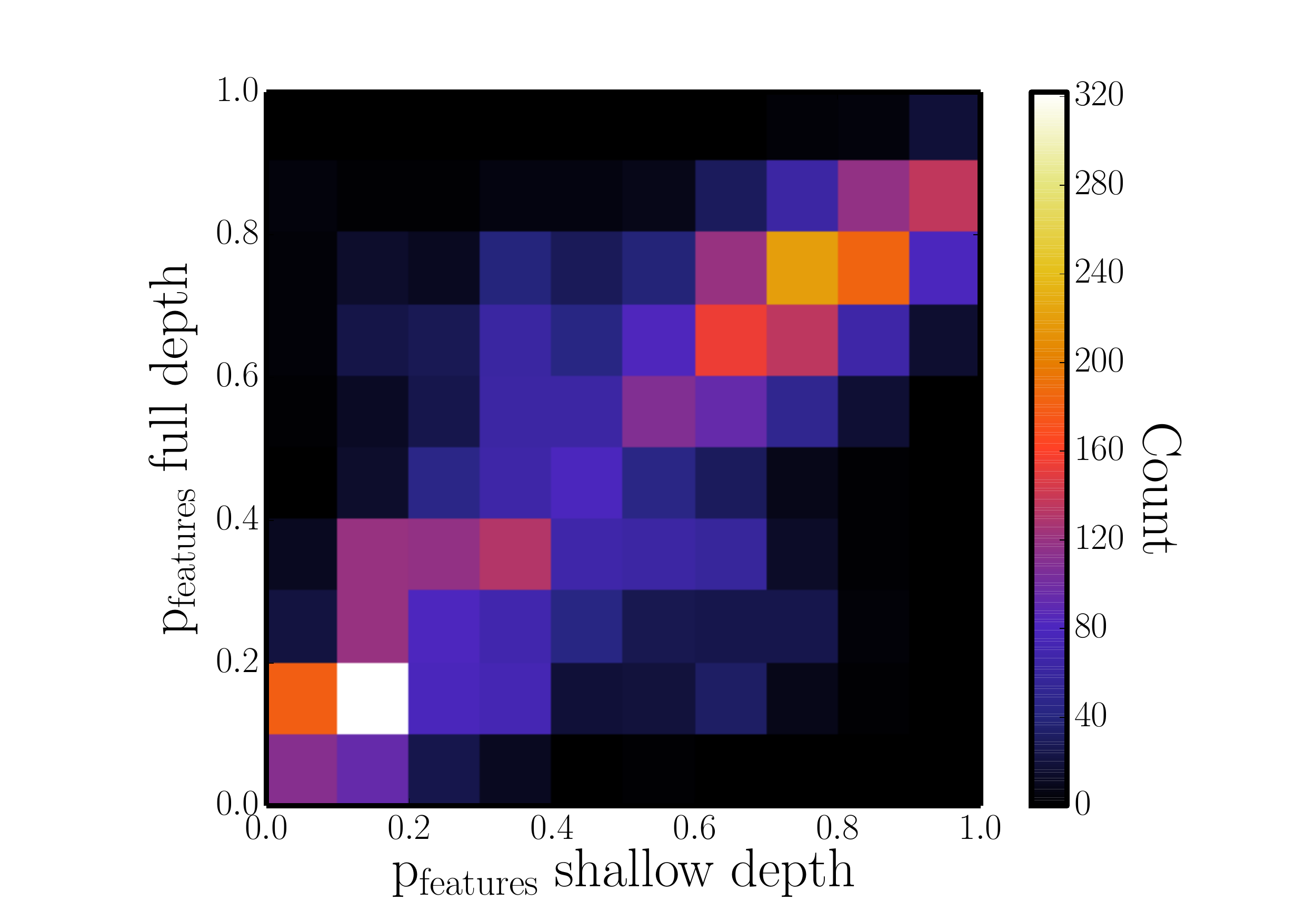}
\caption{Distribution of \ffeatures{} for the 4,460 GOODS galaxies with both
shallow (2-epoch; \goods) and full-depth (5-epoch; \main) images
morphologically classified in GZH. For most galaxies, the value of \ffeatures{}
is consistent ($\Delta f_\mathrm{features}<0.2$) between depths. Examples of
galaxies with sharp changes in \ffeatures, as well as those with little to no
change are shown in
Figures~\ref{fig:shallow_smooth}-\ref{fig:shallow_featured}.}
\label{fig:shallow_vs_full}
\end{center}
\end{figure*}

\begin{figure*}
\centering
\includegraphics[width=\textwidth]{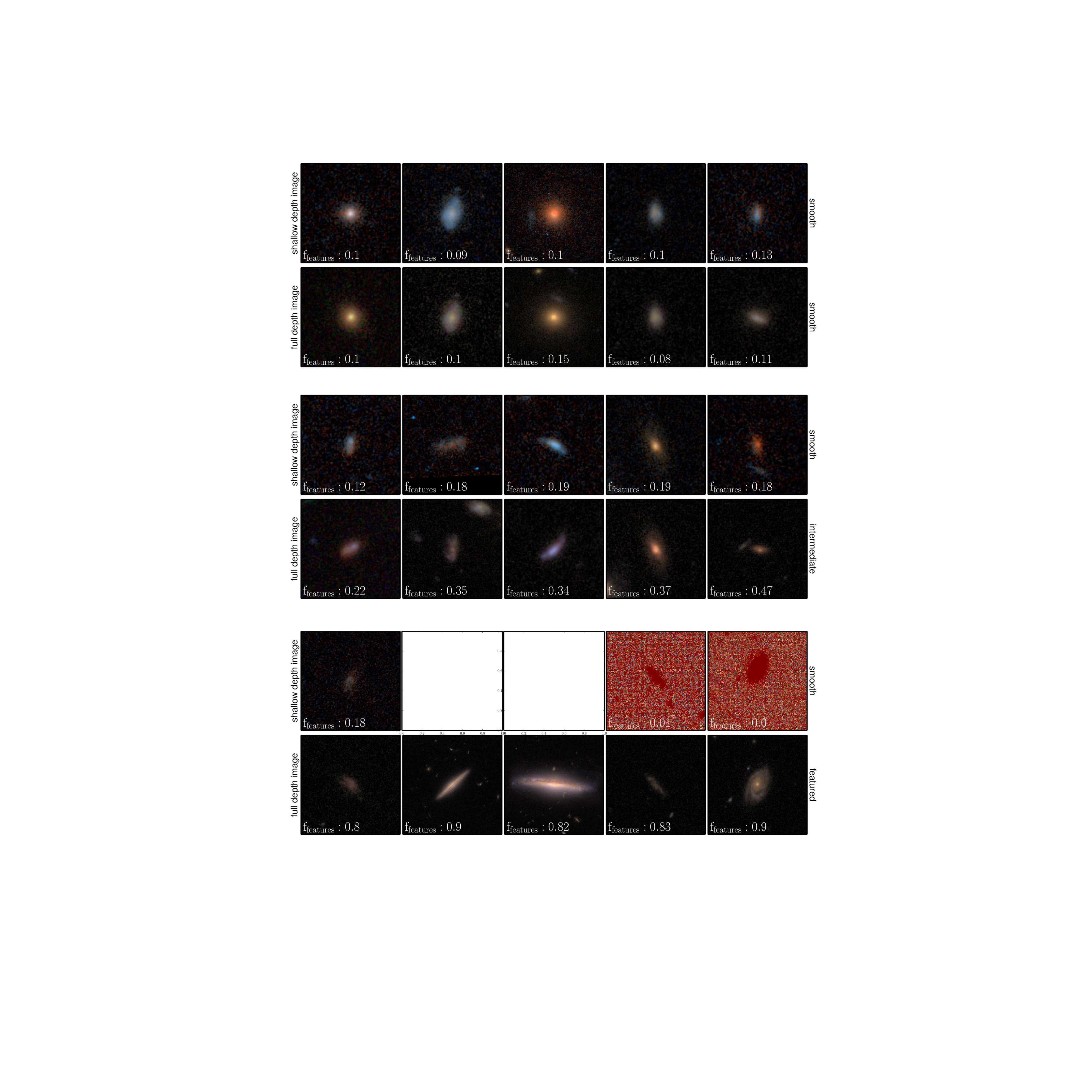}
\caption{Example images of GOODS galaxies that compare GZH morphological
classifications between their 2-epoch and 5-epoch imaging.  The top set
were classified as ``smooth'' in both the 2- and 5-epoch imaging. The
middle set was classified as ``smooth'' in the 2-epoch imaging and as
intermediate between ``smooth'' and ``featured'' in the 5-epoch imaging. The
bottom set was classified as ``smooth'' in the 2-epoch imaging and
``featured'' in the 5-epoch imaging (there are only seven such images
in the sample).}
\label{fig:shallow_smooth}
\end{figure*}

\begin{figure*}
\centering
\includegraphics[width=\textwidth]{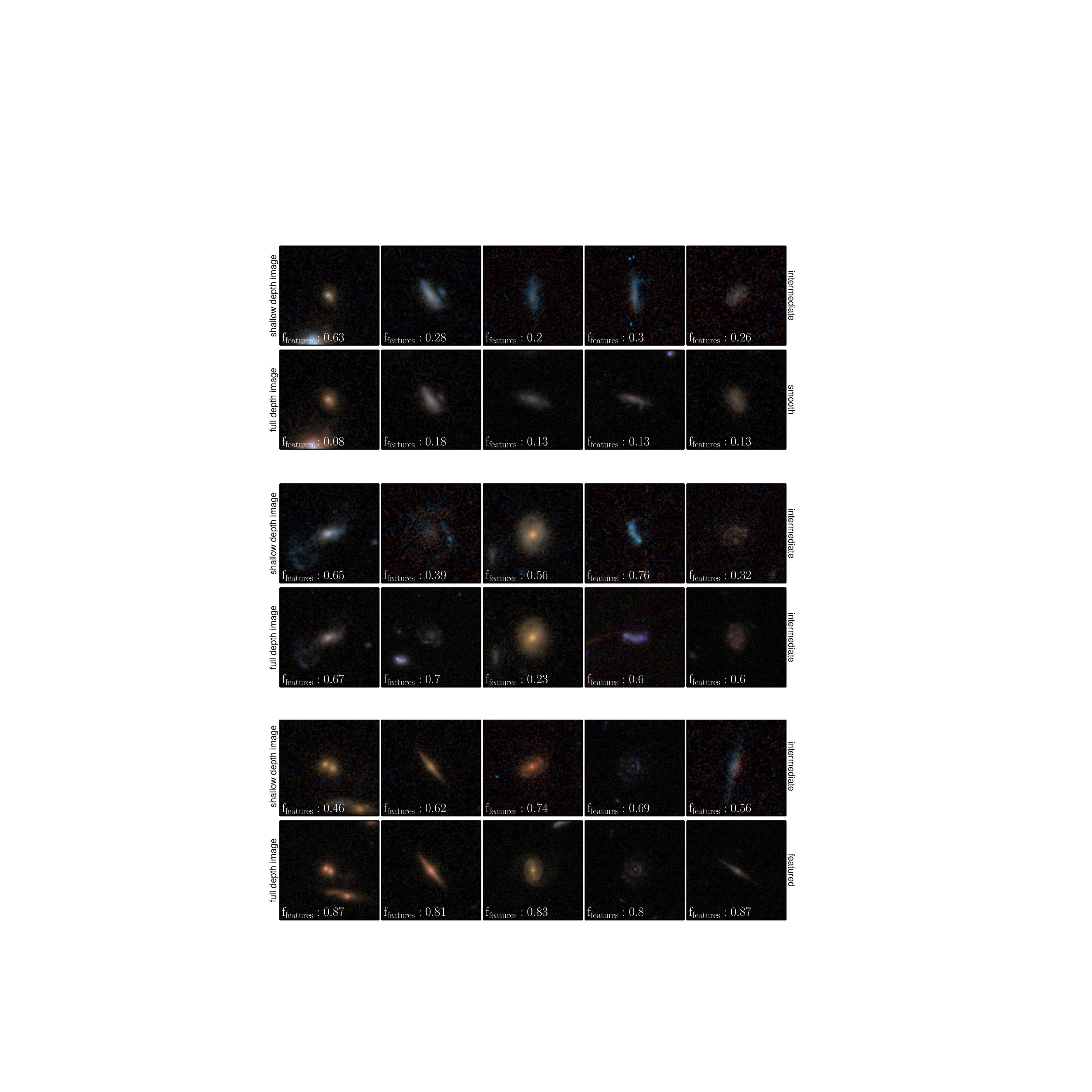}
\caption{Similar to Figure~\ref{fig:shallow_smooth}. The top set (a) were
classified as ``intermediate'' in the 2-epoch imaging and ``smooth'' in the
5-epoch imaging. The middle set (b) was classified as intermediate in both the
2- and 5-epoch imaging. The bottom set (c) was classified as ``intermediate''
in the 2-epoch imaging and ``featured'' in the 5-epoch imaging.}
\label{fig:shallow_intermediate}
\end{figure*}

\begin{figure*}
\centering
\includegraphics[width=\textwidth]{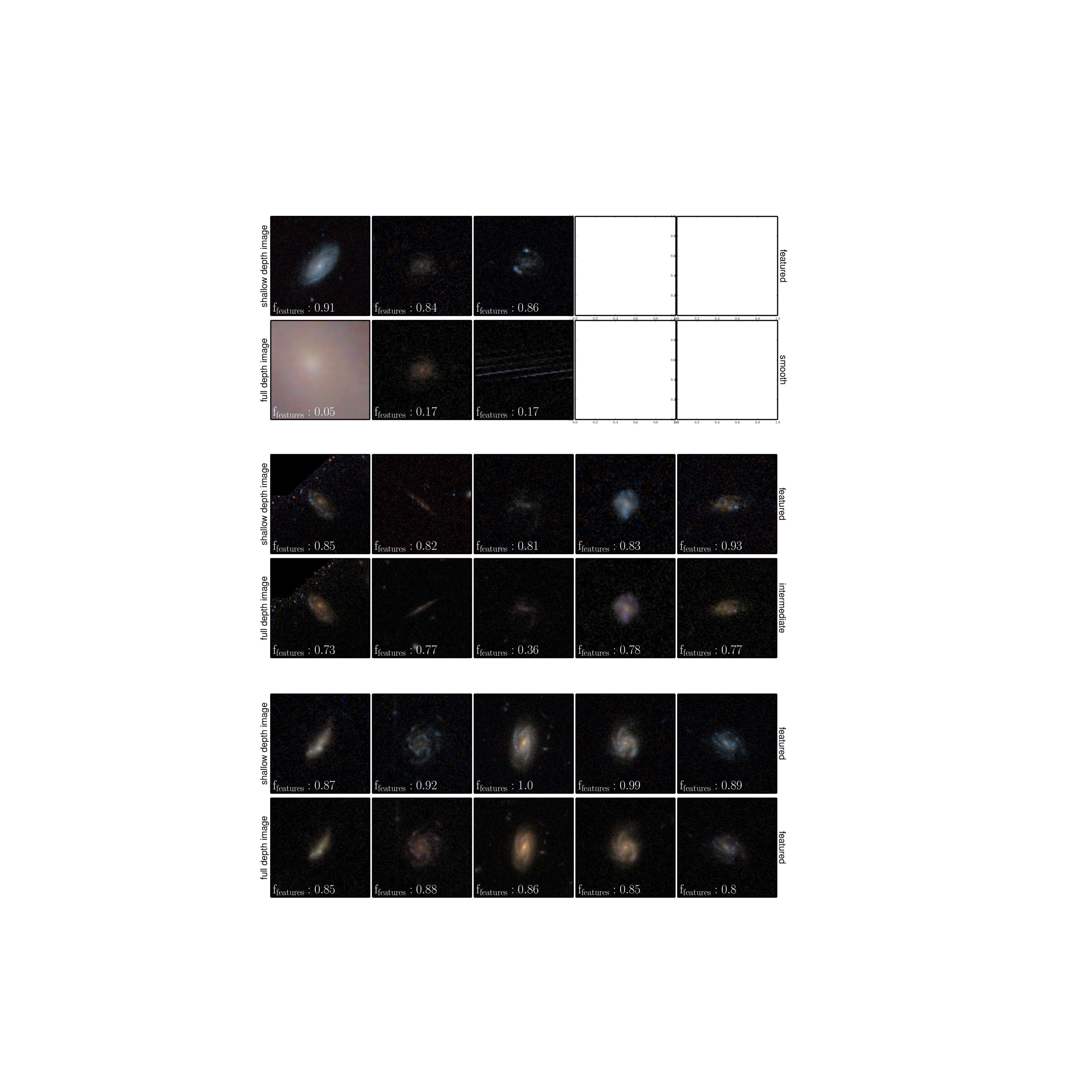}
\caption{Similar to Figure~\ref{fig:shallow_smooth}. The top set (a) were
classified as ``featured'' in the 2-epoch imaging and ``smooth'' in the 5-epoch
imaging (there were only six such images in the sample). The middle set (b) was
classified as ``featured'' in the 2-epoch imaging and ``intermediate'' in the
5-epoch imaging. The bottom set (c) was classified as ``featured'' in both the
2- and 5-epoch imaging.}
\label{fig:shallow_featured}
\end{figure*}

Of the 11,157 galaxies in the GOODS-N and GOODS-S full depth sample, 4,460
are also in the shallow-depth sample. Figure~\ref{fig:shallow_vs_full} shows a
strong correlation between \ffeatures{} for both sets of images. The mean
change in \ffeatures{} from the shallow to full depth images
$f_\mathrm{features,full} - f_{features,shallow} \equiv \Delta f = -0.01$, with
a standard deviation of $\sigma = 0.18$. While there is some variance in
$\Delta f$ in the whole sample, the change is usually small and not often
significant enough to change a morphological classification. Defining a clean
sample of disk galaxies as those with \fbest$>0.8$, elliptical galaxies as
those with $f_\mathrm{smooth,best}<0.2$, and intermediate as those in between,
75\% of the sample would not change morphology. Of the remaining 25\% that
would change morphology, only 0.2\% (9~galaxies)
drastically change morphology either from smooth to featured or vice versa, while the
rest transition to or from the ``intermediate'' morphology. Details can
be seen in Table~\ref{tbl:shallow_to_full_stats} and examples of images
representing the 9 possible changes (or lack of) in morphology are shown in
Figures~\ref{fig:shallow_smooth},\ref{fig:shallow_intermediate}, and
\ref{fig:shallow_featured}.

\begin{table}
\caption{Properties of galaxies whose morphologies changed or stayed the same
in the shallow vs full images. Featured here is defined as \fbest$>0.8$,
intermediate = $0.2<$\fbest$<0.8$, smooth = $f_\mathrm{smooth,best}<0.2$.
\label{tbl:shallow_to_full_stats}}
\begin{tabular}{lrrrrrrrr}
\hline\hline
                              & $N$     & \%       & $<\Delta f>$ & $<z>$   \\
\hline
smooth to smooth              & 708     & 15.9     &    0.00      &  0.72   \\
smooth to intermediate        & 346     & 7.8      &    0.21      &  0.70   \\
smooth to featured            & 6       & 0.1      &    0.80      &  0.45   \\ 
intermediate to smooth        & 266     & 6.0      & $-$0.24      &  0.66   \\
intermediate to intermediate  & 2,363   & 53.0     &    0.00      &  0.75   \\
intermediate to featured      & 121     & 2.7      &    0.17      &  0.82   \\
featured to smooth            & 3       & 0.1      & $-$0.73      &  0.71   \\
featured to intermediate      & 370     & 8.3      & $-$0.15      &  0.69   \\
featured to featured          & 277     & 6.2      & $-$0.05      &  0.71   \\
\hline
Total                         & 4,460   & 100      &              &         \\
\hline\hline
\end{tabular}
\end{table}

\bsp	% typesetting comment
\label{lastpage}
\end{document}